\documentclass{aa}  
\usepackage{natbib}     
\usepackage{graphicx}
\usepackage{txfonts}
\usepackage{lipsum}

\bibpunct{(}{)}{;}{a}{}{,}  
\usepackage{subcaption}        
\usepackage{lscape}            
\usepackage{placeins}           
\usepackage[colorlinks=true, citecolor=blue, linkcolor=blue, urlcolor=blue]{hyperref}
\usepackage{adjustbox}

\begin{document}

\title{Tidal braking does not dominate open-cluster slow rotators}

   \author{Zixun Zhan\inst{1,2}       
        \and Chenyu He\inst{1,2}
        \and Weijia Sun\inst{3}
        \and Chengyuan Li\inst{1,2}\corrauth{lichengy5@mail.sysu.edu.cn} 
       }

   \institute{School of Physics and Astronomy, Sun Yat-sen University, Zhuhai, 519082, China
   \and CSST Science Center for the Guangdong–Hong Kong–Macau Greater Bay Area, Zhuhai, 519082, China
   \and Leibniz-Institut für Astrophysik Potsdam (AIP), An der Sternwarte 16, 14482 Potsdam, Germany}

   \date{Received September 30, 20XX}

\abstract
{Extended main-sequence turn-offs (eMSTOs) observed in open clusters younger than $2~\mathrm{Gyr}$ are commonly attributed to a spread in stellar rotation rates. Tidal synchronization in close binary systems has been proposed as a possible mechanism for producing slowly rotating stars. To test this hypothesis, we measured the radial-velocity (RV) dispersions of slow-rotator samples associated with the eMSTOs of eight open clusters using ESO VLT/GIRAFFE and SALT/RSS spectroscopy. We extended the analysis to 40 open clusters using Gaia DR3 RV measurements. The observed cluster RV dispersions are generally substantially smaller than those predicted for binary populations dominated by close systems with orbital periods for which tidal synchronization is expected to be effective. In particular, all 40 clusters in the Gaia sample have RV dispersions below the prediction of the $P=10$~d binary-population model. These results suggest that close, tidally synchronized binaries are unlikely to be the dominant origin of slow-rotator populations in young and intermediate-age open clusters, although a contribution from 
undetected or wider binary systems cannot be excluded.}


   \maketitle\nolinenumbers

\section{Introduction}

    Extended main-sequence turn-offs (eMSTOs) and split or extended main sequences (MSs) have been widely observed in star clusters younger than $\sim$2 Gyr in both the Milky Way and the Magellanic Clouds (MCs) \citep{2008ApJMackey,2009AJGoudfrooij,2011ApJGoudfrooij,2018MNRASMilone,2018ApJCordoni,2024MNRASCordoni,2019ApJLi}. These features initially challenged the classical picture that star clusters can be described as simple stellar populations with a single age and chemical composition, and are now widely attributed primarily to stellar rotation. 
    \citet{2009MNRASBastian} first proposed that stellar rotation could drive these features, a scenario subsequently supported by photometric studies \citep{2018ApJCordoni,2018MNRASMilone}. Spectroscopic observations then confirmed that stars located in the eMSTO and along split MSs exhibit different projected rotational velocities (e.g. \citealt{2017ApJDupree,2018AJMarinoA,2018MNRASBastian,2019ApJSunA,2019ApJSunB}). 
    In particular, the blue MS (bMS) is generally populated by slowly rotating stars, whereas the red MS (rMS) is dominated by rapidly rotating stars. Stellar rotation is therefore considered a key ingredient in interpreting eMSTOs and split MSs in young and intermediate-age star clusters.

    Early-type stars span a range of rotation rates, so the presence of some slow rotators is not by itself surprising \citep{2012A&AZorec}. The key puzzle is instead that some young clusters exhibit very broad or even bimodal rotational-velocity distributions, in which slow rotators form a substantial, and sometimes distinct, component rather than merely the low-velocity tail of a unimodal population. Spectroscopy has revealed distinct rotational distributions along the split main sequences of NGC~1818 and NGC~1850 \citep{2018AJMarinoA,2021MNRASKamann}, while the split main sequence of NGC~1856 has likewise been interpreted as the coexistence of slow- and fast-rotating populations \citep{2015MNRASMilone,2015MNRASDAntona}. A similar rotational dichotomy has also been identified in the Galactic open cluster NGC~2287 \citep{2019ApJSunB}. Because these stars lie above the Kraft break, ordinary magnetic braking during their main-sequence evolution is not expected to produce such a prominent slow-rotator component \citep{1967ApJKraft}. Whether this slow-rotating population is primordial or produced by subsequent angular momentum evolution remains unclear. Proposed explanations include prolonged disk locking before the main sequence \citep{2020MNRASBastian} and magnetic braking following binary mergers \citep{2022NatAsWang}.

    A third possibility, and the focus of this work, is that the slowly rotating population is produced by tidal interactions in binary systems. \citet{2015MNRASDAntona} suggested that bMS stars may have been initially born as rapid rotators and subsequently spun down by binary companions. In this scenario, tidal interactions can circularize the binary orbit and synchronize the stellar rotation with the orbital motion, thereby reducing the rotation rate of the primary star. Although standard tidal theories predict efficient synchronization mainly in very close binaries, observations indicate that some binaries with relatively wide separations can also have circularized orbits \citep{2004Abt}. \citet{2015MNRASDAntona} therefore argued that tidally interacting binaries with orbital periods of about 4--500 days may contribute to the formation of slow rotators. This idea was further developed by \citet{2017NatAsDAntona}, who proposed that a substantial fraction of bMS stars could be rapidly rotating stars braked by close or intermediate-period companions. Testing this tidal-braking scenario is important not only for understanding the origin of bMS stars, but also for assessing whether binary interactions play a broader role in shaping peculiar stellar populations in star clusters. More importantly, the scenario makes a direct observational prediction: if slow rotators are produced by tidal braking in binaries, they should contain a large fraction of close or intermediate-period companions and therefore exhibit enhanced radial-velocity variations.

    Several studies have attempted to test the tidal-braking scenario in individual star clusters. \citet{2020MNRASKamann,2021MNRASKamann} found no significant difference in the binary fraction between bMS and rMS stars in NGC 1856 and NGC 1850. \citet{2023MNRASHe} measured radial-velocity (RV) variations of slow rotators in NGC 2422 and found that most of them are unlikely to be short-period binaries. \citet{2024MNRASMaurya,2025ApJMaurya} further showed that slow rotators in NGC 2355 and NGC 6067 are not preferentially concentrated toward the cluster centres. This provides another test of the tidal-braking scenario, because bMS stars produced through tidal braking are expected to reside in relatively massive binary systems; in dynamically evolved clusters, such systems should be more centrally concentrated as a result of mass segregation. The lack of such a central concentration therefore disfavors the expectation of the tidal-braking scenario. \citet{2026ApJRao&Chen} investigated 53 Galactic open clusters, measuring the slow-rotator fractions of their MSTO stars using $v\sin i$ from Gaia-ESO spectroscopy and comparing them with the known binary populations of the same clusters.
    Based on 52 known binaries, they found no statistically significant correlation between slow rotators and the binary fraction. However, most of these binaries are eclipsing systems identified in Gaia DR3, which are intrinsically difficult to detect in large numbers and therefore constitute only a limited binary sample. A large-sample test of whether slow rotators are preferentially associated with tidally interacting binaries therefore remains lacking. 
    
    To address this gap, we investigate the slow-rotator populations in 40 Galactic open clusters using radial velocities from Gaia Data Release 3 (Gaia DR3), supplemented by ground-based spectroscopy for eight benchmark clusters. If tidal braking were the dominant formation channel, these populations should contain a large fraction of close binaries and consequently exhibit enhanced RV dispersions. We test this prediction by comparing the observed RV distributions with those expected for binary populations spanning a range of orbital periods. The observed dispersions are generally much smaller than predicted for close binaries capable of efficient tidal synchronization, indicating that tidal braking is unlikely to be the dominant origin of slow rotators in eMSTOs and split main sequences.

    This paper is organized as follows. In Section~\ref{section 2}, we presents the ground-based observations and the RV dispersion measurements of the eight benchmark clusters. In Section~\ref{section 3}, we extends the analysis to 40 open clusters with Gaia DR3, including the validation of the Gaia RVs and the binary-population simulations. In Section~\ref{section 4}, we discuss the implications of our results. Finally, we summarize our conclusions in Section~\ref{section 5}.

\section{Ground-based observations and RV dispersions}
\label{section 2}
\subsection{Data sets}
\label{subsection:data}

Our analysis is based on two complementary RV data sets. The first is a ground-based spectroscopic sample of eight Galactic open clusters observed with the Very Large Telescope (VLT) and the Southern African Large Telescope 
(SALT). The second data set consists of Gaia DR3 Radial Velocity Spectrometer (RVS) measurements 
for 40 Galactic open clusters. The two samples are not independent cluster samples; rather, the ground-based data are used as a benchmark for the Gaia-based RV analysis. After this assessment, we use the Gaia DR3 data to investigate the RV dispersions of blue-edge sample in the full cluster sample.

The ground-based spectra were obtained with two spectroscopic facilities. The VLT observations were taken with the GIRAFFE spectrograph, while the SALT observations were carried out with the Robert Stobie Spectrograph (RSS) in multi-object spectroscopy (MOS) mode. The programme IDs and the main absorption 
lines used for the RV measurements are listed in Table~\ref{fitlines}. For the SALT spectra, RSS was used with the PG2300 grating at a grating angle 
of $34.25^\circ$ and a camera station angle of $68.5^\circ$. With a 1-arcsec slit, this setup gives a central wavelength of 4884.4~\AA, covering approximately 
4345--5373~\AA\ at a spectral resolution of $R \sim 4000$.

\subsection{Cluster parameters and member-star selection}
\label{subsection:cluster_selection}

Cluster members were selected from the catalogue of \citet{Hunt23list}, 
which is based on Gaia DR3 astrometry and photometry \citep{GaiaDR3}. For each cluster, we constructed a colour--magnitude diagram (CMD) using the Gaia photometry of the selected member stars. The cluster ages and distances were then determined by fitting the CMDs with PARSEC version 2.0 
isochrones \citep{2022A&APARSEC}, adopting an age step of 10 Myr. We adopted solar metallicity, i.e. $[\mathrm{Fe/H}]=0.0$ dex, for all clusters in the isochrone fitting. 

To focus on the stellar population relevant to the eMSTO and split-MS phenomenon, we selected stars with masses above $1.5\,M_\odot$. This mass threshold lies above the Kraft break \citep{1967ApJKraft}, where stars are 
expected to have inefficient magnetic braking and can therefore retain information about their rotational histories. This selection also minimizes contamination from lower-mass stars whose rotation may be substantially affected by magnetic braking.
The resulting ages and distances, as well as the RV dispersions of the ground-based benchmark sample are summarized in Table~\ref{cluster param table}.

    \begin{figure*}[]
        \centering
        \includegraphics[width=0.33\hsize]{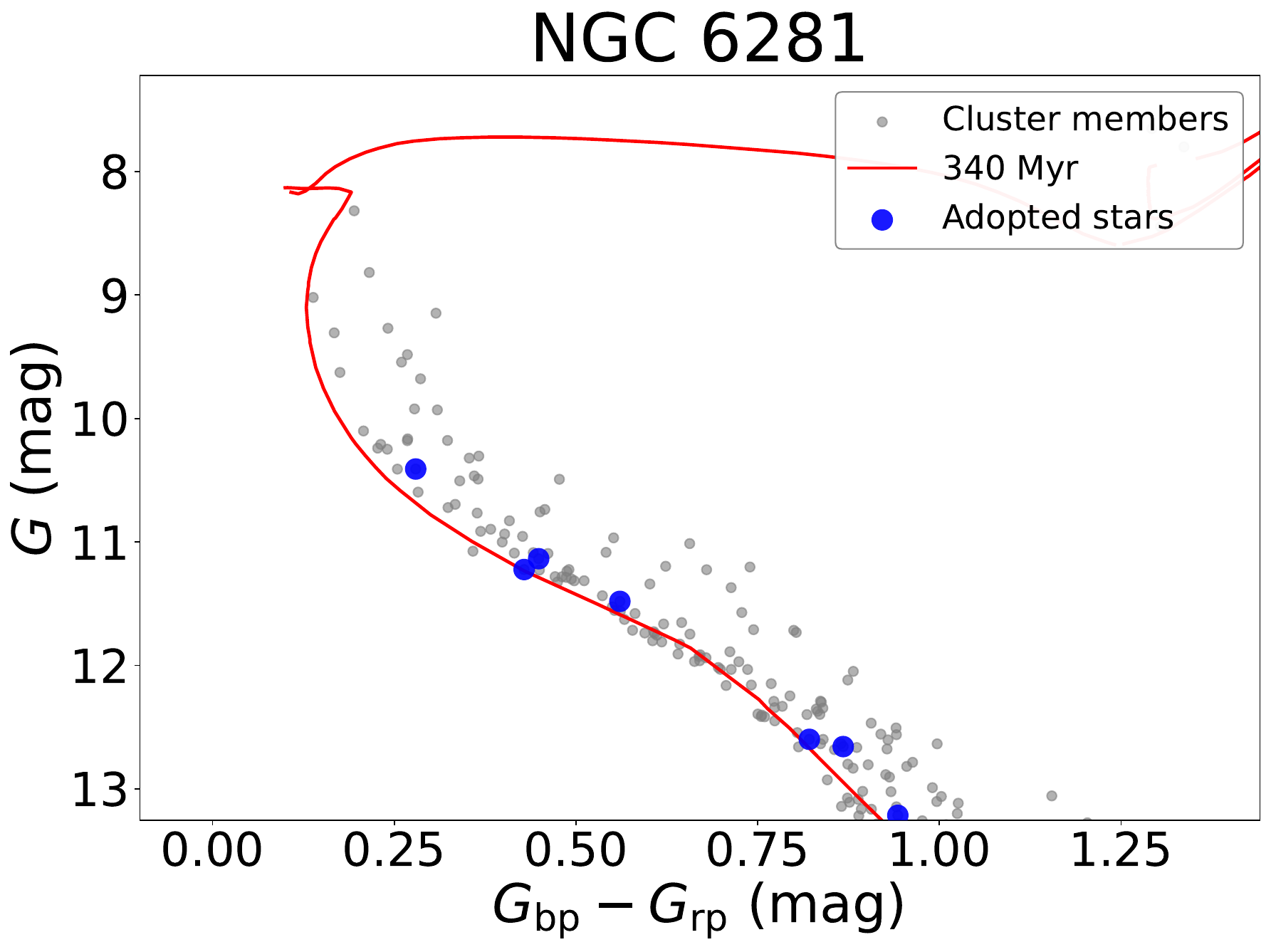}
        \includegraphics[width=0.33\hsize]{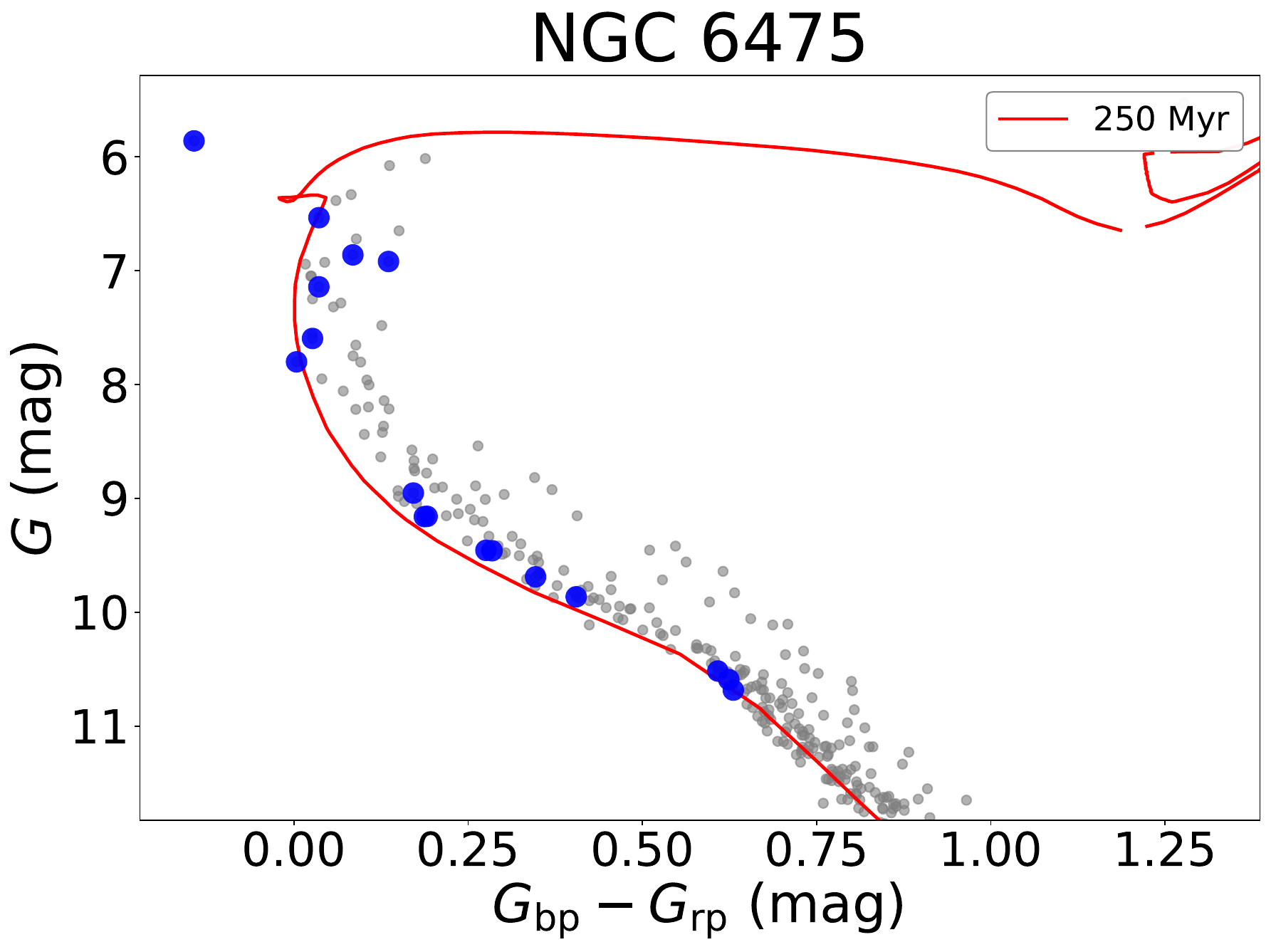}
        \includegraphics[width=0.33\hsize]{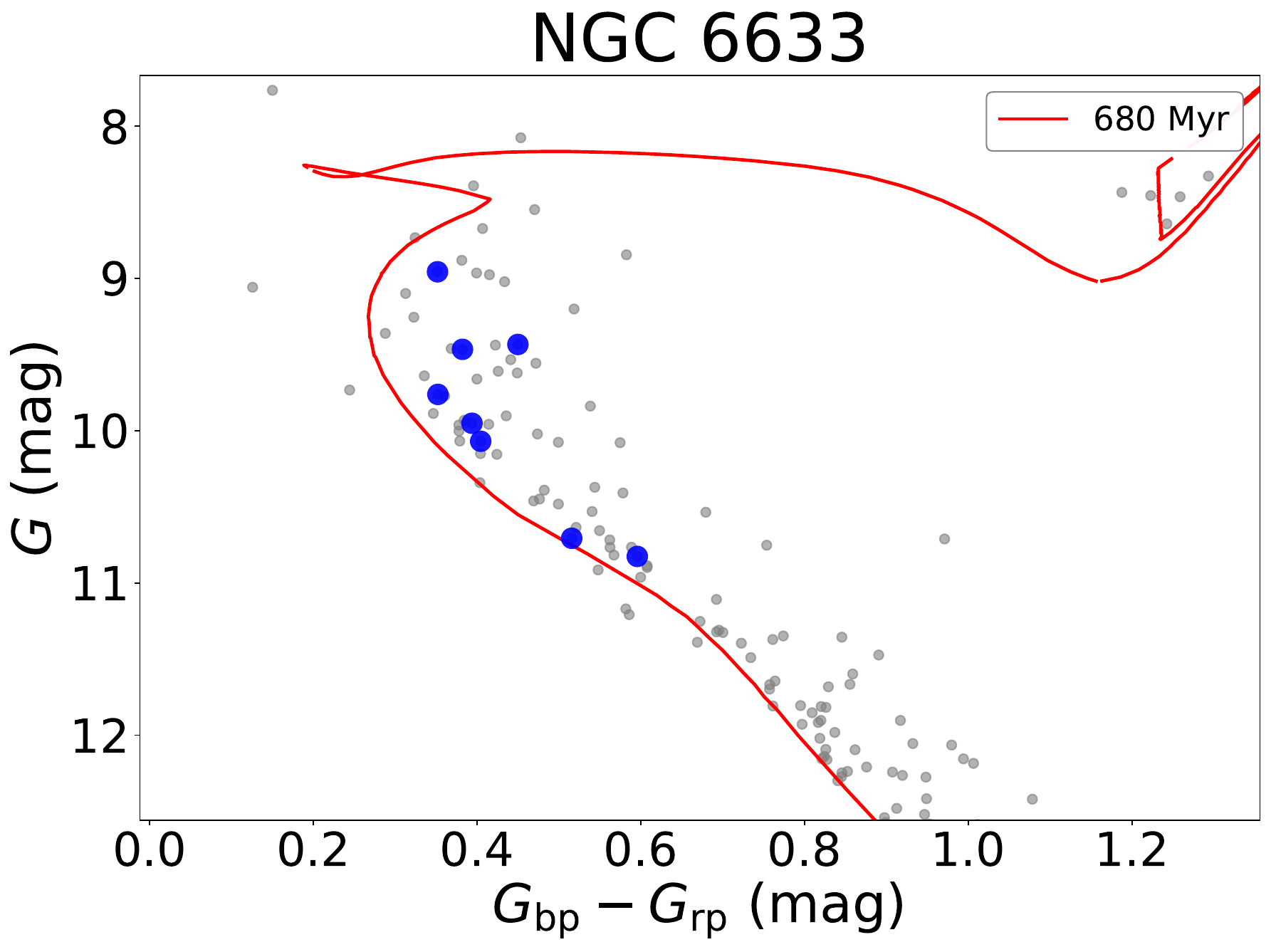}
        \includegraphics[width=0.33\hsize]{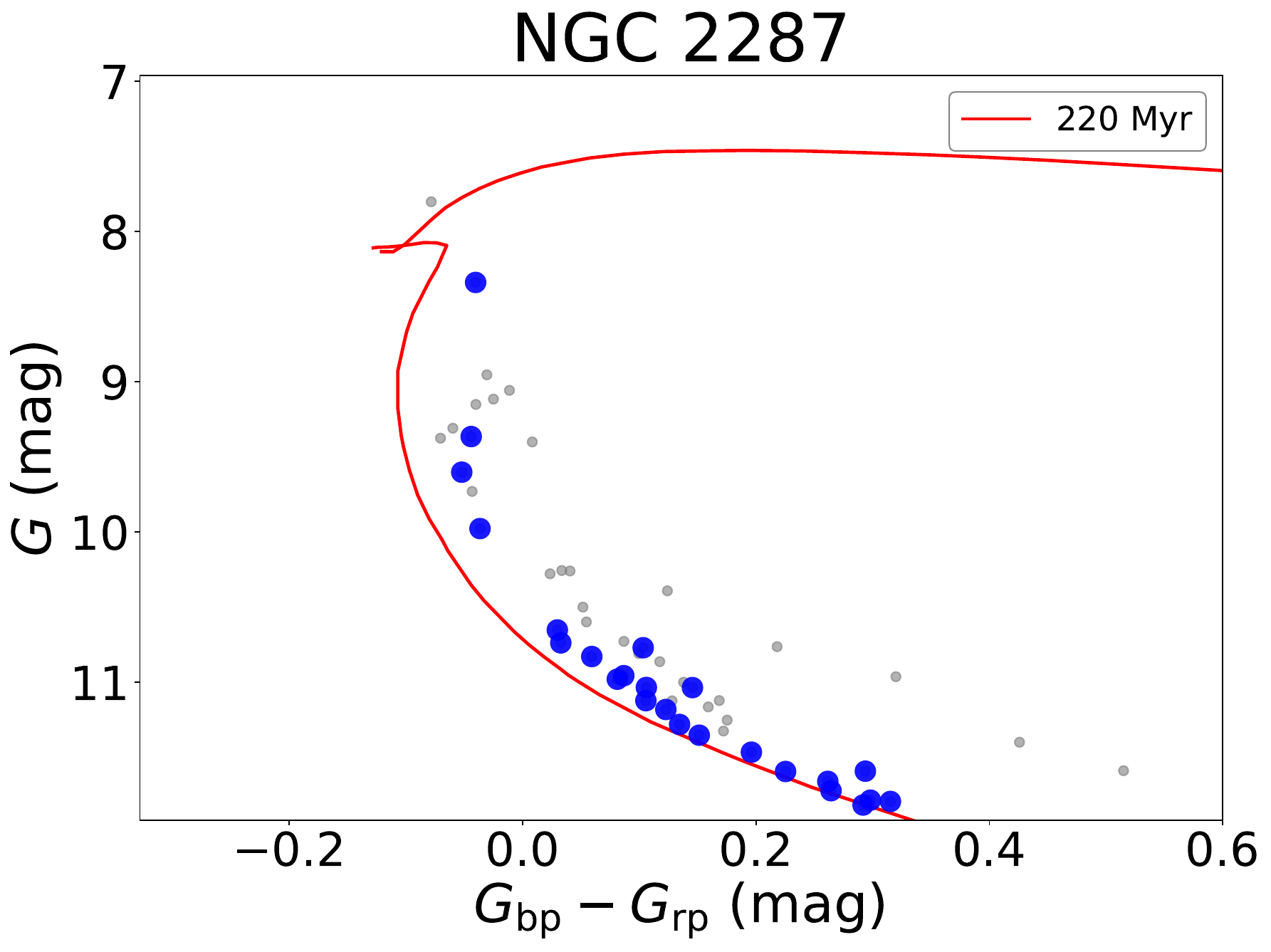}
        \includegraphics[width=0.33\hsize]{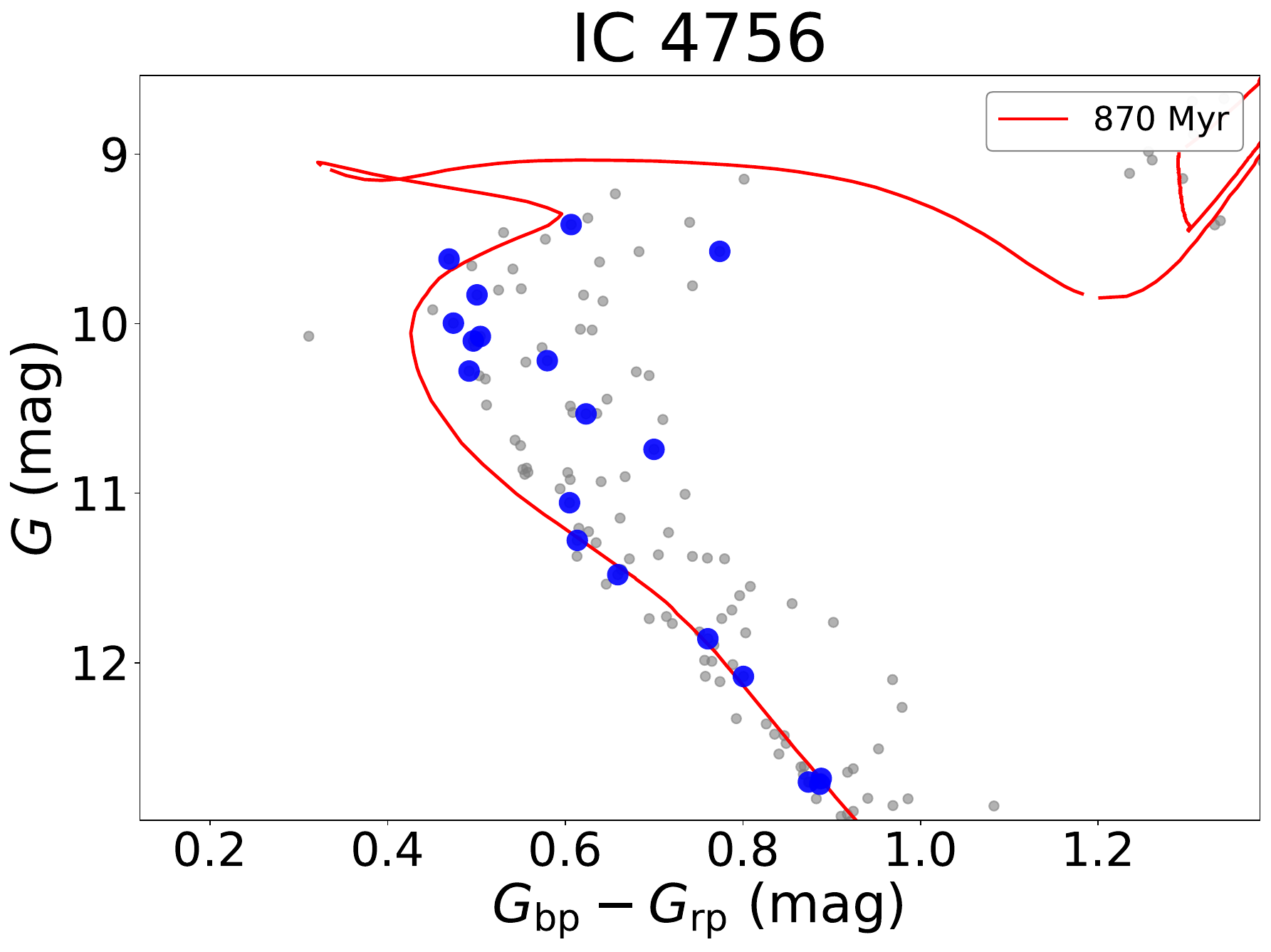}
        \includegraphics[width=0.33\hsize]{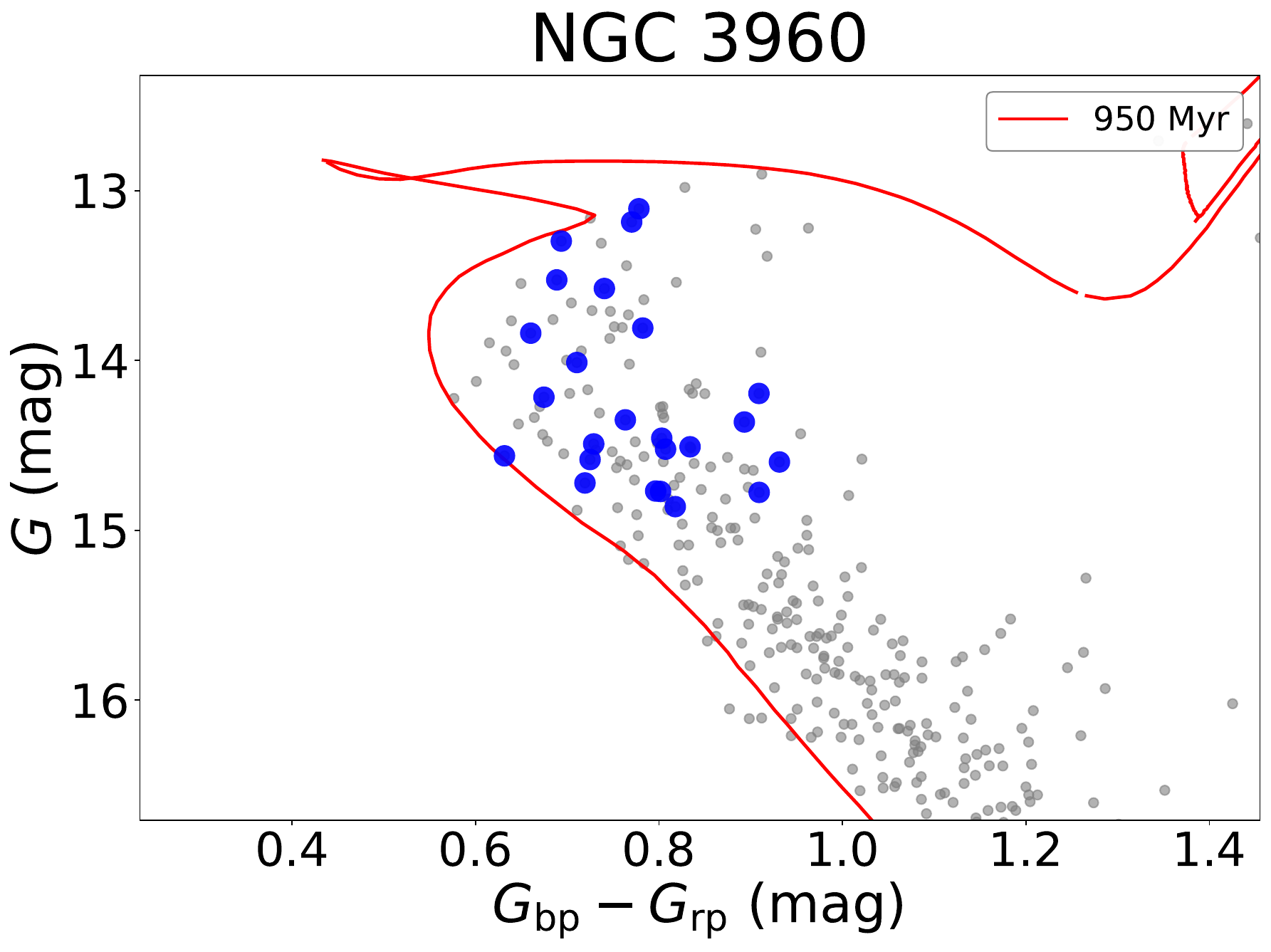}
        \includegraphics[width=0.33\hsize]{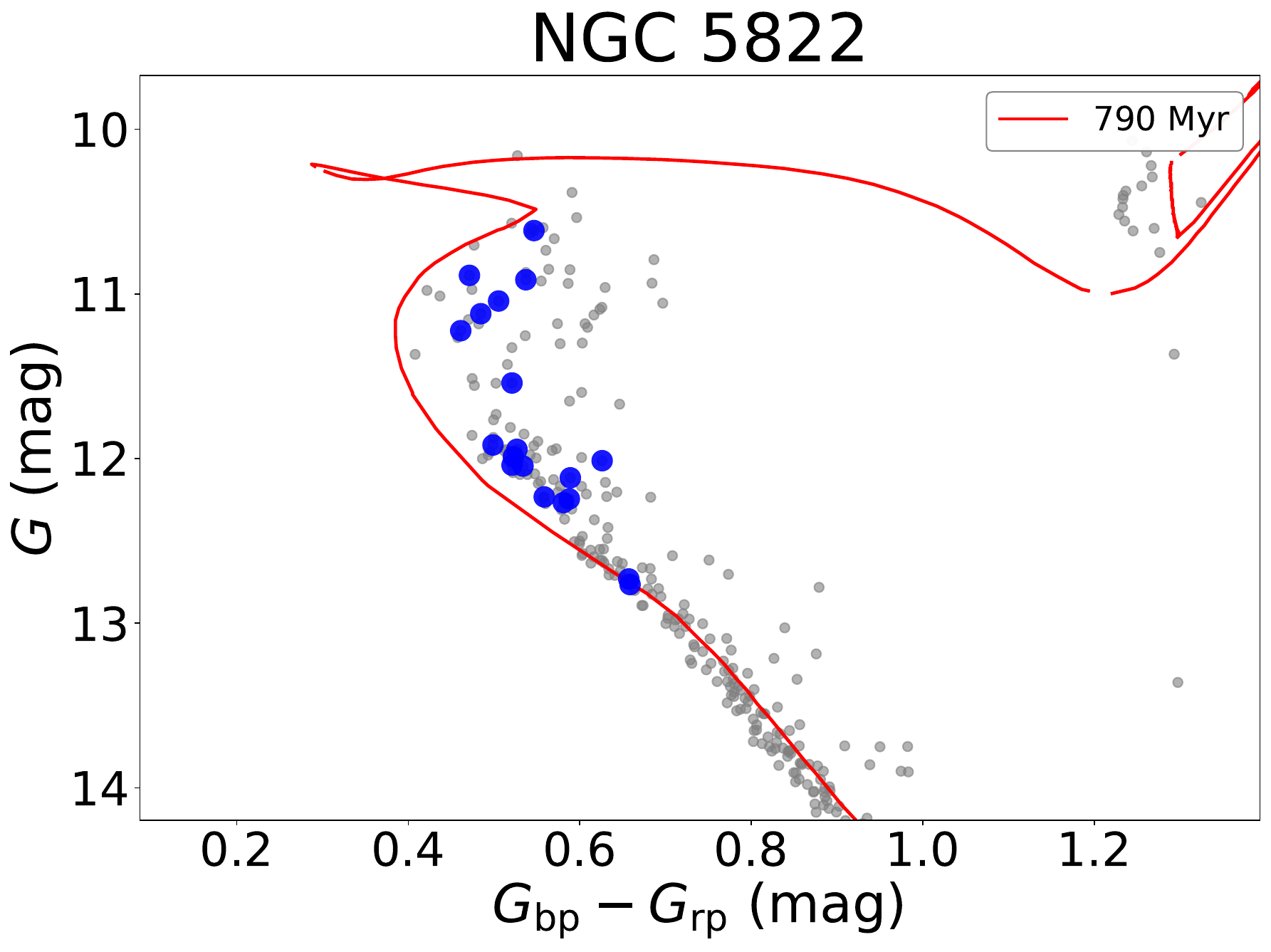}
        \includegraphics[width=0.33\hsize]{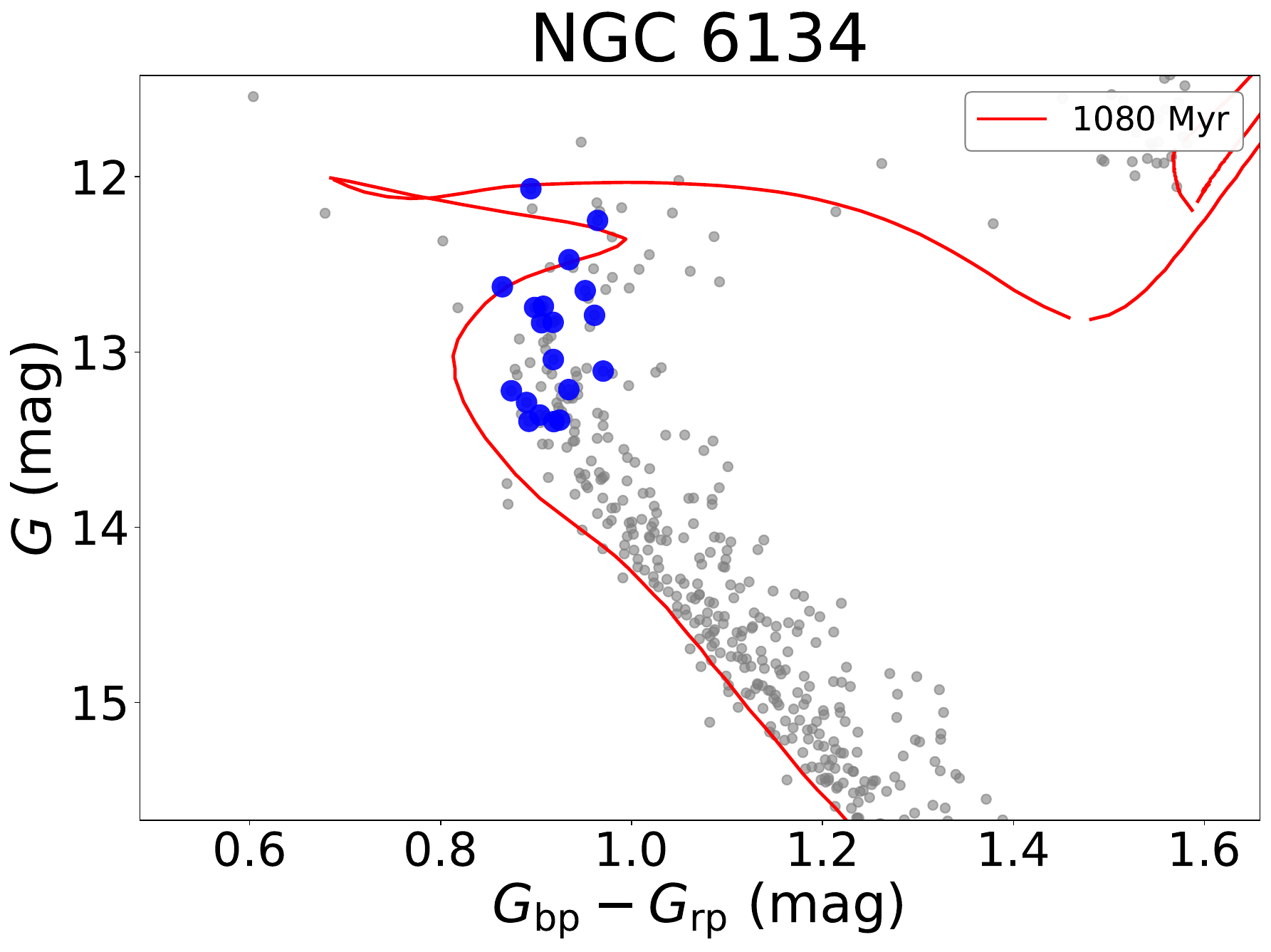}

        \caption{Gaia DR3 colour--magnitude diagrams and isochrone fits for the eight ground-based benchmark clusters observed with VLT/GIRAFFE or SALT/RSS. Grey dots denote cluster members, while blue dots mark the stars adopted for the ground-based spectroscopic analysis. The red lines show the best-fitting PARSEC v2.0 isochrones at the labelled ages, assuming solar metallicity ($[\mathrm{Fe/H}]=0.0$).}
      
        \label{HRdiagram}
    \end{figure*}

\subsection{Radial velocities and projected rotational velocities}
\label{subsection:rv_vsini_measurement}

For the ground-based spectra, we measured the RVs and projected rotational velocities ($v\sin i$) by fitting the observed absorption-line 
profiles with synthetic spectra. The synthetic spectra were taken from the Pollux database \citep{POLLUX2010} and were computed with the SYNSPEC code \citep{Hubeny1992SYNSPEC}, using plane-parallel ATLAS12 model atmospheres under the assumption of local thermodynamic equilibrium \citep{Kurucz2005ATLAS12}. 
A fixed microturbulent velocity of $2~\mathrm{km~s^{-1}}$ was adopted.

The synthetic spectral grid covers effective temperatures from $6000$ to $15\,000$~K in steps of $100$~K, surface gravities from $\log g=3.5$ to $5.0$ in steps of $0.1$ dex, and metallicities from $[\mathrm{Fe/H}]=-1.0$ to $1.0$ dex in steps of $0.5$ dex. In the fitting, we adopted solar metallicity, $[\mathrm{Fe/H}]=0.0$ dex, consistent with the isochrone fitting described above. The synthetic spectra were convolved with instrumental and rotational broadening using PyAstronomy \citep{2019asclPyAstonomy}. The instrumental broadening was set according to the spectral resolution of the observations, and the $v\sin i$ grid was sampled in steps of $5~\mathrm{km~s^{-1}}$.

For the RV measurement, the broadened model spectra were shifted over a velocity range from $-70$ to $50~\mathrm{km~s^{-1}}$ in steps of $1~\mathrm{km~s^{-1}}$, which is broad enough to cover the expected cluster and stellar RVs. The model fluxes were interpolated onto the observed wavelength grid using PySynphot \citep{PySynphot2013}. The best-fitting 
parameters were then determined on a grid of effective temperature, surface gravity, $v\sin i$, and RV, with $[\mathrm{Fe/H}]$ fixed at solar metallicity.

\begin{figure}[!ht]
        \centering
        \includegraphics[width=0.9\hsize]{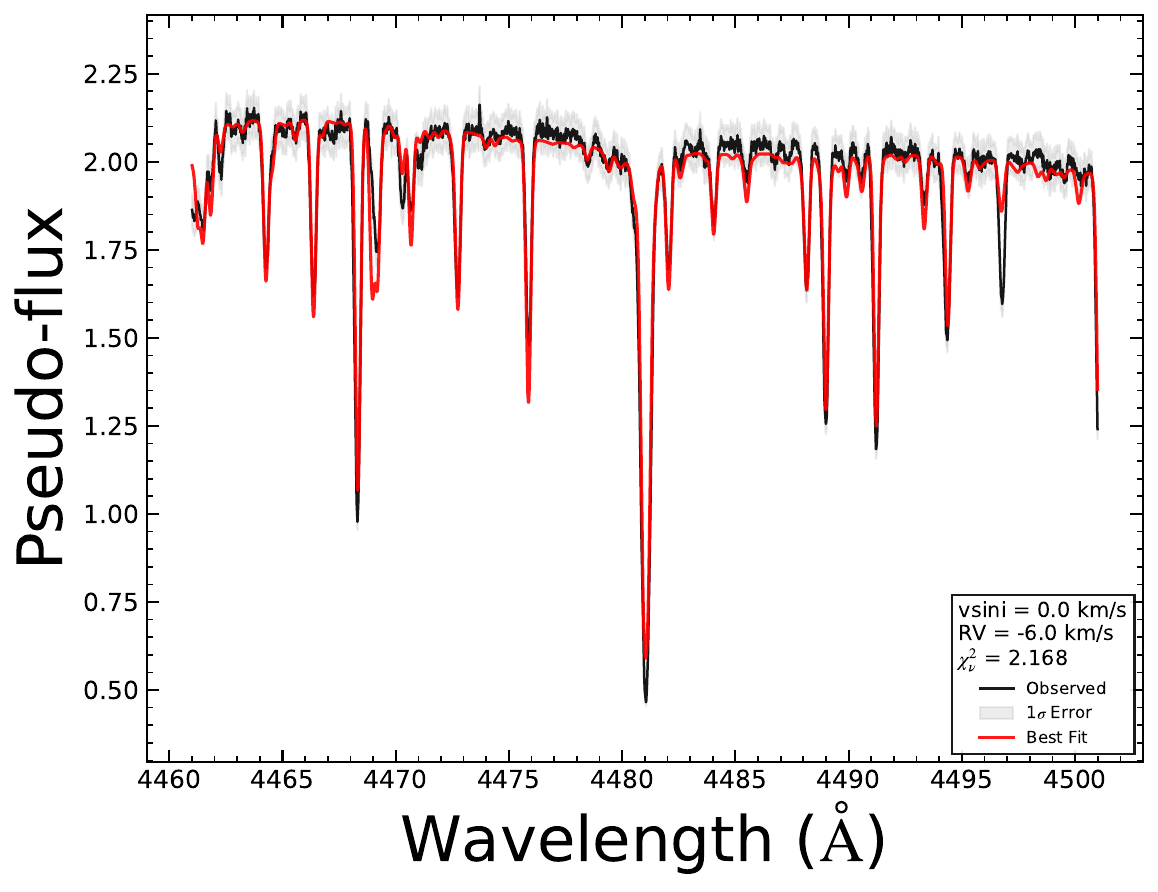}
        \includegraphics[width=0.9\hsize]{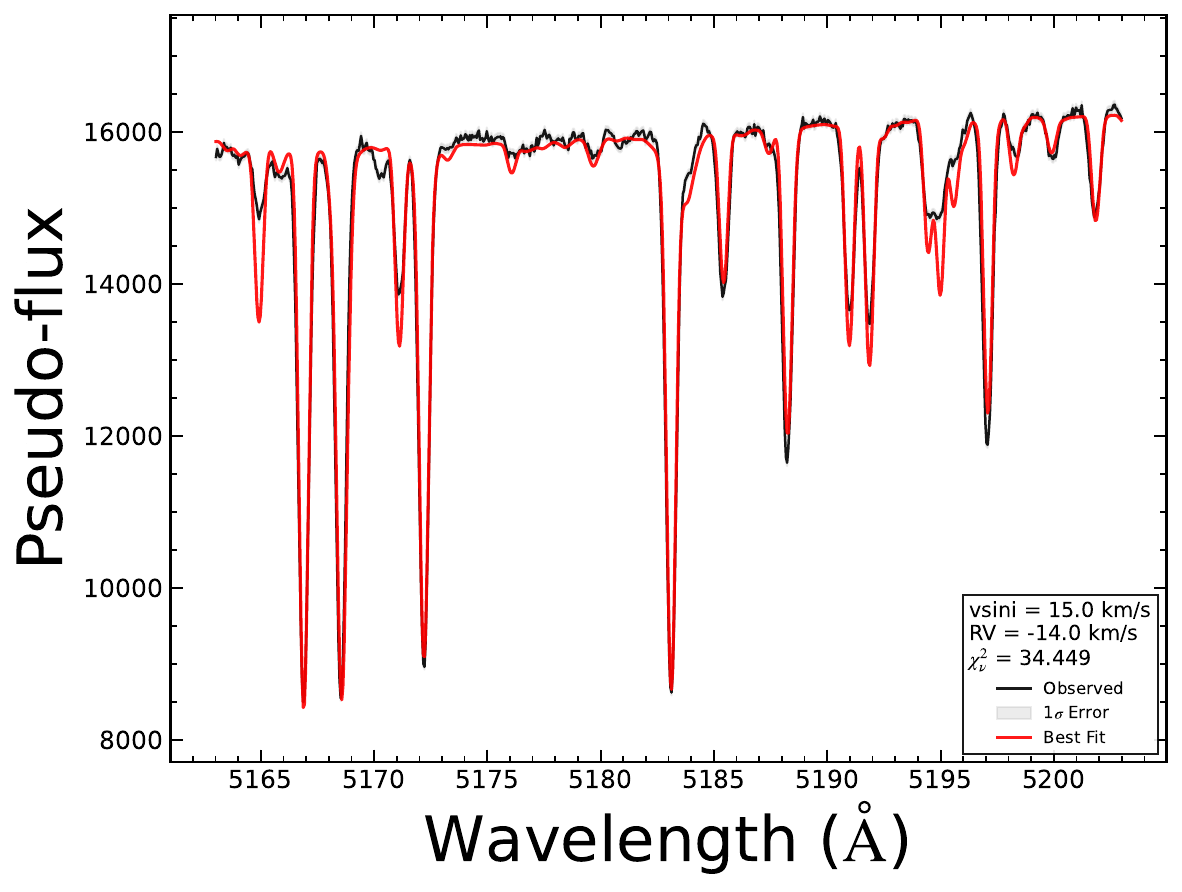}
        \caption{Observed absorption profile (black line) for Gaia ID 4477223820897301248 (left) and 4477212005453986688 (right), with their best-fitting model(red line).
        The gray regions are the 1$\sigma$ range.}
        \label{spectra_eg}
    \end{figure}

The fitting was performed over 40~\AA\ wavelength intervals centred on the absorption-line regions, in order to include multiple absorption features and improve the constraints on both RV and $v\sin i$. The specific absorption lines used for each cluster are listed in Table~\ref{fitlines}. For each observed spectrum, the best-fitting model was determined by minimizing $\chi^2$ between the observed and model line profiles. We retained measurements only when at least three absorption lines with signal-to-noise ratios $\mathrm{S/N}>20$ were available. The resulting RV and $v\sin i$ measurements were then adopted for the subsequent analysis. An example of the spectral fitting is shown in Figure~\ref{spectra_eg}.

\begin{table*}[ht!]
\caption{Ground-based spectroscopic sample and absorption lines used for the RV measurements}
\label{fitlines}
\centering
\small
\begin{tabular}{l l c p{7.0cm} c}
\hline\hline
Cluster & Instrument & Prog. & Fitted absorption lines & Ref. \\
\hline
NGC 6281 
         & VLT/GIRAFFE 
         & P3, P4 
         & Ca~II $\lambda\lambda3934,3968$, Fe~I $\lambda\lambda4025$, 
           Mg~I $\lambda\lambda5167,5173,5184$, Mg~II $\lambda\lambda4481$ 
         & [1] \\

NGC 6475 
         & VLT/GIRAFFE 
         & P4, P5 
         & Ca~II $\lambda\lambda3934,3968$, Fe~I $\lambda\lambda4025$, 
           Mg~I $\lambda\lambda5167,5173,5184$, Mg~II $\lambda\lambda4481$
         & [1] \\
NGC 6633 
         & VLT/GIRAFFE 
         & P3, P6 
         & Mg~I $\lambda\lambda5167,5173,5184$, Mg~II $\lambda\lambda4481$
         & [1] \\
NGC 2287 
         & VLT/GIRAFFE 
         & P2 
         & Mg~I $\lambda\lambda5167,5173,5184$ 
         & [2] \\

IC 4756  
         & SALT/RSS 
         & P1 
         & H$\beta$ $\lambda4861$, Mg~I $\lambda\lambda5167,5173,5184$ 
         & [3] \\

NGC 3960 
         & SALT/RSS 
         & P1 
         & H$\beta$ $\lambda4861$, Mg~I $\lambda\lambda5167,5173,5184$ 
         & [3] \\

NGC 5822 
         & SALT/RSS 
         & P1 
         & H$\beta$ $\lambda4861$, Mg~I $\lambda\lambda5167,5173,5184$ 
         & [3] \\

NGC 6134 
         & SALT/RSS 
         & P1 
         & H$\beta$ $\lambda4861$, Mg~I $\lambda\lambda5167,5173,5184$ 
         & [3] \\

\hline
\end{tabular}
\tablefoot{
Programme IDs: 
P1 = 2017-2-SCI-038, 2018-1-SCI-006, 2018-2-SCI-002; 
P2 = 380.D-0161; 
P3 = 193.B-0936; 
P4 = 095.C-0818; 
P5 = 71.D-0127; 
P6 = 079.D-0178. 
References: [1] This work; [2] \citet{2020IAUSSun}; [3] \citet{2021MNRASSun}.
}
\end{table*}

\begin{table*}[ht!]
\caption{Cluster parameters and RV dispersions of the ground-based benchmark sample}
\label{cluster param table}
\centering
\small
\begin{tabular}{l c c c c}
\hline\hline
Cluster & Age & Distance & Mean RV & $\sigma_{\rm RV,GB}$ \\
        & (Myr) & (pc) & (km~s$^{-1}$) & (km~s$^{-1}$) \\
\hline
NGC 6281 & $340\pm30$ & $540\pm18$   & $-7.62\pm0.62$  & $0.46\pm0.11$  \\
NGC 6475 & $250\pm32$ & $275\pm6$    & $-15.08\pm0.25$ & $2.92\pm0.52$  \\
NGC 6633 & $680\pm43$ & $389\pm12$   & $-27.87\pm0.43$ & $6.33\pm1.69$  \\
NGC 2287 & $220\pm25$ & $734\pm38$   & $22.78\pm0.28$  & $1.79\pm0.29$  \\
IC 4756  & $870\pm45$ & $429\pm25$   & $-24.95\pm0.49$ & $14.69\pm2.45$ \\
NGC 3960 & $950\pm54$ & $1862\pm202$ & $-25.16\pm0.55$ & $11.84\pm1.75$ \\
NGC 5822 & $790\pm49$ & $780\pm45$   & $-28.94\pm0.50$ & $16.54\pm2.68$ \\
NGC 6134 & $1080\pm55$ & $1138\pm112$ & $-26.46\pm0.89$ & $19.94\pm3.32$ \\

\hline
\end{tabular}
\tablefoot{
The ages and distances were derived by fitting the cluster CMDs with PARSEC version 2.0 isochrones \citep{2022A&APARSEC}, adopting solar metallicity for all clusters. 
The mean RVs were calculated from the member-star RVs in \citet{Hunt23list}. 
$\sigma_{\rm RV,GB}$ denotes the RV dispersion measured from the ground-based spectroscopic data.
}
\end{table*}

\subsection{Sample refinement and RV-dispersion measurements}
\label{subsection:rv_dispersion}

Based on the measured RVs and $v\sin i$ values, we further refined the ground-based stellar sample. We adopted $v\sin i \leq 200~\mathrm{km~s^{-1}}$ as the upper limit for the blue-edge sample in all clusters. This threshold was chosen because, in our measurements, the RV uncertainty increases rapidly to more than $5~\mathrm{km~s^{-1}}$ for stars with 
$v\sin i \gtrsim 200~\mathrm{km~s^{-1}}$.

We excluded candidate equal-mass-ratio binaries identified from the CMD. Stars with $v\sin i > 200~\mathrm{km~s^{-1}}$ and spectroscopic binaries were also removed from the sample. After these selections, the final ground-based 
sample contains 10 stars in NGC~6281, 17 stars in NGC~6475, and 8 stars in NGC~6633, corresponding to 15.63\%, 18.28\%, and 30.77\% of the photometric member
stars within the defined eMSTO region, respectively. 
The median uncertainties of our measurements are approximately $5~\mathrm{km~s^{-1}}$ for $v\sin i$. 
For the RVs, the median uncertainty is approximately $1~\mathrm{km~s^{-1}}$ for the VLT/GIRAFFE spectra (R $~\sim$ 20000) and $5-8~\mathrm{km~s^{-1}}$ for the lower-resolution SALT/RSS spectra (R $~\sim$ 4000).

For each cluster, we subtracted the mean cluster RV from the individual stellar RVs and calculated the RV dispersion from the residual RVs. The RV dispersion of each population was measured with the biweight midvariance, and its uncertainty was estimated with the bootstrap method: we resampled the RVs with replacement 1000 times and adopted the standard deviation of the resampled dispersions as the 1$\sigma$ error. No $\sigma$-clipping was applied to the observed samples. We relied on the robustness of the biweight midvariance, which is insensitive to extreme outliers such as residual non-members or stars with poor measurements, while remaining responsive to a genuinely broadened distribution.
In addition to the three clusters measured in this work, we also used five clusters from \citet{2020IAUSSun,2021MNRASSun}, namely NGC~2287, IC~4756, NGC~3960, NGC~5822, and NGC~6134, to validate our analysis method. The CMDs of these clusters are shown in Figure~\ref{HRdiagram}. Their ages and distances 
are listed in Table~\ref{cluster param table}. Together, these eight clusters form the ground-based benchmark sample used to assess the reliability of the Gaia DR3 RV measurements.

The low resolution of SALT/RSS leads to large per-star RV uncertainties that can significantly inflate the measured dispersions; we therefore carry out a noise-budget analysis for each cluster. Holding the per-star RV uncertainties fixed, we draw 20000 noise-only Monte Carlo realisations from Gaussians, and adopt the median of their dispersion distribution as the noise dispersion $\sigma_{noise}$. The noise-corrected intrinsic dispersion is computed as $\sigma_{int}^2 = \sigma_{obs}^2~-~\sigma_{noise}^2$. The result is shown in Table~\ref{noise}. The comparison shows that the large observed dispersion of NGC~3960 is dominated by measurement noise. For the other three clusters, the noise contribution is minor and the measured dispersions predominantly reflect the RV dispersion of their member stars.

\begin{table}[ht!]
\caption{Noise-budget analysis of the four SALT/RSS clusters.}
\label{noise}
\centering
\small
\begin{tabular}{l c c c c}
\hline\hline
Cluster & N & $\sigma_{obs}$ & $\sigma_{noise}$ & $\sigma_{int}$ \\
        &  & (km~s$^{-1}$) & (km~s$^{-1}$) & (km~s$^{-1}$) \\
\hline
IC 4756  & 19 & 14.69   & 5.33 & $13.69\pm2.98$ \\
NGC 3960 & 24 & 11.84 & 10.71 & $5.05\pm4.18$ \\
NGC 5822 & 20 & 16.54   & 7.27 & $14.86\pm3.36$ \\
NGC 6134 & 19 & 19.94 & 7.23 & $18.58\pm3.23$ \\

\hline
\end{tabular}
\tablefoot{
$\sigma_{obs}$ is the observed dispersion (not noise-corrected); $\sigma_{noise}$ is the median dispersion of 20000 noise-only Monte Carlo realisations;  $\sigma_{int}$ is the intrinsic dispersion after noise subtraction via $\sigma_{int}^2 = \sigma_{obs}^2~-~\sigma_{noise}^2$.
}
\end{table}

\section{Gaia DR3 extension and binary-population models}
\label{section 3}
\subsection{Gaia DR3 cluster sample}
\label{subsection:gaia_cluster_sample}

We constructed the Gaia DR3 cluster sample from the open-cluster catalogue of \citet{Hunt23list}. We first selected 202 clusters with more than 300 candidate members having membership probabilities greater than 0.8. From these clusters, we retained 40 systems younger than 2~Gyr that show eMSTOs, and the Gaia DR3 RVs of selected stars were filtered using standard quality cuts, requiring \texttt{rv\_expected\_sig\_to\_noise} $>10$, and \texttt{radial\_velocity\_error} $<20~\mathrm{km~s^{-1}}$ \citep{2023A&AGaiaRVKatz,2023A&AGaiaRVBabusiaux}.

\begin{figure}
        \includegraphics[width=\hsize]{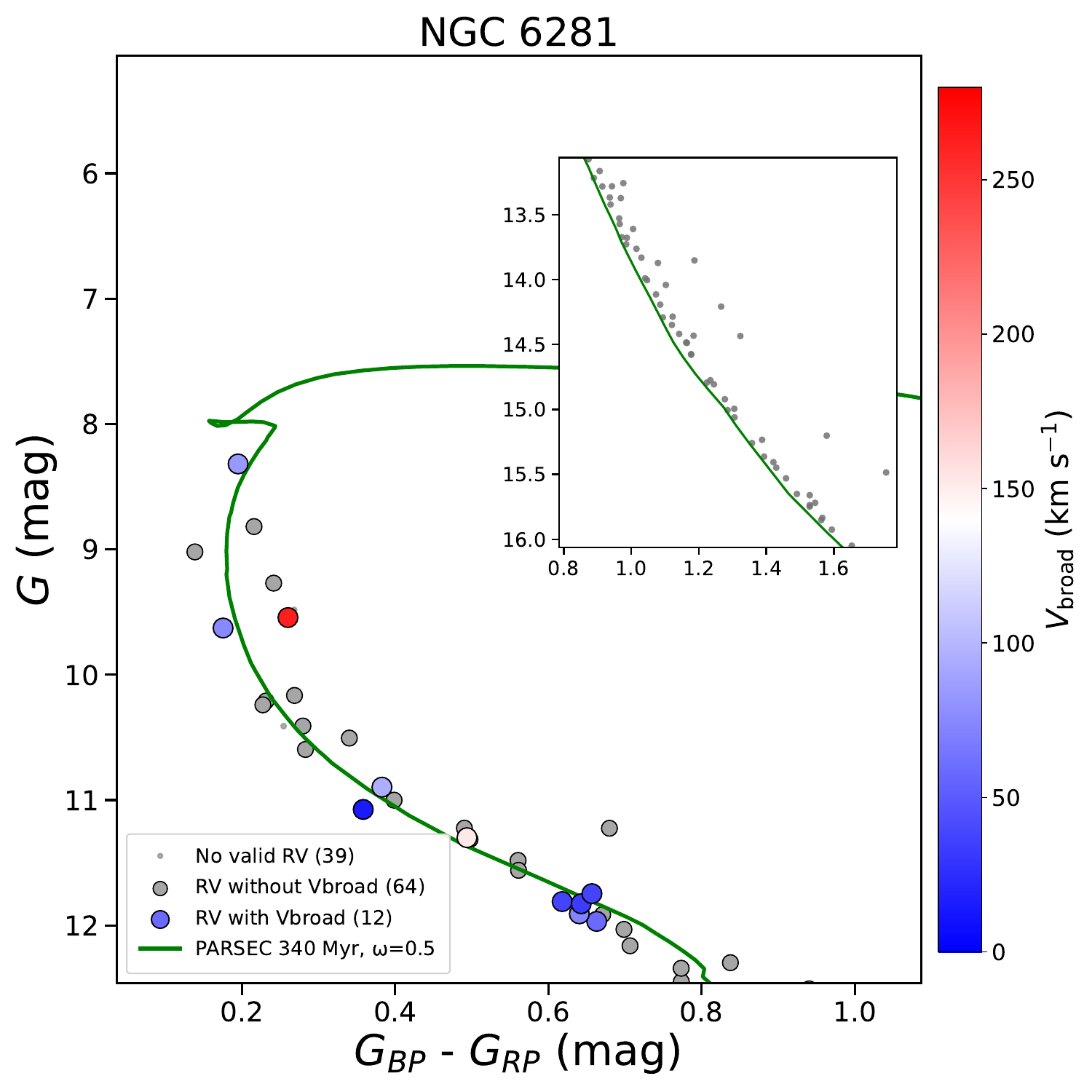}
        \caption{Sample selection for the blue-edge sample. Gray dots are member stars. The circles are the stars have RV value provided by Gaia DR3 and colored according to their $V_{broad}$ value. The green curve is the isochrone with an age of 340~Myr and a rotation parameter of $\omega = 0.5$, which we adopt as selected boundary. The blue-edge sample is  selected on the left of the boundary. The inset shows a zoom-in of the lower main sequence below the turn-off, whose narrow width indicates that differential reddening within the cluster is small compared with the width of the selection band.}
        \label{selected sample}
    \end{figure}

For each of the 40 clusters, we selected candidate blue-edge sample in the CMD. We considered stars with masses between $1.5$ and $3.0\,M_\odot$, located between the blue edge of the MS and the rotational-broadening contour of the isochrone  with $\omega=0.5$ of PARSEC v2.0, corresponding to $V_{\rm broad}\approx165~\mathrm{km~s^{-1}}$. This CMD selection is intended to retain stars with relatively low projected rotational broadening while excluding the rapidly rotating population. An example of this selection is shown in Figure~\ref{selected sample}. We note that our blue-edge selection is a proxy for, rather than a direct identification of, slow rotators: the edge is populated by genuinely slowly rotating stars seen at any inclination, mixed with moderate rotators viewed at low $i$, so individual blue-edge stars cannot be classified as slow rotators from their $v\sin i$ alone.
Differential reddening within a cluster could broaden the main sequence and scatter stars across the blue-edge selection boundary. We assessed its effect using the lower main sequence below the turn-off. As shown in the inset of Figure~\ref{selected sample}, the lower main sequence forms a narrow, well-defined sequence, indicating that differential reddening is small compared with the width of the selection band and has a negligible impact on the selected sample.
Because the tidal-braking hypothesis predicts that the edge population is dominated by synchronized binaries, this distinction does not affect our binary-based test of its origin.
The number of selected stars with Gaia DR3 RVs entering the dispersion analysis ranges from 10 to 64 per cluster, with a median of 22; the full per-cluster numbers, ages, and resulting dispersions are listed in Table~\ref{table:gaia}. Relative to the eMSTO stars, the median selected fraction of our fiducial blue-edge sample is 53\%. This is only moderately larger than the median $v\sin i<100$ km s$^{-1}$ fraction of 0.41 reported by \citet{2026ApJRao&Chen}, as expected because our $\omega=0.5$ boundary corresponds approximately to $V_{\rm broad}\simeq165$ km s$^{-1}$.
We also construct a narrow blue-edge sample containing approximately the bluest 30\% of the eMSTO stars as a robustness test, applying it to the 26 clusters with at least ten usable Gaia RV measurements.

We then used the Gaia DR3 RVs of the selected stars to calculate the RV dispersion for each cluster. Only stars brighter than \texttt{G\_RVS} = 12 mag (\texttt{rv\_method\_used}=1) were retained, for which the published Gaia DR3 RV is the median of the individual epoch measurements \citep{2023A&AGaiaRVKatz}.  
The Gaia DR3 $V_{\rm broad}$ values were used only to define the blue-edge selection boundary and were not treated as direct measurements of $v\sin i$.  After subtracting the mean cluster RV, the dispersion was measured as for the ground-based sample (Section~\ref{subsection:rv_dispersion}), and its 1$\sigma$ uncertainty was derived from 1000 bootstrap resamplings. The observed dispersion was then corrected for the two variance contributions that do not arise from binary orbital motion: $\sigma_{\rm corr}^2 = \sigma_{\rm obs}^2 - \sigma_{\rm Gaia}^2 -\sigma_{\rm cl}^2$, where $\sigma_{\rm Gaia}$ is the contribution of the Gaia per-star RV uncertainties (typically $\sim$0.5~km~s$^{-1}$; \citealt{2023A&AGaiaRVKatz}) and $\sigma_{\rm cl}$ is the intrinsic cluster velocity dispersion ($\sim$0.6~km~s$^{-1}$;  \citealt{2003A&AMadsen}). The resulting corrected dispersions and their errors are listed in Table~\ref{table:gaia}.

\subsection{Validation of Gaia DR3 radial velocities}
\label{subsection:gaia_rv_validation}

We compared the ground-based RVs with those provided by Gaia DR3 to assess whether Gaia RVs can be used for the subsequent RV-dispersion analysis. The validation uses the stars common to the refined ground-based sample (Section~\ref{subsection:rv_dispersion}) and the Gaia DR3 catalogue; for these stars we applied the same quality cuts as in Section~\ref{subsection:gaia_cluster_sample}.
 
\begin{figure}
    \centering
    \includegraphics[width=\hsize]{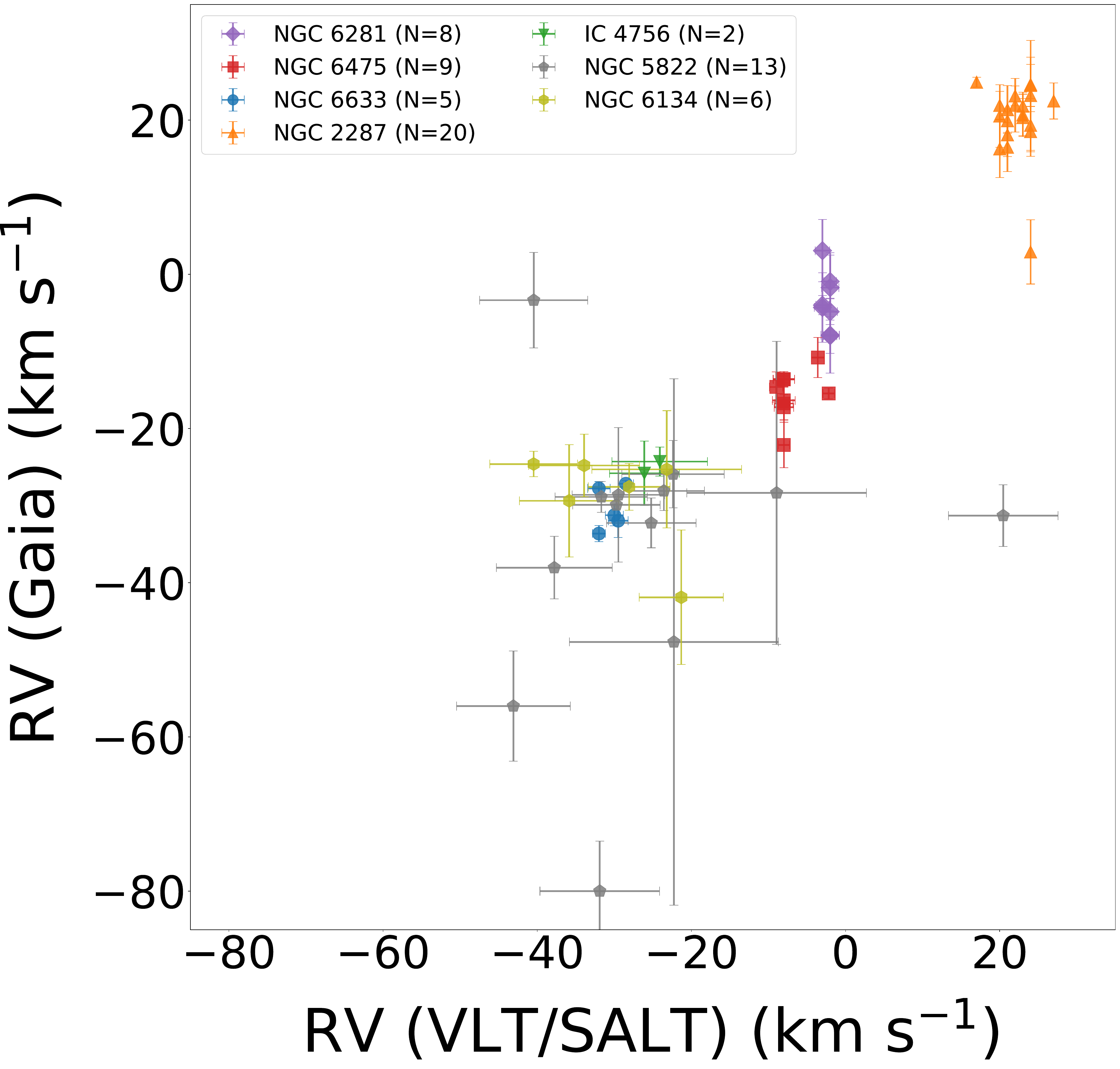}

    \caption{Comparison of individual stellar RVs measured from the ground-based spectra with the corresponding Gaia DR3 RVs for the same member stars in the eight benchmark clusters. Excepted NGC~3960 lacks efficient Gaia RV data, seven clusters are shown with a different colour and symbol, with the number of member stars indicated in the legend. Error bars represent the measurement uncertainties. }
    \label{RV_comparision}
\end{figure}

Figure~\ref{RV_comparision} compares the individual stellar RVs measured from the ground-based spectra with the corresponding Gaia DR3 RVs for the same member stars in the eight benchmark clusters. Overall, the ground-based and Gaia DR3 RVs show good agreement, with most stars lying close to the one-to-one relation. Some individual stars, especially in NGC~5822, show noticeable discrepancies between the two measurements. This is because close binaries were not explicitly removed at this stage, and their RVs can vary significantly between observations taken at different epochs. Despite these star-to-star differences, both the ground-based measurements and Gaia DR3 indicate that NGC~5822 has a relatively large RV dispersion. This agreement at the cluster level suggests that Gaia DR3 RVs are suitable for identifying clusters with enhanced RV dispersions, although individual stellar RVs may be affected by epoch-dependent variations.

For the eight ground-based clusters, we compared their RV dispersions with those derived from the Gaia DR3 RVs. Seven clusters have Gaia measurements, while NGC~3960 lacks efficient Gaia RV data to measure dispersion. For the 3 clusters (NGC~6281, NGC~6475, NGC~2287) with $\sigma_{\rm Gaia} < 5$~km~s$^{-1}$, the Gaia dispersion is  larger than or consistent with the ground-based value, which can be attributed to the much larger Gaia samples and the resulting statistical fluctuations. For the clusters with $\sigma_{\rm Gaia} > 5$~km~s$^{-1}$, the Gaia dispersion is smaller than the ground-based one, because the Gaia combined RV is the median of multiple epochs over the DR3 baseline, which suppresses the dispersion. A fair comparison with binary population models therefore requires the simulations to reproduce the observing strategy of each data set, and we ran single-epoch and Gaia-like simulations separately.

\subsection{Tidal-synchronization and binary simulations}
\label{subsection:binary_simulation}

To assess whether tidal interactions can produce the slowly rotating stars in our sample, we estimated the tidal-synchronization timescale for binaries with A- and F-type primaries. We adopted the analytical approximation of 
\citet{2002MNRASHurleytimescale} and assumed circular orbits:
\begin{equation}
\label{eq:tidal_synchronization}
\frac{1}{\tau_{\rm sync}} =
5 \times 2^{5/3}
\left(\frac{GM}{R}\right)^{1/2}
\frac{MR^2}{I}
q^2(1+q)^{5/6}E_2
\left(\frac{R}{a}\right)^{17/2},
\end{equation}
where $\tau_{\rm sync}$ is the synchronization timescale; $M$, $R$, and $I$ are the mass, radius, and moment of inertia of the primary star, respectively; $q$ is the mass ratio between the secondary and primary stars; $a$ is the orbital semi-major axis; $G$ is the gravitational constant; and $E_2$ is the tidal coefficient, for which we adopt the fit $E_2 = 1.592 \times 10^{-9} M^{2.84}$ (with $M$ in solar masses; \citealt{1975A&AZahn}).

Using stellar parameters appropriate for stars with masses between $1.5$ and $3.0\,M_\odot$, we find that the synchronization timescale increases rapidly with increasing orbital separation. For representative systems, $\tau_{\rm sync}$ increases from approximately $0.1$~Myr to $1000$~Myr as $a/R$ increases from 2.5 to 8.0. These separations correspond to short-period binaries, with orbital periods of order $\lesssim10$~d. Therefore, only close 
binaries are expected to undergo substantial tidal synchronization within the ages of the clusters considered here.

We performed Monte Carlo simulations to estimate the RV dispersions expected for binary populations with different orbital periods. For each period model, we generated $10\,000$ binary systems with primary masses between $1.5$ and $3.0\,M_\odot$ and mass ratios in the range $0.3 \leq q \leq 0.9$. We assumed circular orbits and isotropically distributed orbital orientations, such that $\cos i_{\rm orb}$ follows a uniform distribution between 0 and 1, where 
$i_{\rm orb}$ is the orbital inclination relative to the line of sight \citep{2019Lim_cosj}.
Because the two observational samples probe the binary signal in fundamentally different ways, we simulated them separately. For the eight ground-based benchmark clusters, each simulated binary was assigned a random orbital phase, representing a single-epoch RV measurement of the primary star. For the Gaia DR3 sample, we reproduced the observing strategy of the Gaia RVS: the Gaia DR3 published RV is not a single-epoch measurement, but is combined from individual transit measurements obtained over the 34-month DR3 baseline, with a strongly clustered temporal sampling imposed by the Gaia scanning law \citep{2023A&AGaiaRVKatz}. For each cluster, we drew the number of epochs of each simulated binary at random from the observed distribution of the number of RVS transits (\texttt{rv\_nb\_transits}) of that cluster's member stars. The epochs were distributed in clusters over a 1034-d baseline to mimic the scanning-law sampling, the instantaneous RV was computed at each epoch, and the median of the epoch RVs was taken, corresponding to the published value for stars brighter than \texttt{G\_RVS} = 12 mag \citep{2023A&AGaiaRVKatz}.

We considered orbital periods of 10, 100, 500, and 1000~d. The 10-d model represents close binaries for which tidal synchronization may occur on timescales comparable to, or shorter than, the cluster ages. The 100-d and 500-d models represent wider binary populations and provide intermediate 
comparison cases. These periods are also relevant to the observed connection between binary orbital period and stellar rotation \citep{2015MNRASDAntona,2004Abt}. The 1000-d model represents a weakly interacting binary population, for which tidal effects on stellar rotation are expected to be small.

For each simulated binary, we calculated the instantaneous RV of the primary star and derived the RV dispersion of the full simulated population. 
The dispersion was measured with the biweight midvariance, the same robust estimator used for the observed samples (Section~\ref{subsection:rv_dispersion}), and its uncertainty was estimated with the bootstrap method: we resampled the RVs with replacement 1000 times and adopted the standard deviation of the resampled dispersions as the 1$\sigma$ error.

We note that double-lined spectroscopic binaries were removed from the observational sample (Section~\ref{subsection:rv_dispersion}). Since the recognition of a secondary set of lines requires a flux ratio close to unity, this cut approximately corresponds to near-equal-mass systems with $q \gtrsim 0.8$, i.e. those with the largest orbital velocity
amplitudes, and hence biases the observed dispersions slightly low. To quantify this effect, we repeated the simulations with the mass-ratio upper limit reduced from 0.9 to 0.8, which removes the same $\sim$16\% of systems. The resulting model dispersions decrease by only $\sim$5\% at all periods. We further checked how this shift propagates to the classification of clusters relative to the model predictions in Section~\ref{section 3}: with the lowered model values, the number of clusters falling in each interval changes by no more than one in either sample, while the overall distribution across intervals remains unchanged. Our conclusions are therefore  unaffected by this selection effect.
The resulting simulated dispersions were compared with those measured for the eight ground-based benchmark clusters (Figure~\ref{RVdispersion_obs}) and the 40-cluster Gaia DR3 sample (Figure~\ref{RVdispersion_all}).
If the observed blue-edge samples were dominated by tidally synchronized close binaries, their RV dispersions would be expected to be comparable to those predicted by the short-period binary model.

\subsection{Comparison with observed RV dispersions}
\label{subsection:comparison result}

\begin{figure*}[ht!]
    \centering
    \includegraphics[width=0.24\hsize]{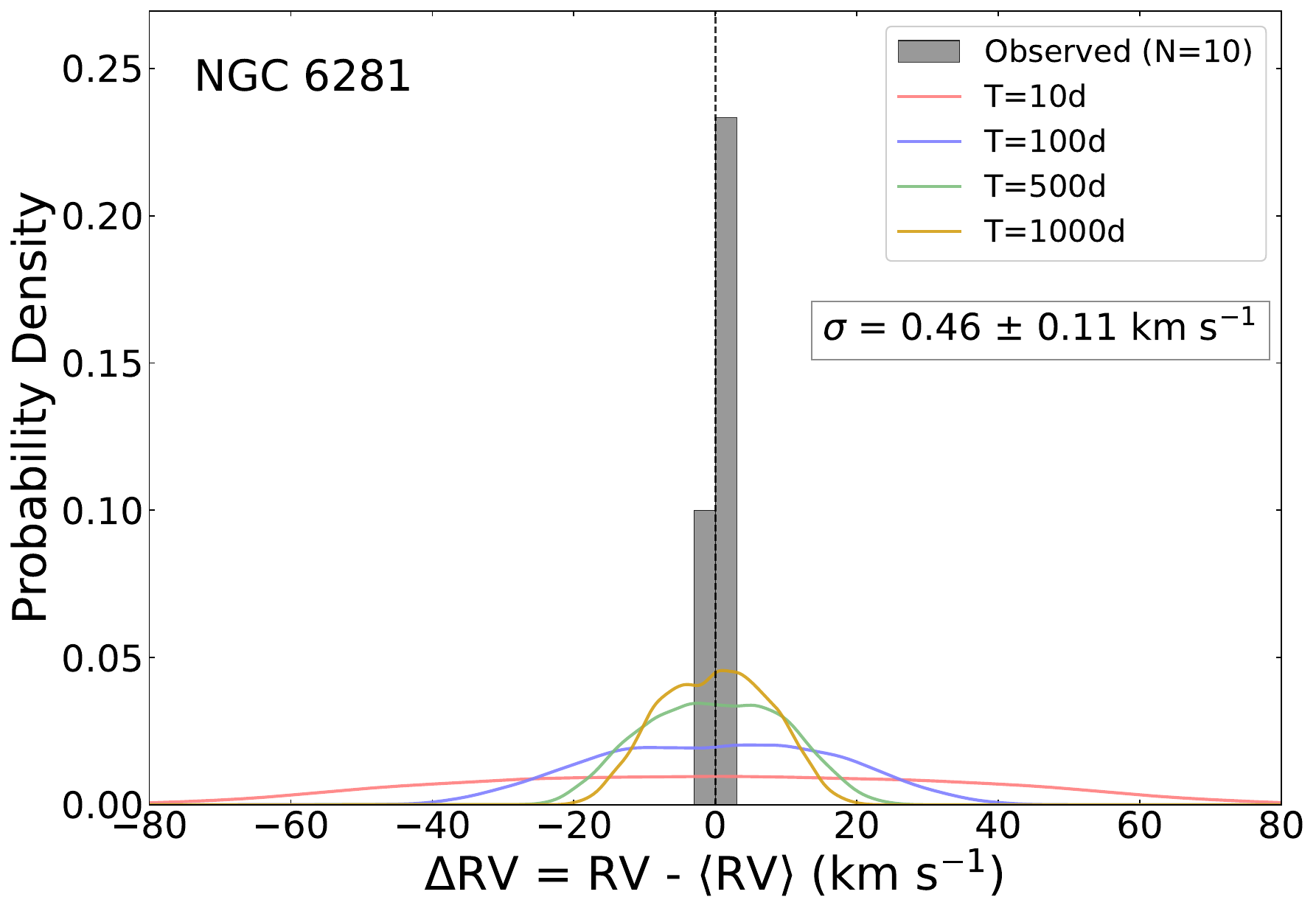}
    \includegraphics[width=0.24\hsize]{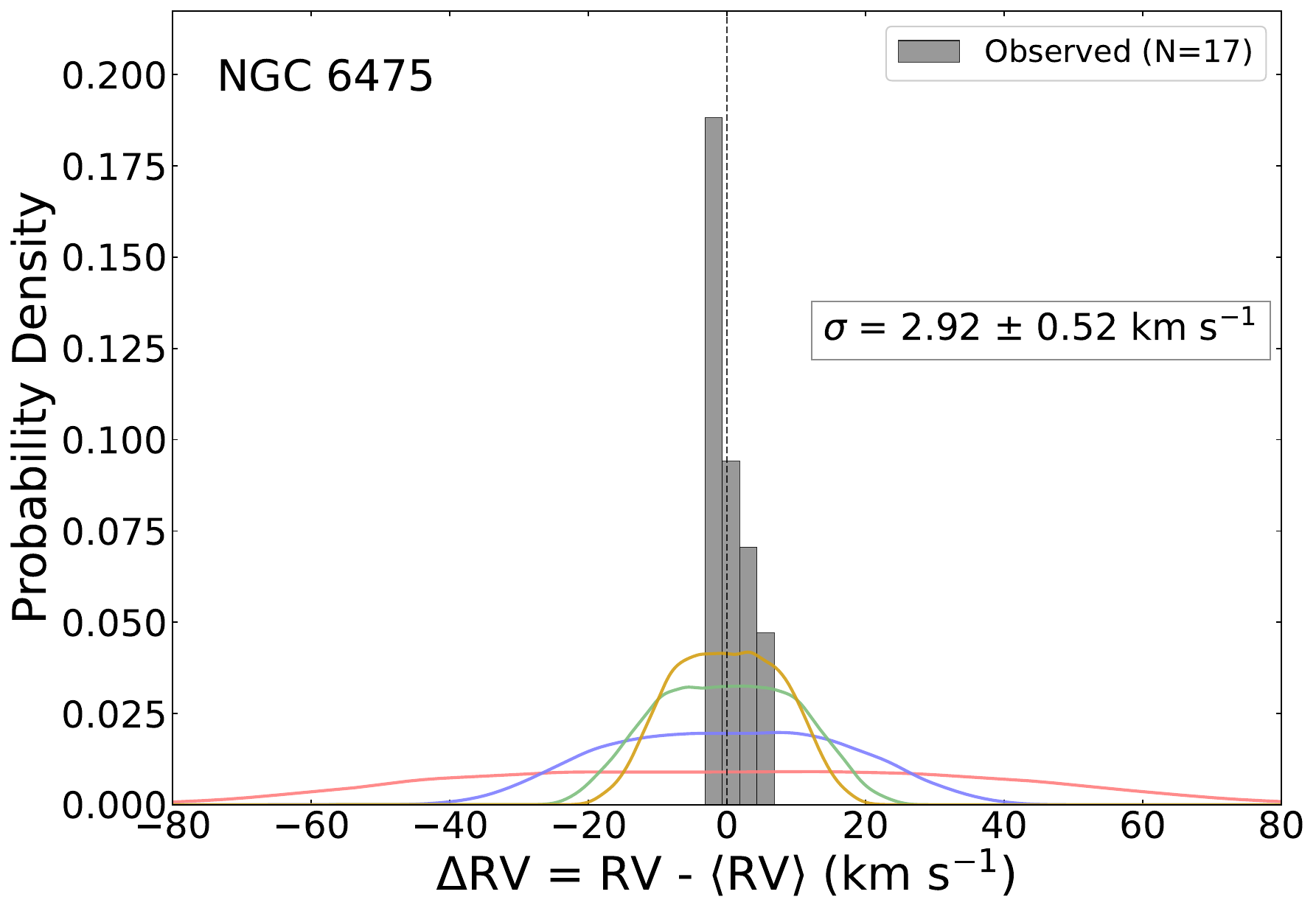}
    \includegraphics[width=0.24\hsize]{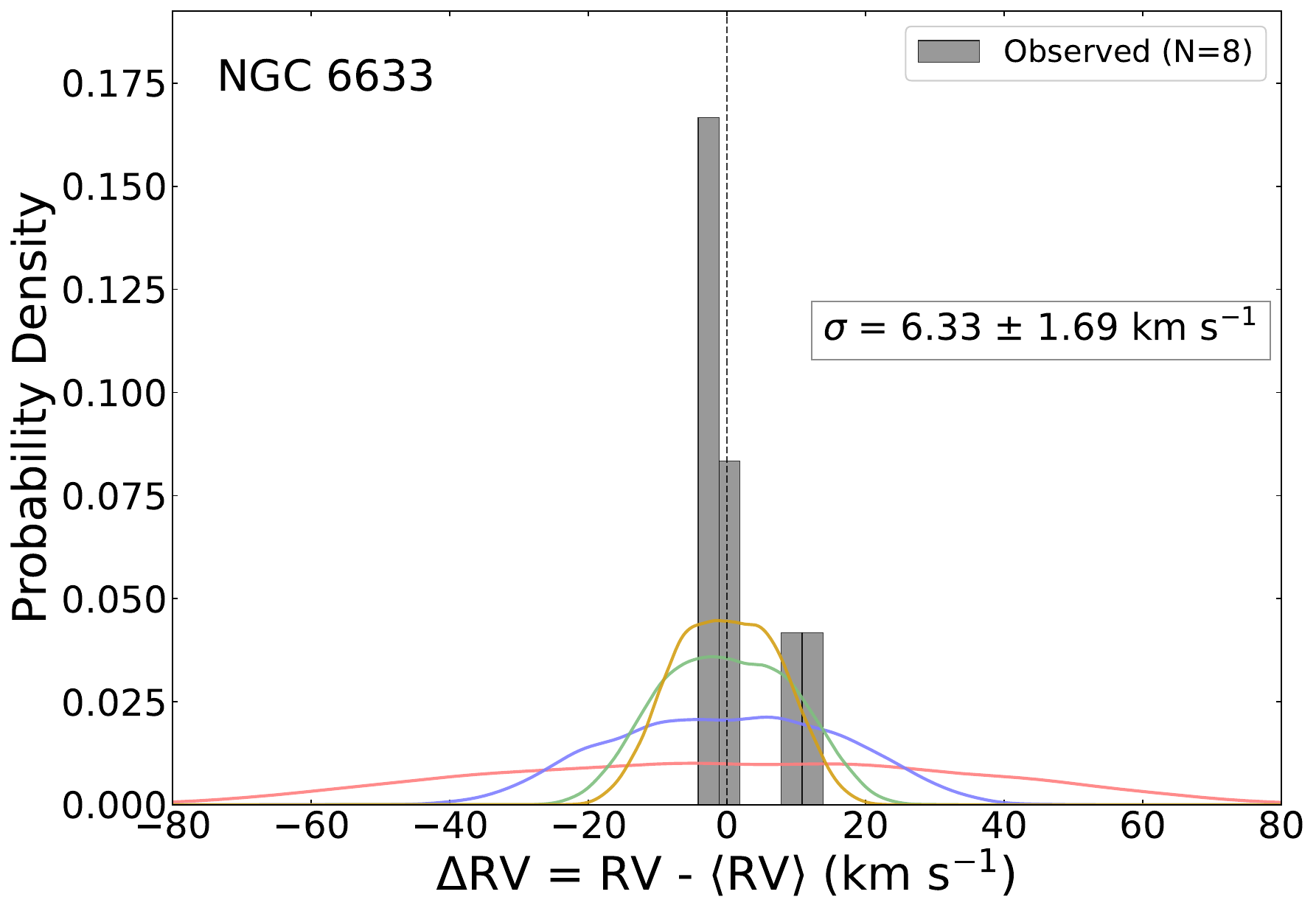}
    \includegraphics[width=0.24\hsize]{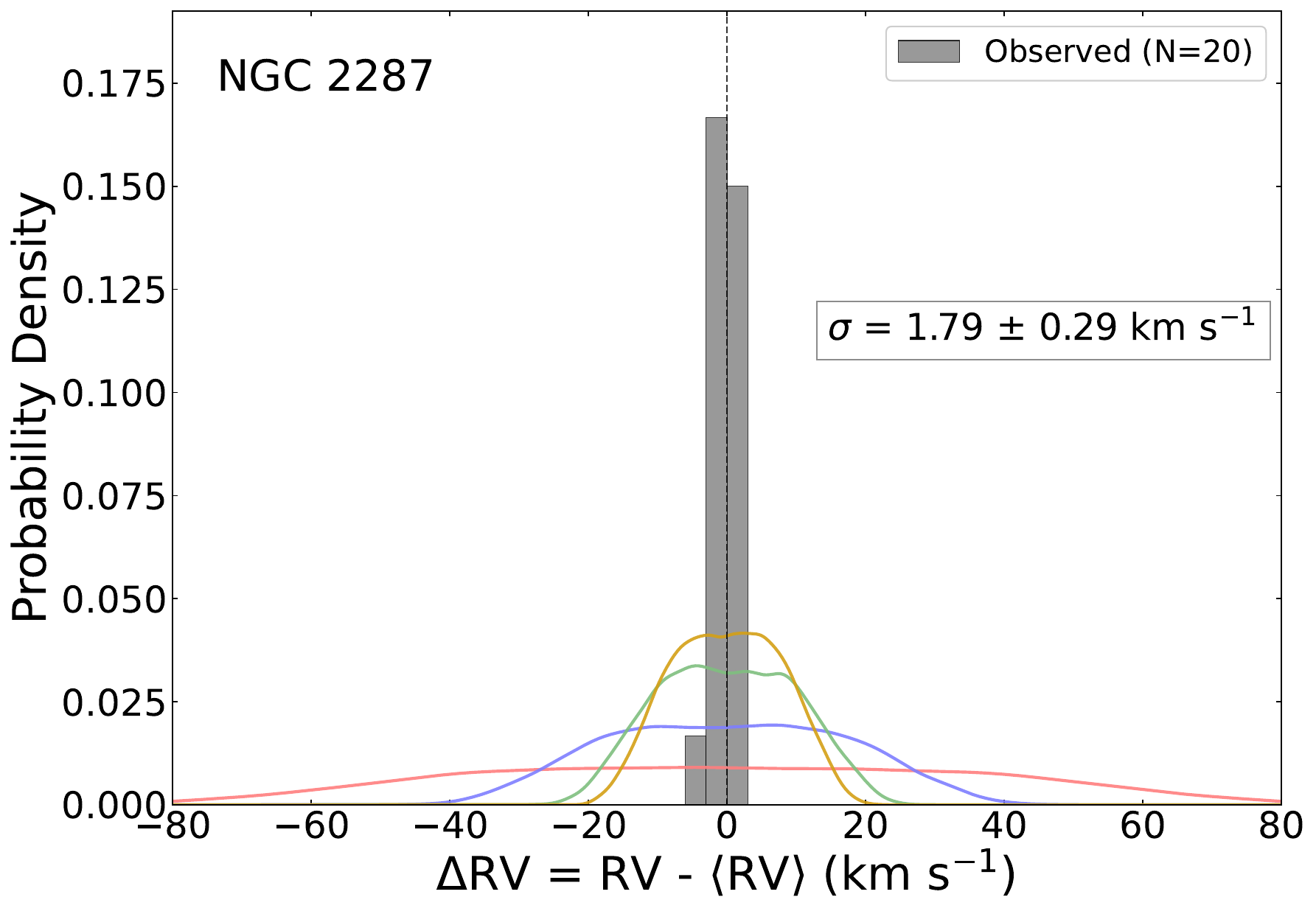}
    \includegraphics[width=0.24\hsize]{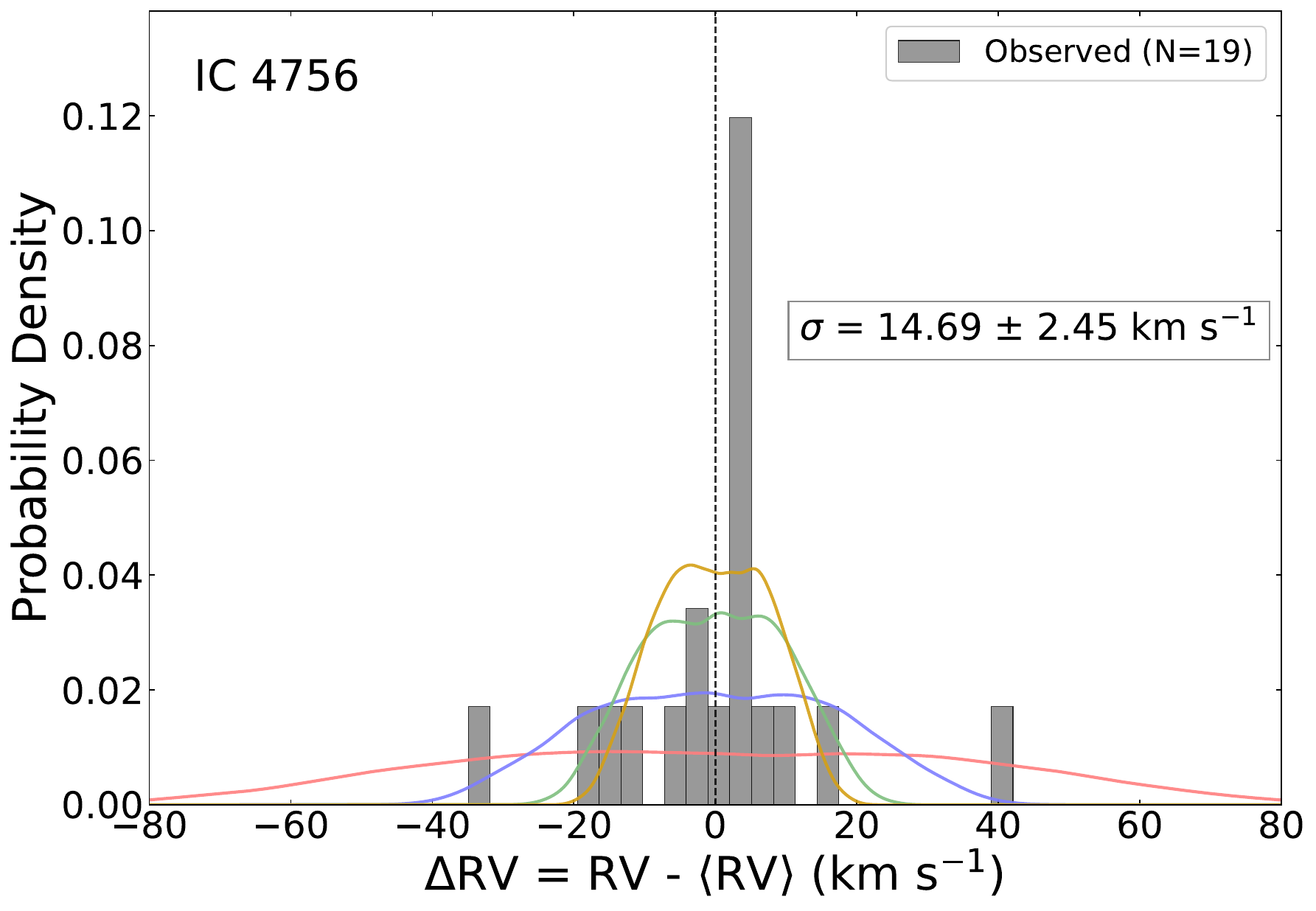}
    \includegraphics[width=0.24\hsize]{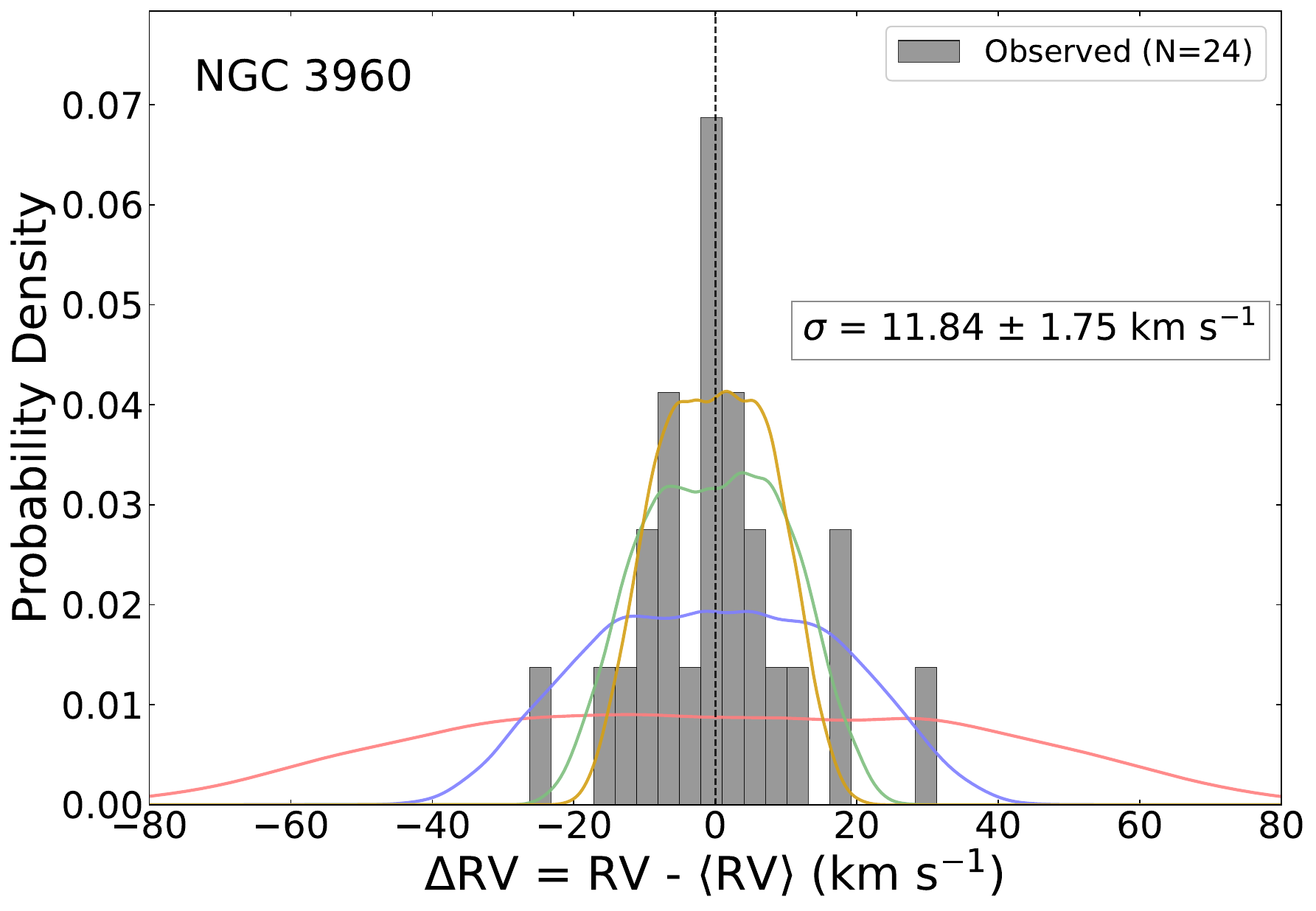}
    \includegraphics[width=0.24\hsize]{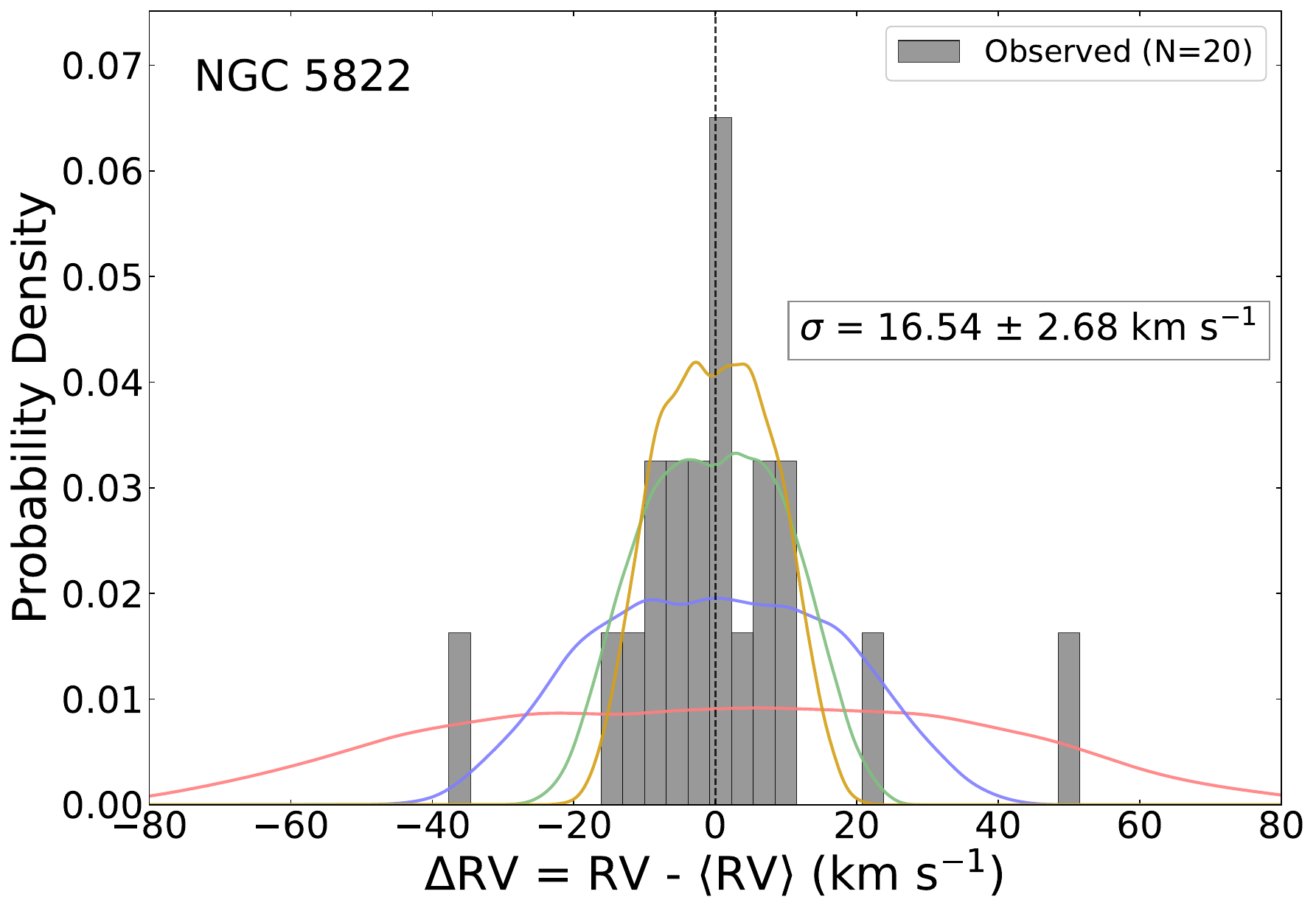}
    \includegraphics[width=0.24\hsize]{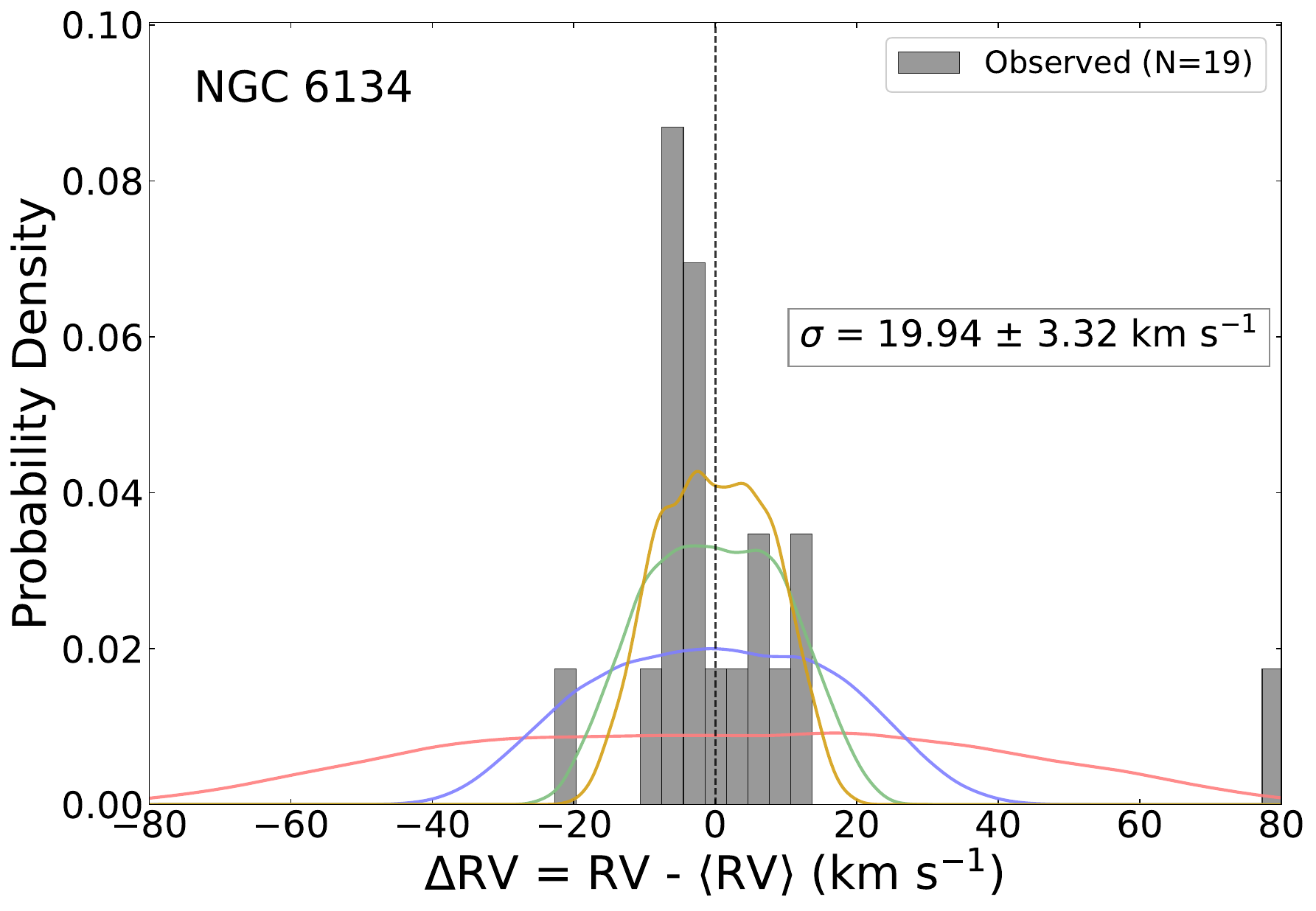}
    
    \caption{RV distributions of the blue-edge samples in the eight benchmark clusters. The grey histograms show the observed stellar RV distributions, with a bin width of $3~\mathrm{km~s^{-1}}$. The coloured curves show the simulated RV distributions for binary populations with fixed orbital periods of $P=10$~d (red), $P=100$~d (blue), $P=500$~d (green), and 
    $P=1000$~d (orange). The $P=10$~d model represents close binaries for which tidal synchronization may be effective, whereas the longer-period models 
    provide comparisons with wider binary populations.
    }
    \label{RVdispersion_obs}
\end{figure*}

We first compare the simulated dispersions with the ground-based measurements and then with the Gaia sample. Figure~\ref{RVdispersion_obs} shows the RV distributions of the blue-edge samples in the eight benchmark clusters with ground-based spectroscopic RV measurements. We then extended the analysis to the 40 open clusters selected from \citet{Hunt23list} (Section~\ref{section 2}), using Gaia DR3 RVs and $V_{\rm broad}$ measurements to identify the blue-edge samples. The RV dispersions of these clusters were compared with Monte Carlo simulations of 
binary populations with orbital periods of $P=10$, 100, 500, and 1000~d.

For single-epoch measurements, the simulated RV dispersions are 
$\sigma_{\rm 10\,d}=36.01~\mathrm{km~s^{-1}}$,
$\sigma_{\rm 100\,d}=16.70~\mathrm{km~s^{-1}}$,
$\sigma_{\rm 500\,d}=9.78~\mathrm{km~s^{-1}}$, and
$\sigma_{\rm 1000\,d}=7.76~\mathrm{km~s^{-1}}$.
These values provide reference predictions for binary populations with the corresponding fixed orbital periods. For the Gaia-like multi-epoch case, in which the published value is the median of the epoch RVs, the corresponding dispersions are reduced to $\sigma_{\rm 10\,d}~\sim~18.0~\mathrm{km~s^{-1}}$, $\sigma_{\rm 100\,d}~\sim~12.1~\mathrm{km~s^{-1}}$, $\sigma_{\rm 500\,d}~\sim~7.8~\mathrm{km~s^{-1}}$, and $\sigma_{\rm 1000\,d}~\sim~6.2~\mathrm{km~s^{-1}}$; for each cluster the simulation adopts the observed \texttt{rv\_nb\_transits} distribution of its own members, and the values quoted here are representative averages. The difference between the two sets of model dispersions arises from the different observing modes: the ground-based simulations represent single-epoch measurements, whereas the Gaia simulations reproduce the temporal averaging of the Gaia DR3 RV measurements. Figure~\ref{RVdispersion_all} presents the RV-dispersion distribution for the 40-cluster Gaia DR3 sample.

\begin{figure*}[]
    \centering
    \includegraphics[width=0.45\hsize]{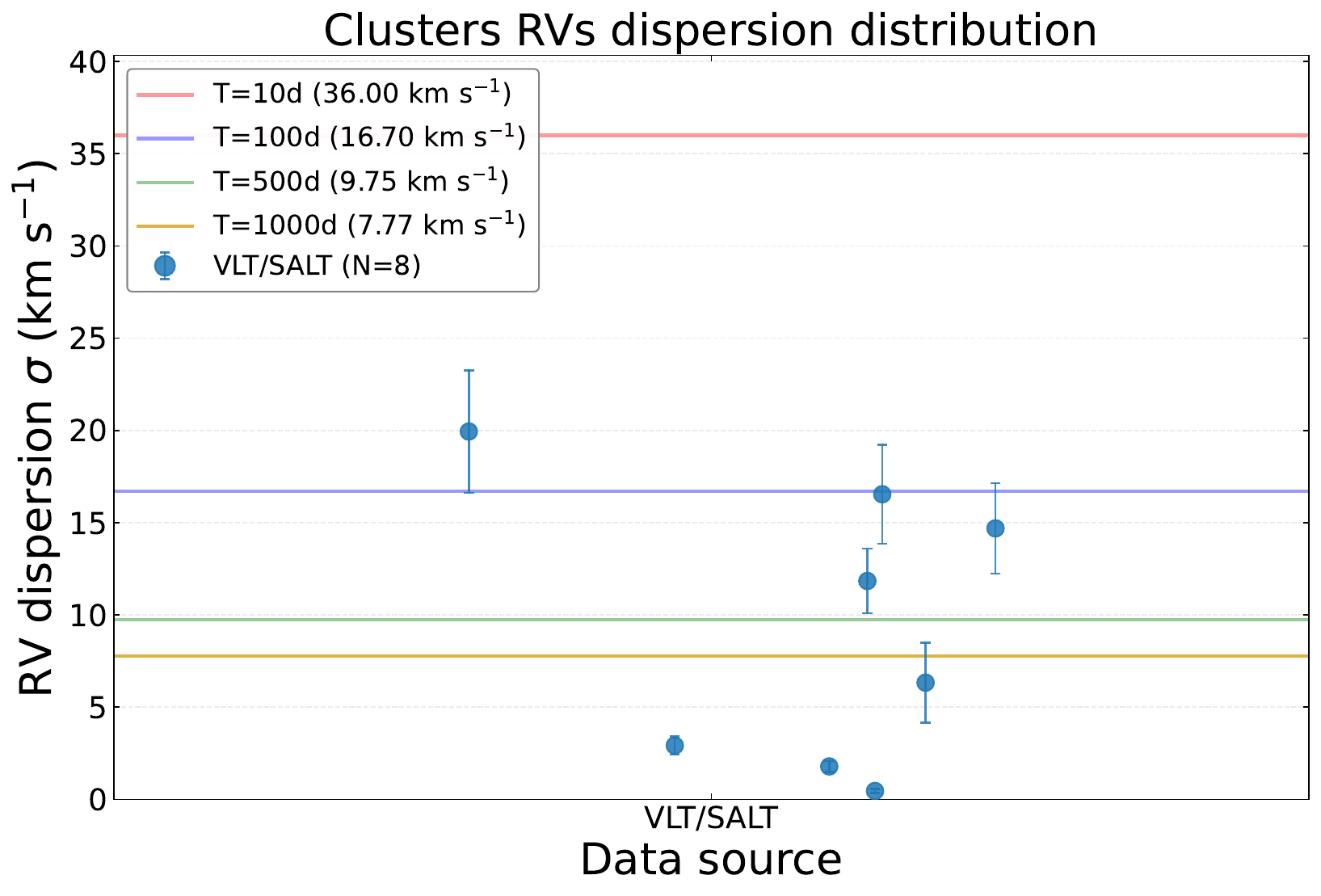}
    \includegraphics[width=0.45\hsize]{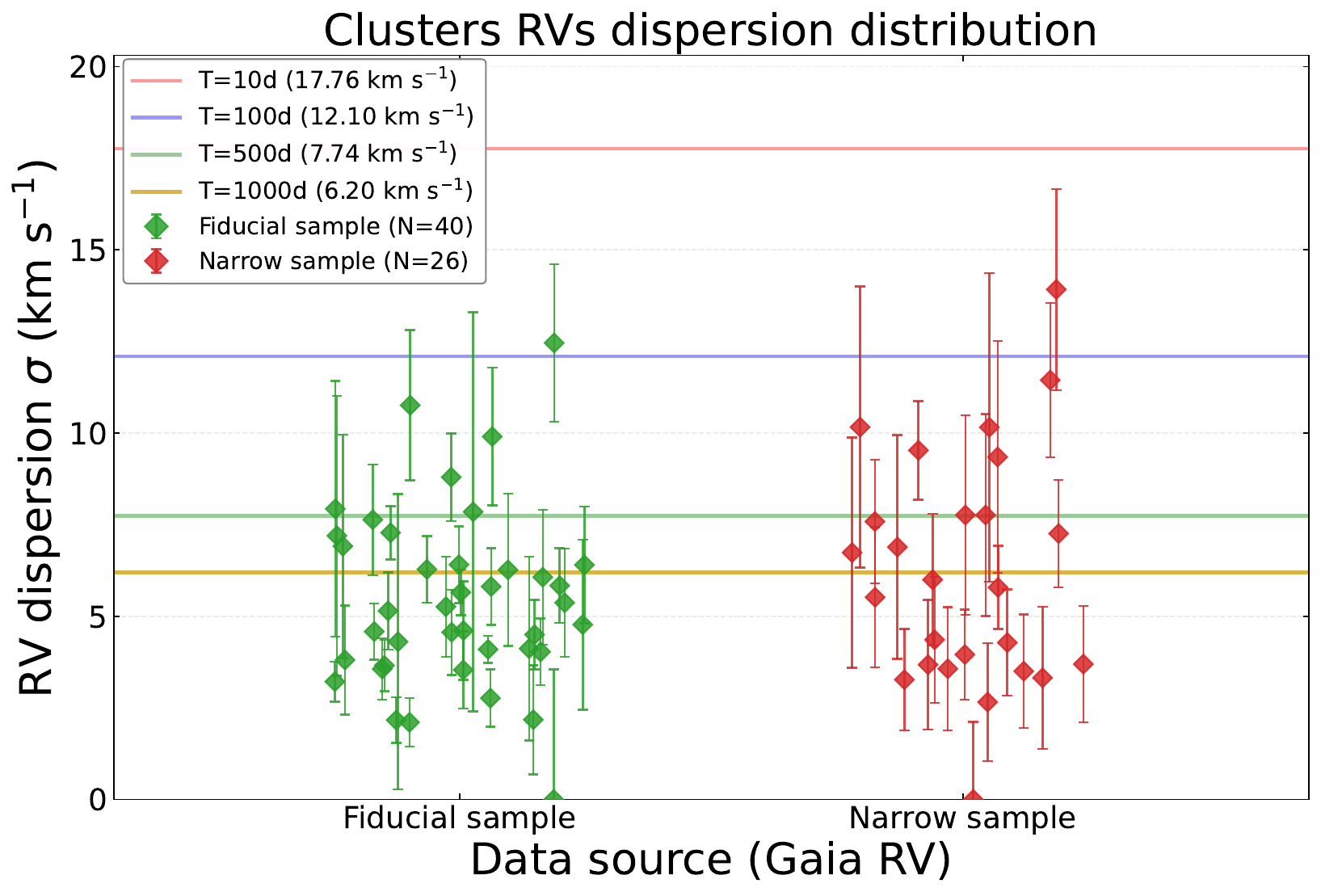}
    \caption{
    RV dispersions ($\sigma_{\rm RV}$) of the cluster samples. Left: RV 
    dispersions derived from ground-based spectroscopic measurements for the 
    eight benchmark clusters, shown as blue circles with error bars. The plotted dispersions $\sigma_{\rm obs}$ are not corrected for measurement noise and thus represent conservative upper limits; the noise-corrected intrinsic dispersions $\sigma_{\rm int}$, on which our discussion is based, are listed in Table~\ref{noise}. The horizontal lines show the simulated single-epoch RV dispersions of binary populations with fixed orbital periods of     
    $P=10$~d (red; $\sigma_{\rm RV}=36.00~\mathrm{km~s^{-1}}$), $P=100$~d 
    (blue; $\sigma_{\rm RV}=16.70~\mathrm{km~s^{-1}}$), $P=500$~d (green; 
    $\sigma_{\rm RV}=9.75~\mathrm{km~s^{-1}}$), and $P=1000$~d (orange; 
    $\sigma_{\rm RV}=7.77~\mathrm{km~s^{-1}}$).
    Right: 
    RV dispersions derived from Gaia DR3 measurements for
    the fiducial sample (green diamonds; $N=40$) and the narrow sample (red diamonds; $N=26$). 
    The narrow sample includes only clusters retaining at least ten usable Gaia RV measurements after the stricter selection.
    Error bars show the bootstrap uncertainties. 
    The horizontal lines show the simulated dispersions for the same orbital periods as combined in the
    Gaia DR3 observing mode of
    $P=10$~d (red; $\sigma_{\rm RV}=17.76~\mathrm{km~s^{-1}}$), $P=100$~d 
    (blue; $\sigma_{\rm RV}=12.10~\mathrm{km~s^{-1}}$), $P=500$~d (green; 
    $\sigma_{\rm RV}=7.74~\mathrm{km~s^{-1}}$), and $P=1000$~d (orange; 
    $\sigma_{\rm RV}=6.20~\mathrm{km~s^{-1}}$).
    }
    \label{sigmadistribution}
\end{figure*}

Figure~\ref{sigmadistribution} compares the observed cluster RV dispersions with those obtained from the simulated binary populations. For the eight benchmark clusters with ground-based spectroscopic measurements, 
four clusters have RV dispersions even smaller than that predicted by the $P=1000$~d binary-population model. The RV dispersions of the IC~4756, NGC~3960 and NGC~5822 lie approximately between the predictions of the $P=500$~d and $P=100$~d models. The RV dispersion of NGC~6134 lies approximately between the predictions of the $P=100$~d and $P=10$~d models. None of the benchmark clusters has an RV dispersion larger than that predicted by the $P=10$~d model. 
The dispersions quoted above are observed values including measurement noise. The noise-budget analysis of Section~\ref{subsection:rv_dispersion} (Table~\ref{noise}) shows that the large dispersions of IC~4756 and NGC~5822 remain significant after noise subtraction, with intrinsic dispersions of 13.7 and 14.9~km~s$^{-1}$, respectively. The large dispersion of NGC~6134 is dominated by a single star offset by $\approx$81~km~s$^{-1}$ from the systemic velocity; this star is a strong binary candidate, but the cluster-wide dispersion is consistent with noise within $\sim$2$\sigma$. The dispersion of NGC~3960 does not significantly exceed the noise-only expectation. Since measurement noise can only inflate $\sigma_{\rm obs}$, the comparisons with the model predictions based on the uncorrected values are conservative.

For the 40-cluster Gaia DR3 sample, one cluster (NGC~1912; $2.5\%$) has an RV dispersion comparable to that predicted by the $P=100$~d model. Nine clusters ($22.5\%$) have RV dispersions lying approximately between the predictions of the $P=500$~d and $P=100$~d models, while five clusters ($12.5\%$) have RV dispersions between those predicted by the $P=1000$~d and $P=500$~d models. The remaining 25 clusters ($62.5\%$) have RV dispersions smaller than those predicted for binary populations with the corresponding orbital periods.

The narrow sample shows a broadly similar distribution, listed in Table~\ref{table:gaia} and shown in Figure~\ref{sigmadistribution}.
Among the 26 clusters with $N_{\rm narrow}\geq10$, one cluster ($3.8\%$) lies between the $P=100$~d and $P=10$~d models, 5 clusters ($19.2\%$) lie between the $P=500$~d and $P=100$~d models, 5 clusters ($19.2\%$) lie between the $P=1000$~d and $P=500$~d models, and 15 clusters ($57.7\%$) lie below the $P=1000$~d model.
All 26 clusters remain below the prediction of the $P=10$~d model. The similarity between the fiducial and narrow-sample distributions indicates that the measured RV dispersions are not significantly affected by the width of the blue-edge selection, supporting the robustness of the statistical result obtained from the fiducial sample.

\section{Discussion}
\label{section 4}

\subsection{Implications for the origin of slow rotators}

The RV-dispersion results indicate that the blue-edge populations in most of
the examined open clusters are unlikely to be dominated by tidally
synchronized close binaries. In particular, the observed RV dispersions are
generally much smaller than the prediction for a binary population with
$P=10$~d, for which tidal synchronization can be effective according to
Eq.~\ref{eq:tidal_synchronization}. This result suggests that tidal
synchronization in close binaries is not the dominant channel responsible for
the formation of slow rotators in these clusters.

The narrow blue-edge sample provides a robustness check on the width
of the fiducial selection. If the broader fiducial sample had diluted a
population dominated by close synchronized binaries, the narrower
sample should have shifted toward larger RV dispersions. The similar
distributions of the fiducial and narrow samples indicate that the
measured RV dispersions are not significantly affected by the blue-edge
selection width, supporting the robustness of the statistical result
obtained from the fiducial sample.

Our results do not imply that binaries are absent from the slow-rotator
samples. Individual stars with RVs significantly offset from the cluster mean
may still host undetected companions, particularly companions with low mass
ratios. However, tidal interactions in such systems
are expected to be weak, and they are unlikely to produce substantial
rotational braking. Thus, the possible presence of these binaries does not
alter the conclusion that close, tidally synchronized binaries cannot dominate
the observed slow-rotator populations.

NGC~6134 is an exception among the ground-based benchmark clusters, showing a relatively large RV dispersion. This dispersion is driven primarily by one star whose RV differs from the cluster mean by approximately $81~\mathrm{km~s^{-1}}$. The star may be a spectroscopic-binary candidate or a contaminant, although the available single-epoch measurement does not allow its nature to be determined. Excluding this object would substantially reduce the inferred cluster RV dispersion. Therefore, the large dispersion of NGC~6134 should not be interpreted as evidence that its slow-rotator population is dominated by close binaries; multi-epoch spectroscopy is required to establish whether the RV offset is caused by orbital motion.

\subsection{Limitations of the Gaia DR3 RV analysis}

The large-sample analysis is based on Gaia DR3 RV measurements, whose
uncertainties are substantially larger than those of high-resolution
ground-based spectroscopy. In our sample, the typical Gaia RV uncertainty is
approximately $0.5~\mathrm{km~s^{-1}}$ for bright stars and can reach
approximately $1.5~\mathrm{km~s^{-1}}$ for fainter stars(\texttt{G\_RVS} = 12 mag) \citep{2023A&AGaiaRVKatz}. These measurement
uncertainties contribute to the observed cluster RV dispersions and should be
considered when comparing them with the simulated binary-population models.

This issue is particularly relevant when distinguishing between the
$P=500$~d and $P=1000$~d models, whose predicted RV dispersions differ by only
$\sim1.6~\mathrm{km~s^{-1}}$. The classification of an individual cluster
relative to these two reference models may therefore be affected by Gaia RV
uncertainties, especially for clusters with relatively few blue-edge stars.
Nevertheless, the difference between the observed RV dispersions of most
clusters and the prediction of the $P=10$~d model is much larger than the
typical measurement uncertainties. The main conclusion concerning the lack
of a dominant close-binary population is therefore not sensitive to this
limitation.

A related limitation concerns the rotational information available for defining the sample. For the ground-based benchmark clusters, the blue-edge sample was selected directly by $v\sin i$ ($v\sin i\leq200$ km s$^{-1}$; Section~\ref{subsection:rv_dispersion}). For the Gaia sample, a corresponding split based on $V_{\rm broad}$ is not feasible: $V_{\rm broad}$ is available for only a fraction of the selected stars, and a statistical correction for inclination cannot assign individual stars to the slow or fast population. The CMD-based blue-edge selection is itself a proxy for the slow-rotator population, and our population-level RV-dispersion test does not require classifying individual stars.

Our simulations assume a single orbital period for all binaries,
while real clusters contain a mixture of periods. We therefore tested
50/50 mixtures of $P = 10$, 100, and 500~d with $P = 1000$~d: under
Gaia-like sampling these give $\sigma \approx 11.4$, 9.2,
and 7.0~km~s$^{-1}$, respectively.
Using these values together with the $P = 1000$~d case to bin the
40 Gaia clusters, 62.5\% of the clusters fall below the
$P = 1000$~d line, 17.5\% lie between the 100-d and 500-d mixture
lines, 10\% between the 10-d and 100-d mixture lines, and only
10\% require a mixture containing close ($P \sim 10$~d) binaries.
The period mixing shifts the exact boundaries but leaves the
overall distribution trend, and our conclusion that most clusters do
not require a significant binary population, unchanged.

In addition, the Gaia DR3 non-single-star information has limited sensitivity
to some long-period spectroscopic binaries. The Gaia DR3 time baseline is
approximately 34 months, and stars observed at similar orbital phases may not
be identified as non-single systems even when multiple RVS transits are
available \citep{2023A&AGaiaRVKatz}. This limitation is particularly relevant
for binaries with orbital periods comparable to, or longer than, the
$\sim1000$~d timescale considered here.

One possible example is the NGC~5822 member Gaia DR3
5887671534558688640. Despite having 20 RVS transits, its combined Gaia RV
uncertainty is $6.49~\mathrm{km~s^{-1}}$, substantially larger than the
expected precision of approximately $1~\mathrm{km~s^{-1}}$ for a star of
similar brightness. The scatter among its individual transit RV measurements
is approximately $29~\mathrm{km~s^{-1}}$. Furthermore, its Gaia RV of
$-80.0~\mathrm{km~s^{-1}}$ differs by approximately
$55~\mathrm{km~s^{-1}}$ from the RV of $-25~\mathrm{km~s^{-1}}$ reported by
\citet{2021MNRASSun}. This difference could plausibly arise if the two data sets sampled the star at 
different orbital phases, although the available measurements do not uniquely 
establish this interpretation.
Similar effects may contribute to the relatively large RV scatter among
NGC~5822 members shown in Figure~\ref{RV_comparision}.

Such long-period binary systems may contribute to the observed RV dispersion,
but they are not expected to undergo efficient tidal synchronization.
Consequently, their presence does not provide a mechanism for producing the
majority of slow rotators through tidal braking.

\subsection{Comparison with previous work}
\label{sec:comparison}

Our results are consistent with, and extend in both sample size and
methodology, a series of cluster-by-cluster tests of the tidal-braking
scenario. We compare our results with each of these studies below.

To investigate the occurrence of detected short-period binary systems in our
sample, we cross-matched the blue-edge samples with the Gaia DR3
\texttt{nss\_two\_body\_orbit} catalogue. Of the 691 stars in the 29 clusters
covered by the match, 43 (6.2\%) have a detected Gaia NSS orbital solution
with $P<10$~d. The matched systems are distributed among 25 clusters, with
only NGC~6940 (6 systems) and Stock~2 (4 systems) contributing more than
three systems. Thus, only a small fraction of the matched stars have detected
Gaia NSS orbital solutions at $P<10$~d. This result is consistent with our
RV-dispersion analysis, although the Gaia NSS catalogue is subject to
selection effects and does not provide a complete census of close binaries.

\citet{2020MNRASKamann,2021MNRASKamann} used multi-epoch MUSE integral-field
spectroscopy of several hundred main-sequence stars in the Magellanic Cloud
clusters NGC~1856 and NGC~1850 to measure the detected binary fraction among
blue and red main-sequence stars. They found no significant difference
between the two populations, arguing against tidal braking as the origin of
the slow rotators. Our RV-dispersion analysis of the 40-cluster Gaia sample
reaches a similar conclusion, while the eight ground-based benchmark clusters
provide an independent check of the Gaia-based measurements.

\citet{2023MNRASHe} measured RV variations of slow rotators in the Galactic
open cluster NGC~2422 and showed that most of them are unlikely to be
short-period binaries. This is methodologically the closest study to ours,
but it is limited to a single cluster. NGC~2422 is included in our Gaia
sample, where its RV dispersion of
$3.69 \pm 0.98~\mathrm{km~s^{-1}}$ falls below even the prediction of the
$P=1000$~d model. The agreement between the two independent analyses for this
common cluster supports the extension of this test to a larger cluster
sample.

\citet{2026ApJRao&Chen} examined 53 Galactic open clusters using projected rotational velocities from the Gaia-ESO Survey and Gaia DR3 line broadening. They found no statistically significant correlation between the slow-rotator fraction and either the binary fraction or cluster age.
Their binary analysis includes 52 known MSTO binaries, 30 of which have
available $v\sin i$ measurements. Our RV-dispersion test is complementary
to their rotational analysis, and the two studies reach consistent
conclusions regarding the limited role of close binaries.

\subsection{Outlook: Gaia DR4 epoch radial velocities}
\label{sec:dr4}

We note two limitations of the present Gaia-based analysis that can be
addressed with the forthcoming Gaia DR4. First, the DR3 published RV is the
median of the epoch measurements for bright sources
($G_{\rm RVS}<12~\mathrm{mag}$), which suppresses part of the binary orbital
signal (Section~\ref{subsection:binary_simulation}). Second, our analysis was
restricted to bright stars with
\texttt{rv\_method\_used}=1. For fainter sources
($G_{\rm RVS}\gtrsim12~\mathrm{mag}$), DR3 provides only an RV derived from
the combined cross-correlation function, which is equivalent to a mean-like
combination of the individual transits. Because our test relies on RV
variability as a tracer of binarity, this combination largely erases the
differences between the period models, and these sources were therefore
excluded.

The individual epoch RVs expected from DR4 will allow RV variability to be
measured directly for a substantially larger fraction of the cluster sample.
The same method can then be applied more homogeneously to fainter stars,
increasing the usable sample size and improving the statistical significance
of the test. Repeating the present analysis with DR4 epoch RVs will therefore
provide a more direct and stringent test of the role of binary systems in the
formation of slow rotators.

\section{Conclusions}
\label{section 5}

Using ESO and SALT spectroscopy for eight benchmark clusters, together with
Gaia DR3 RV measurements for a larger sample of 40 open clusters younger than
$2~\mathrm{Gyr}$, we investigated the RV dispersions of stars selected from the blue-edge 
slow-rotator population, defined as stars with $v\sin i \leq 200~\mathrm{km~s^{-1}}$ or selected using the rotational-contour boundary, associated with
the eMSTOs of open clusters. 
All 40 clusters have RV dispersions below the $P=10$ d prediction, the regime in which tidal synchronization can be effective within the cluster ages. Within the wider period range proposed for the tidal-braking scenario, 30 clusters (75\%) fall below the $P=500$ d model.

We conclude that tidal braking in binaries is unlikely to be the dominant origin of the slow-rotator populations associated with the eMSTOs of young and intermediate-age open clusters.
This conclusion does not exclude
the presence of undetected binaries or a contribution from wider and
long-period systems.

\begin{acknowledgements}
      C.L. acknowledges support from the Key Project of the National Natural Science Foundation of China (Grant No. 12233013). C.H. acknowledges support from the National Natural Science Foundation of China (Grant No. 12503045).
      
      Based on observations collected at the European Southern Observatory under ESO programmes 380.D-0161, 193.B-0936, 095.C-0818, 071.D-0127, and 079.D-0178. Some of the observations reported in this paper were obtained with the Southern African Large Telescope (SALT) under programmes 2017-2-SCI-038, 2018-1-SCI-006, and 2018-2-SCI-002. This work has made use of data from the European Space Agency (ESA) mission Gaia (\url{https://www.cosmos.esa.int/gaia}), processed by the Gaia Data Processing and Analysis Consortium (DPAC, \url{https://www.cosmos.esa.int/web/gaia/dpac/consortium}). 
\end{acknowledgements}

  \bibliographystyle{aa} 
  \bibliography{bibtex/bibref} 

\begin{appendix}

\onecolumn
\section{Detailed parameters and radial velocity distribution plots for each star cluster.}

\begin{table}[ht!]
\caption{Cluster parameters and RV dispersions of the Gaia sample}
\label{table:gaia}
\centering
\small
\begin{tabular}{l c c c c c c}
\hline\hline
Cluster & Age& N & $\sigma_{\rm corr}$ & Period range & N$_{\rm narrow}$ & $\sigma_{\rm narrow,corr}$ \\
        & (Myr) &  & (km~s$^{-1}$) &  & & (km~s$^{-1}$)\\
\hline
Alessi\_37 & 180 & 11(55\%) & $7.09\pm1.95$ &  $P$=1000--500~d& 6(30\%) & --\\
ASCC\_113 & 90 & 24(58\%) & $5.54\pm1.12$ & $<~P$=1000~d& 13(32\%) & $2.53\pm0.76$\\
COIN-Gaia\_13 & 190 & 12(67\%) & $4.63\pm1.23$ & $<~P$=1000~d & 5(28\%) & --\\
Collinder\_463 & 290 & 31(33\%) & $6.73\pm0.94$ & $P$=1000--500~d& 29(31\%) & $6.72\pm1.04$\\
IC\_2602 & 220 & 20(83\%) & $2.42\pm0.61$ &  $<~P$=1000~d& 7(29\%) & --\\
IC\_4651 & 1780 & 18(57\%) & $5.03\pm1.96$ & $<~P$=1000~d& 11(31\%) & $6.64\pm2.45$\\
IC\_4725 & 90 & 30(58\%) & $10.26\pm1.36$ & $P$=500--100~d& 20(38\%) & $10.54\pm1.77$\\
Mamajek\_4 & 650 & 14(52\%) & $4.49\pm1.13$ & $<~P$=1000~d& 8(30\%) &--\\
NGC\_1039 & 240 & 22(58\%) & $2.86\pm0.52$ & $<~P$=1000~d& 12(32\%) & $3.36\pm0.65$\\
NGC\_1647 & 200 & 24(57\%) & $4.48\pm0.66$ & $<~P$=1000~d& 13(31\%) & $4.13\pm0.93$\\
NGC\_1912 & 220 & 13(46\%) & $14.55\pm4.56$ & $P$=100--10~d& 8(29\%) &--\\
NGC\_2099 & 355 & 10(26\%) & $10.67\pm2.71$ & $P$=500--100~d& 9(23\%) &--\\
NGC\_2168 & 180 & 24(61\%) & $11.77\pm2.33$ & $P$=500--100~d& 12(31\%) & $12.13\pm3.66$\\
NGC\_2281 & 610 & 28(67\%) & $3.11\pm1.32$ & $<~P$=1000~d& 14(33\%) & $0.70\pm1.77$\\
NGC\_2287 & 220 & 28(43\%) & $4.08\pm1.04$ &$<~P$=1000~d& 21(33\%) & $3.20\pm0.79$\\
NGC\_2301 & 220 & 10(50\%) & $10.31\pm3.00$ & $P$=500--100~d& 6(30\%) &--\\
NGC\_2360 & 450 & 14(47\%) & $8.16\pm2.17$ & $P$=500--100~d& 9(30\%) &--\\
NGC\_2422 & 130 & 23(69\%) & $3.69\pm0.98$ &  $<~P$=1000~d& 9(27\%) &--\\
NGC\_2423 & 1060 & 14(50\%) & $2.14\pm1.03$ &  $<~P$=1000~d& 8(29\%) &-- \\
NGC\_2437 & 220 & 12(38\%) & $6.13\pm1.95$ &  $<~P$=1000~d& 9(28\%) &--\\
NGC\_2447 & 480 & 27(60\%) & $7.48\pm1.48$ &$P$=1000--500~d & 12(27\%) & $10.82\pm2.54$\\
NGC\_2516 & 300 & 31(67\%) & $5.40\pm0.84$ &  $<~P$=1000~d& 14(30\%) & $3.88\pm1.45$\\
NGC\_2527 & 780 & 33(60\%) & $3.63\pm0.44$ &  $<~P$=1000~d& 16(29\%) & $4.01\pm1.05$\\
NGC\_2539 & 540 & 10(50\%) & $2.07\pm1.78$ &  $<~P$=1000~d& 6(30\%) &--\\
NGC\_2546 & 140 & 28(62\%) & $8.98\pm2.34$ &  $P$=500--100~d& 13(29\%) & $9.33\pm2.94$\\
NGC\_2548 & 530 & 34(65\%) & $5.43\pm1.10$ & $<~P$=1000~d& 17(33\%) & $3.24\pm1.25$\\
NGC\_3114 & 200 & 62(34\%) & $9.21\pm0.98$ &  $P$=500--100~d& 50(29\%) & $8.53\pm0.85$\\
NGC\_5822 & 790 & 36(59\%) & $5.41\pm1.47$ & $<~P$=1000~d & 19(31\%) & $6.67\pm1.38$\\
NGC\_6025 & 150 & 13(45\%) & $8.89\pm1.76$ & $P$=500--100~d& 8(28\%) &--\\
NGC\_6281 & 340 & 17(53\%) & $3.30\pm1.12$ & $<~P$=1000~d & 10(31\%) & $4.61\pm1.61$\\
NGC\_6405 & 110 & 26(68\%) & $4.20\pm1.14$ & $<~P$=1000~d & 11(29\%) & $5.58\pm1.55$\\
NGC\_6475 & 250 & 41(72\%) & $2.63\pm0.61$ &$<~P$=1000~d& 17(30\%) & $3.51\pm1.06$\\
NGC\_6494 & 330 & 17(47\%) & $7.55\pm1.41$ & $P$=1000--500~d& 10(28\%) & $7.16\pm2.52$\\
NGC\_6633 & 680 & 33(71\%) & $5.31\pm1.57$ &$<~P$=1000~d&  14(30\%) & $6.49\pm4.05$\\
NGC\_6811 & 960 & 12(34\%) & $9.24\pm2.51$ & $P$=500--100~d& 11(31\%) & $8.64\pm3.00$\\
NGC\_6940 & 960 & 13(37\%) & $4.89\pm1.37$ &  $<~P$=1000~d& 10(29\%) & $3.30\pm1.22$\\
NGC\_7209 & 630 & 27(49\% & $4.90\pm0.88$ & $<~P$=1000~d& 16(29\%) &  $5.39\pm1.49$\\
Stock\_2 & 280 & 29(43\%) & $3.39\pm1.15$ &  $<~P$=1000~d& 18(26\%) & $3.08\pm1.22$\\
Theia\_274 & 130 & 22(67\%) & $6.22\pm1.37$ & $P$=1000--500~d& 10(30\%) & $4.96\pm2.90$ \\
Trumpler\_2 & 90 & 14(52\%) & $6.14\pm1.88$ &  $<~P$=1000~d& 8(30\%) &--\\

\hline
\end{tabular}
\tablefoot{ {N and N$_{\rm narrow}$ are the numbers of stars in the fiducial and narrow blue-edge samples, respectively, with Gaia DR3 RVs entering the dispersion analysis
($G_{\rm RVS}<12$~mag, \texttt{rv\_method\_used}=1). The values in parentheses give the percentages of the retained eMSTO stars by each selection (Section~\ref{subsection:gaia_cluster_sample}).
$\sigma_{\rm corr}$ and $\sigma_{\rm narrow,corr}$ are the RV dispersions corrected in quadrature for Gaia per-star RV uncertainties and the intrinsic cluster velocity dispersion \ref{subsection:gaia_cluster_sample},
with bootstrap uncertainty. Period range is the pair of simulated
binary-population models ($\sigma_{\rm RV} = 17.76$, 12.10, 7.74, and
6.20~km~s$^{-1}$ for $P = 10$, 100, 500, and 1000~d; Section~\ref{subsection:binary_simulation})
between whose predicted dispersions the measured $\sigma_{\rm corr}$
falls.
For clusters with $N_{\rm narrow}<10$, the actual number of usable RV measurements is listed, but $\sigma_{\rm narrow}$ is not reported because such small samples do not provide reliable dispersion estimates. These clusters are marked with `--' and excluded from the narrow-sample comparison.}
Ages are from the PARSEC isochrone fits (Section~\ref{subsection:cluster_selection}).
}
\end{table}
    
    \begin{figure}[h]
    \centering
        \includegraphics[width=0.325\hsize]{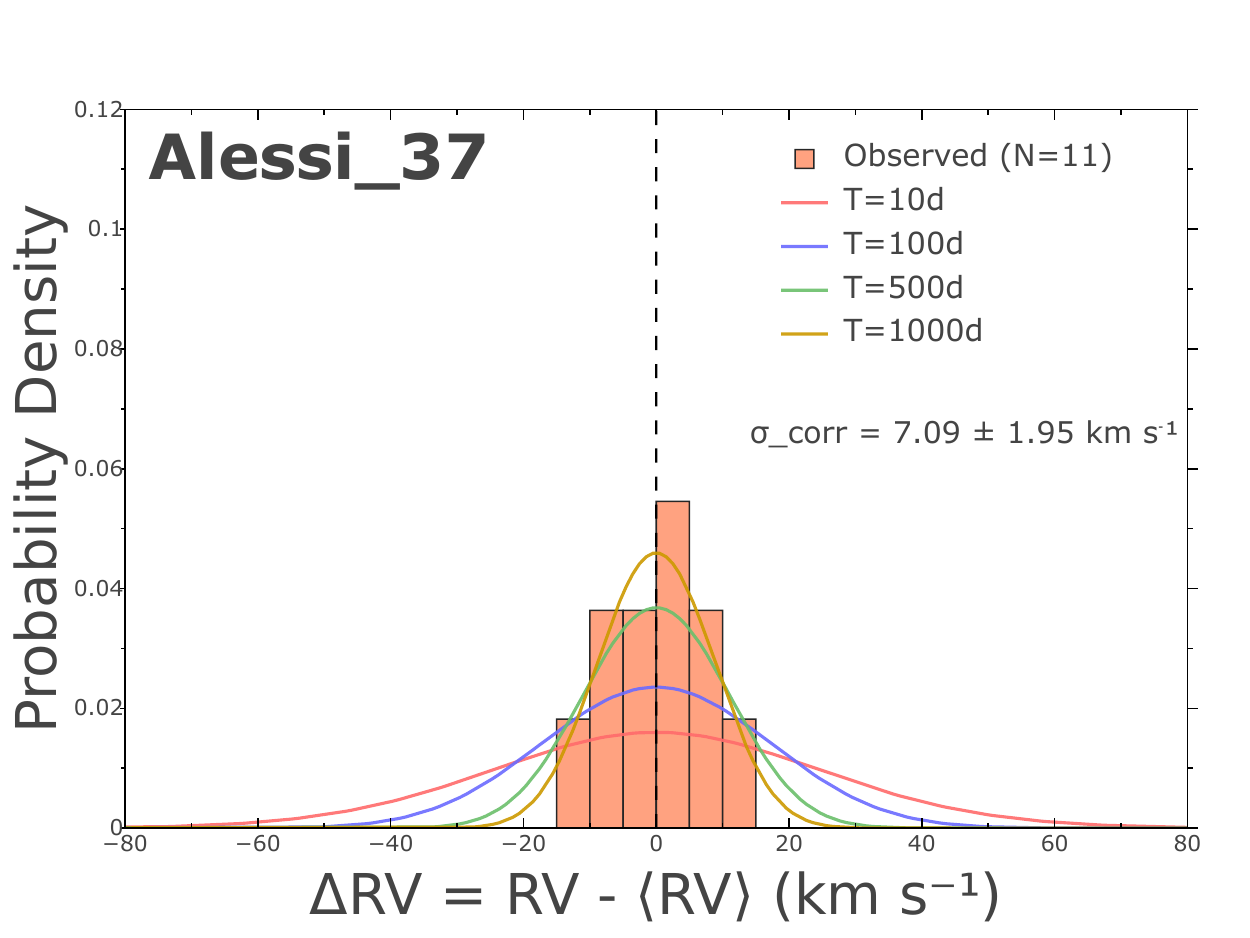}
        \includegraphics[width=0.325\hsize]{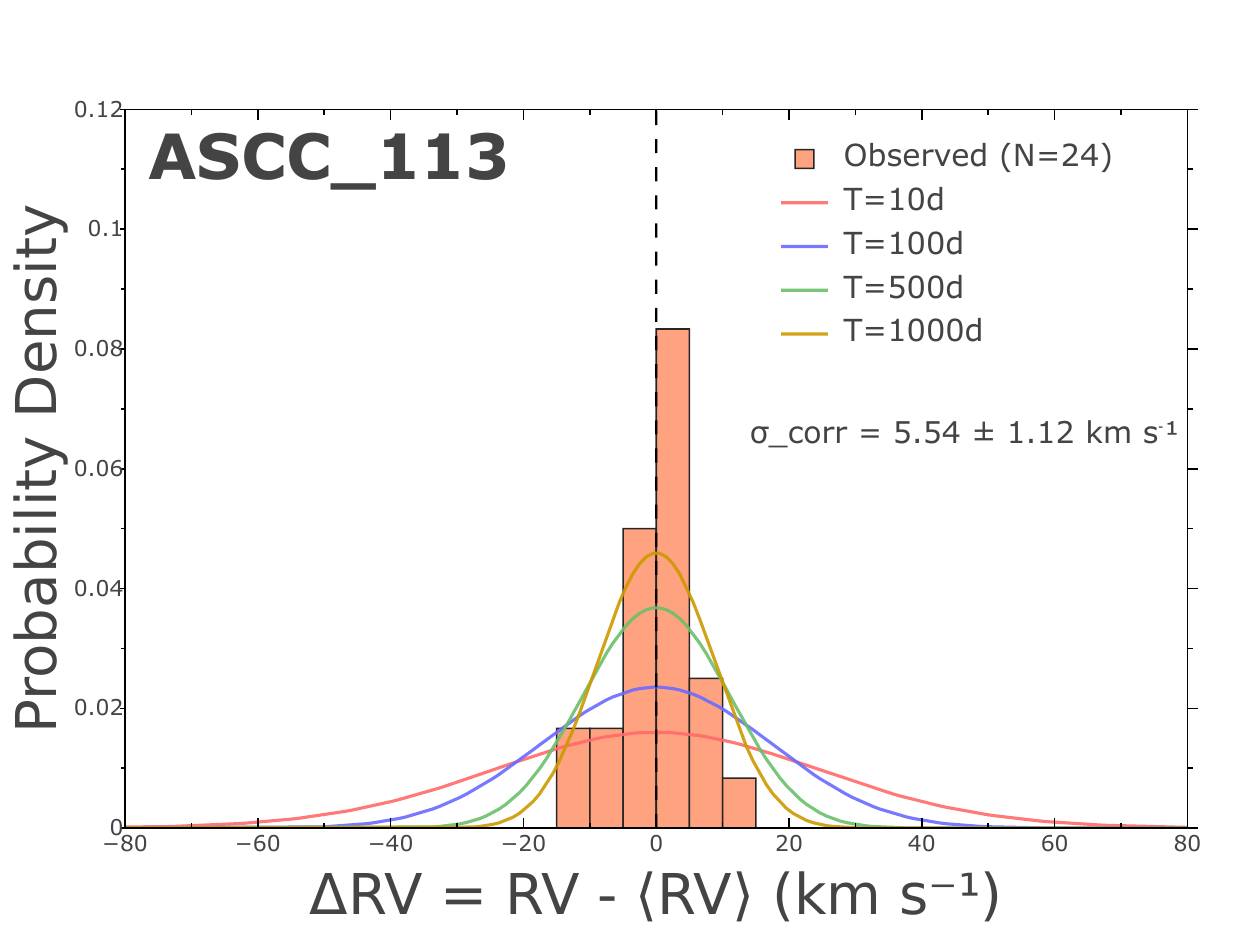}
        \includegraphics[width=0.325\hsize]{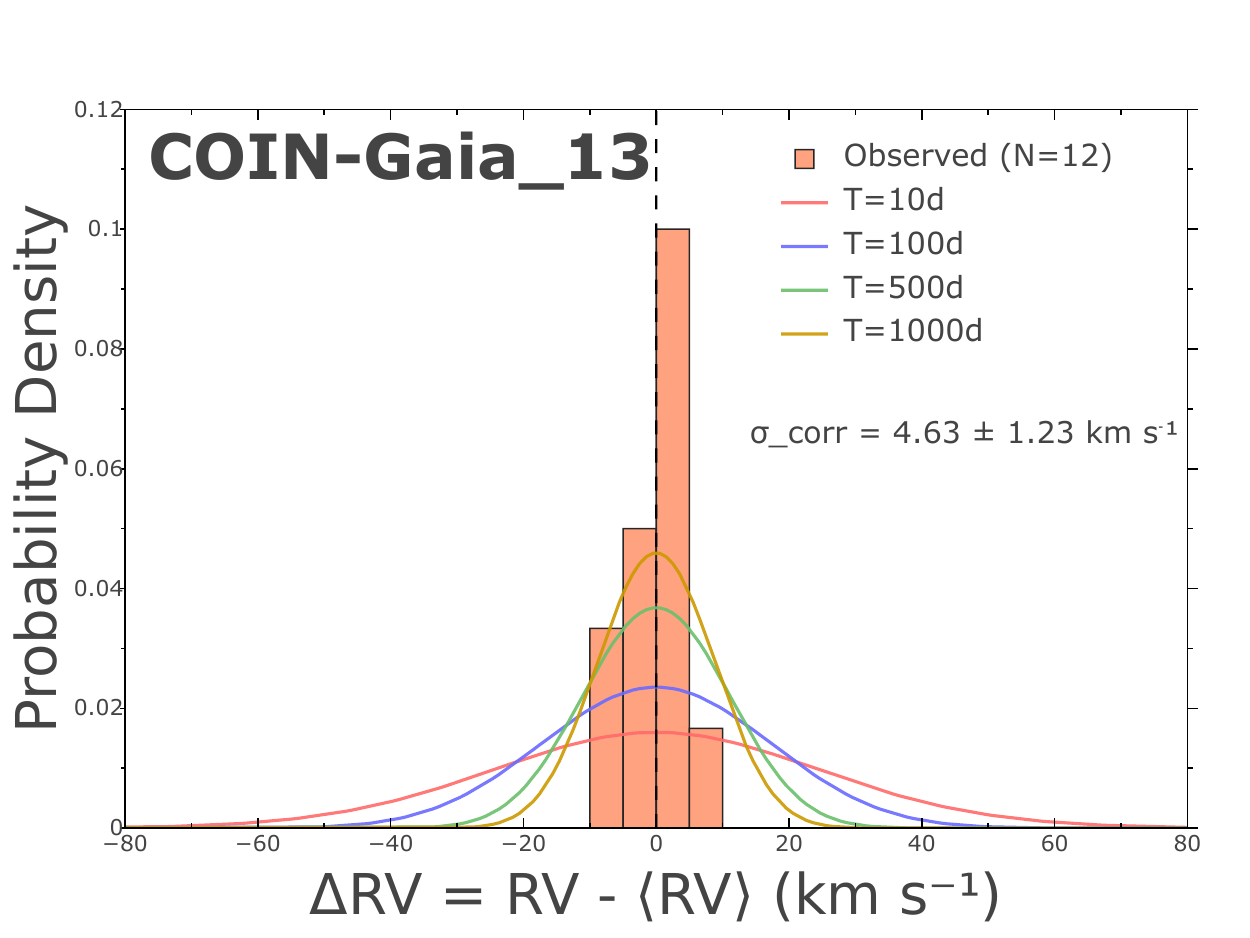}
        \includegraphics[width=0.325\hsize]{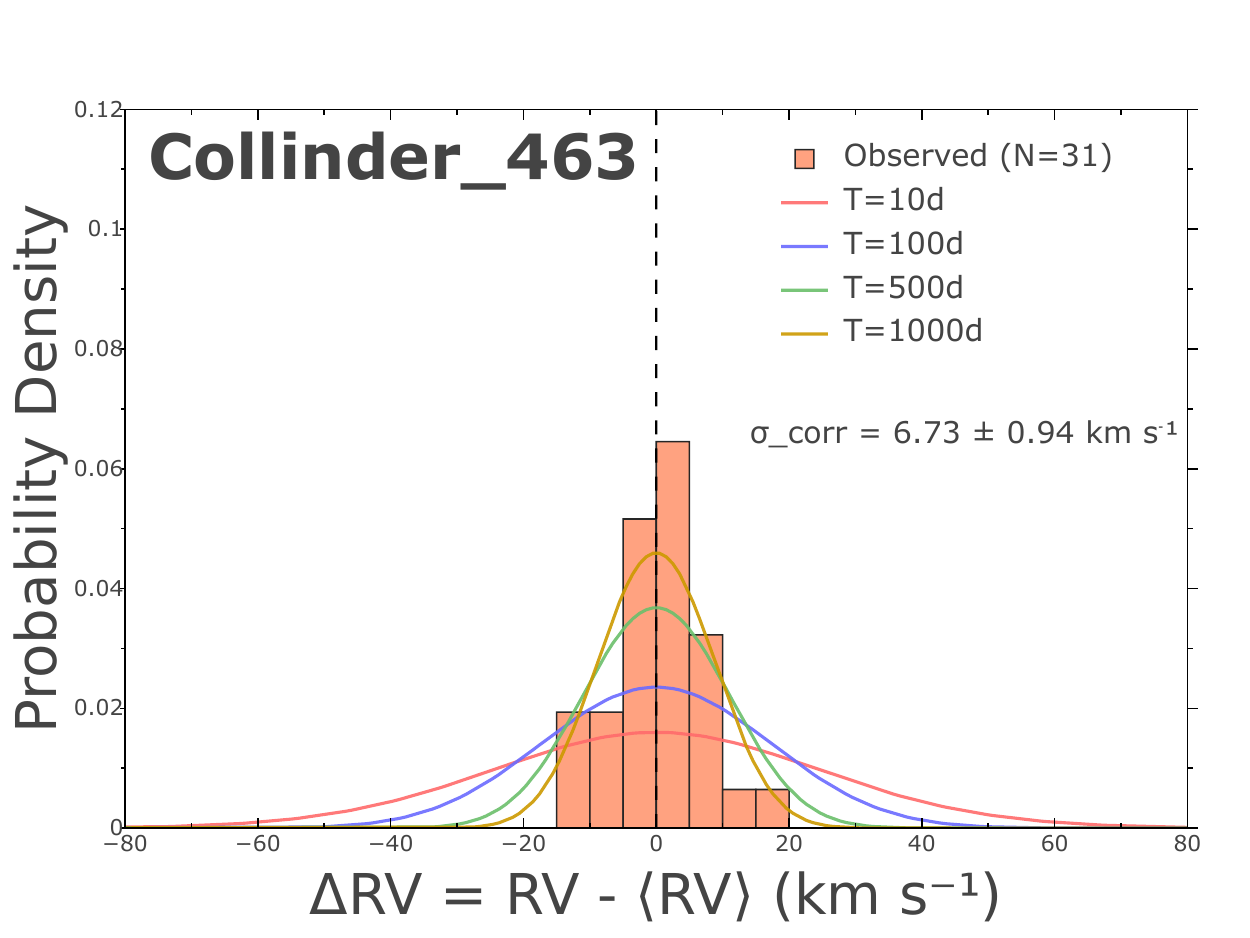}    \includegraphics[width=0.325\hsize]{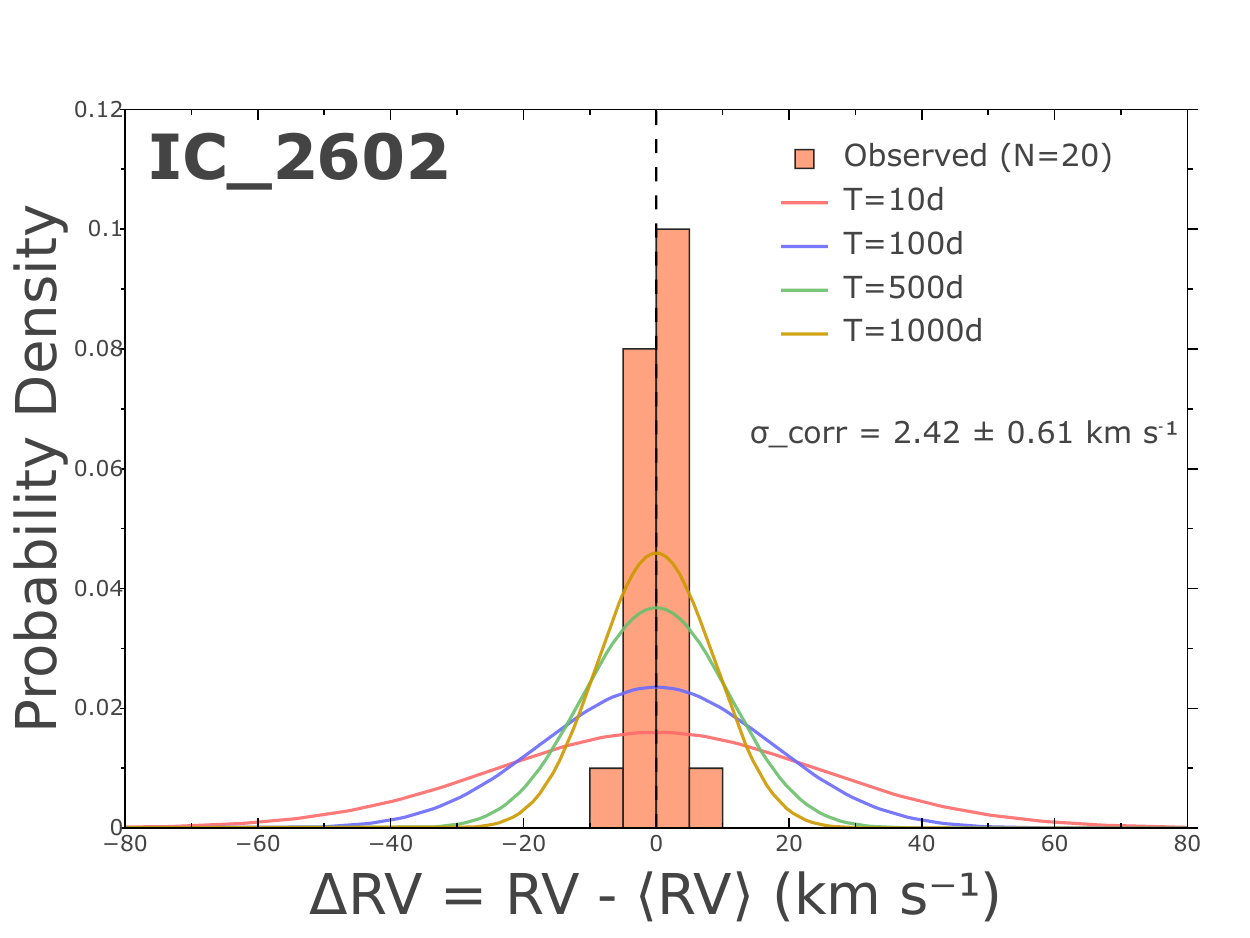}
        \includegraphics[width=0.325\hsize]{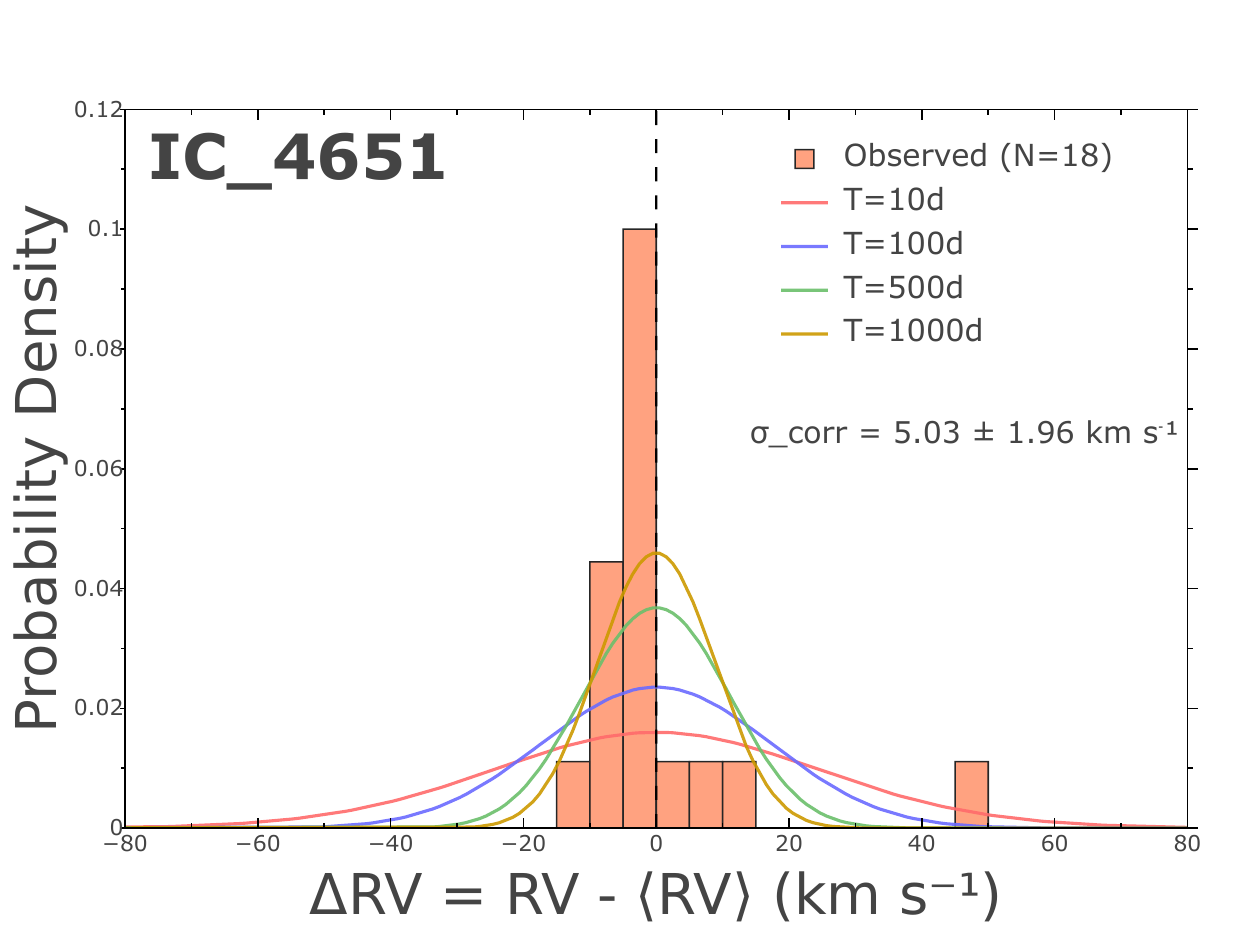}
        \includegraphics[width=0.325\hsize]{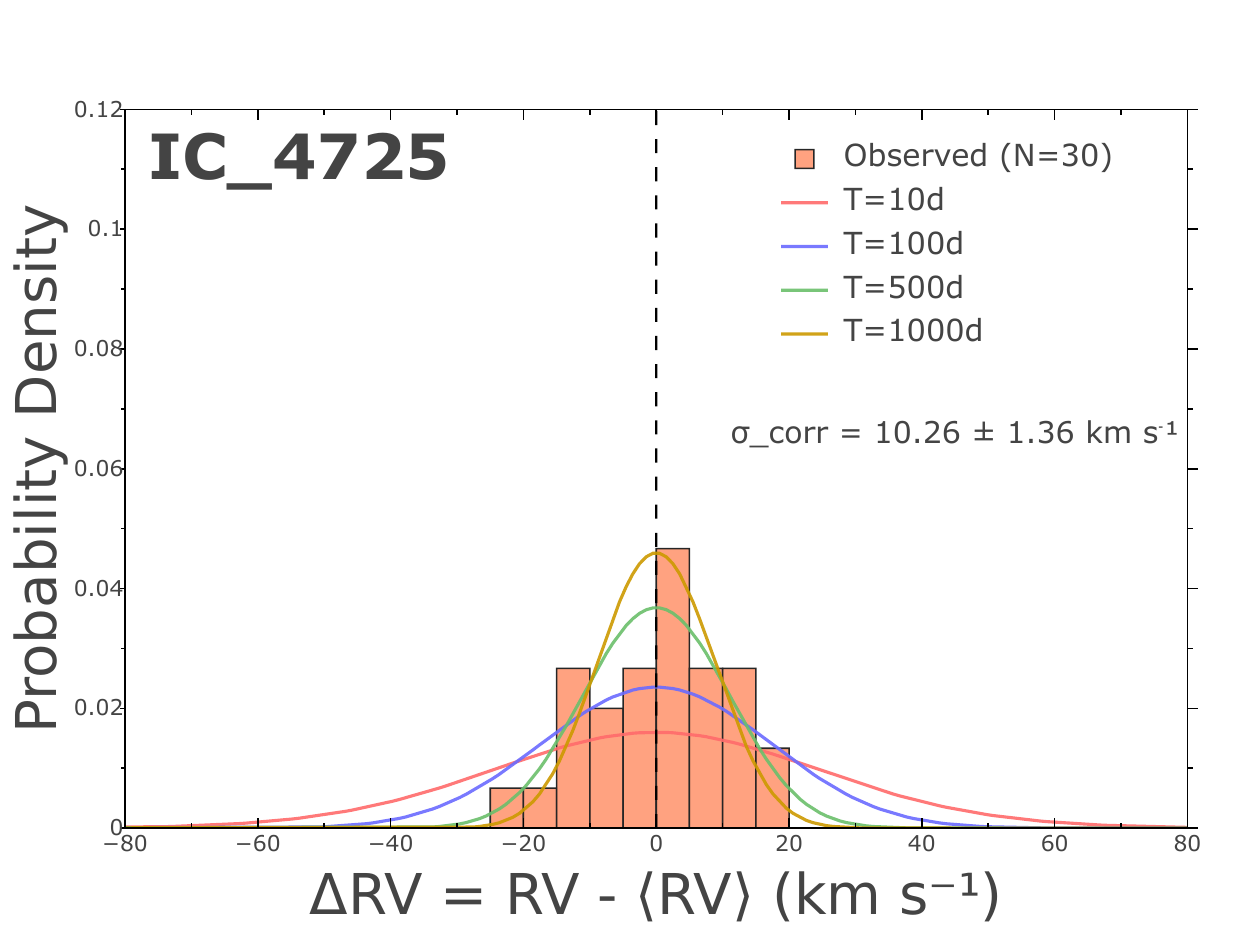}
        \includegraphics[width=0.325\hsize]{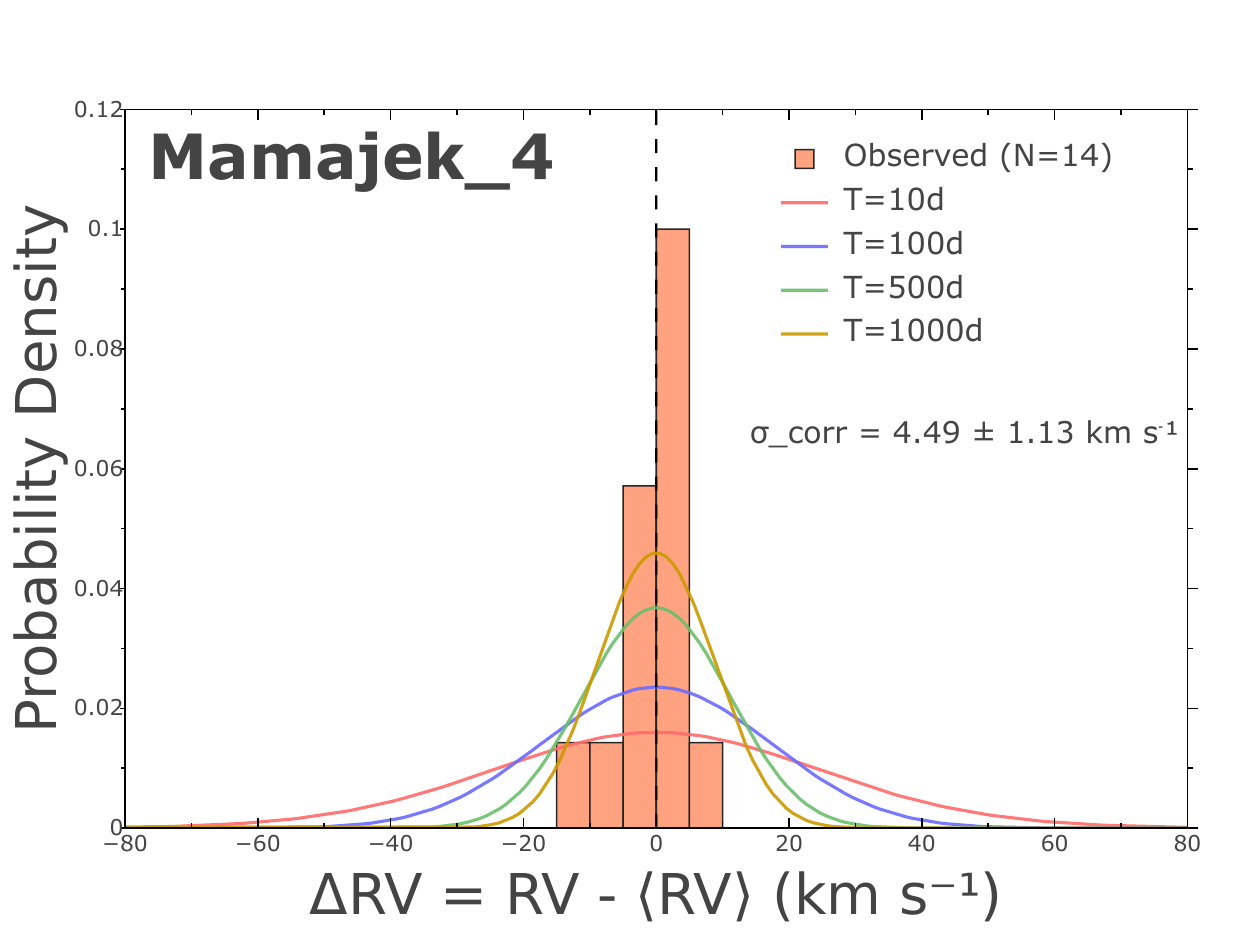}
        \includegraphics[width=0.325\hsize]{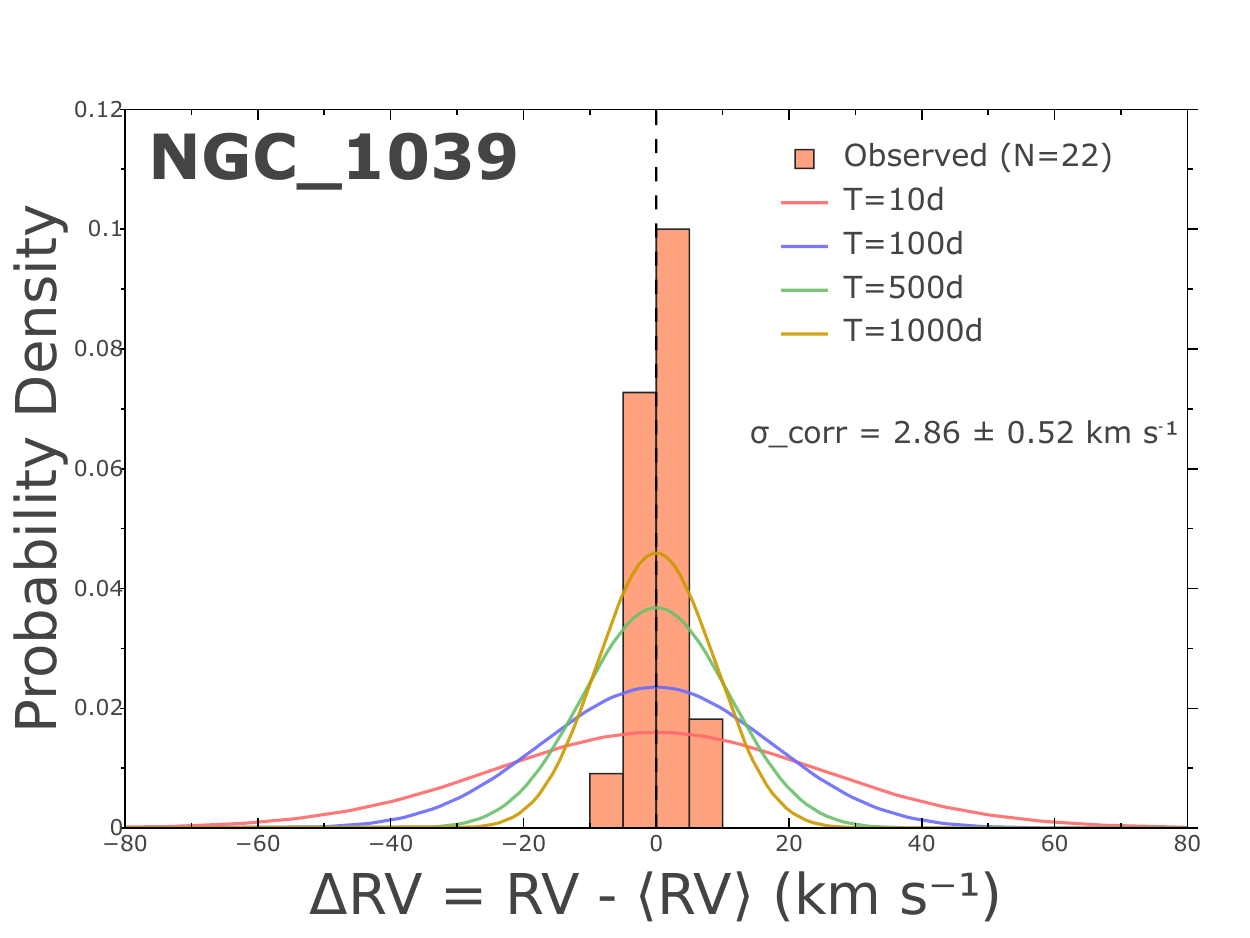}
        \includegraphics[width=0.325\hsize]{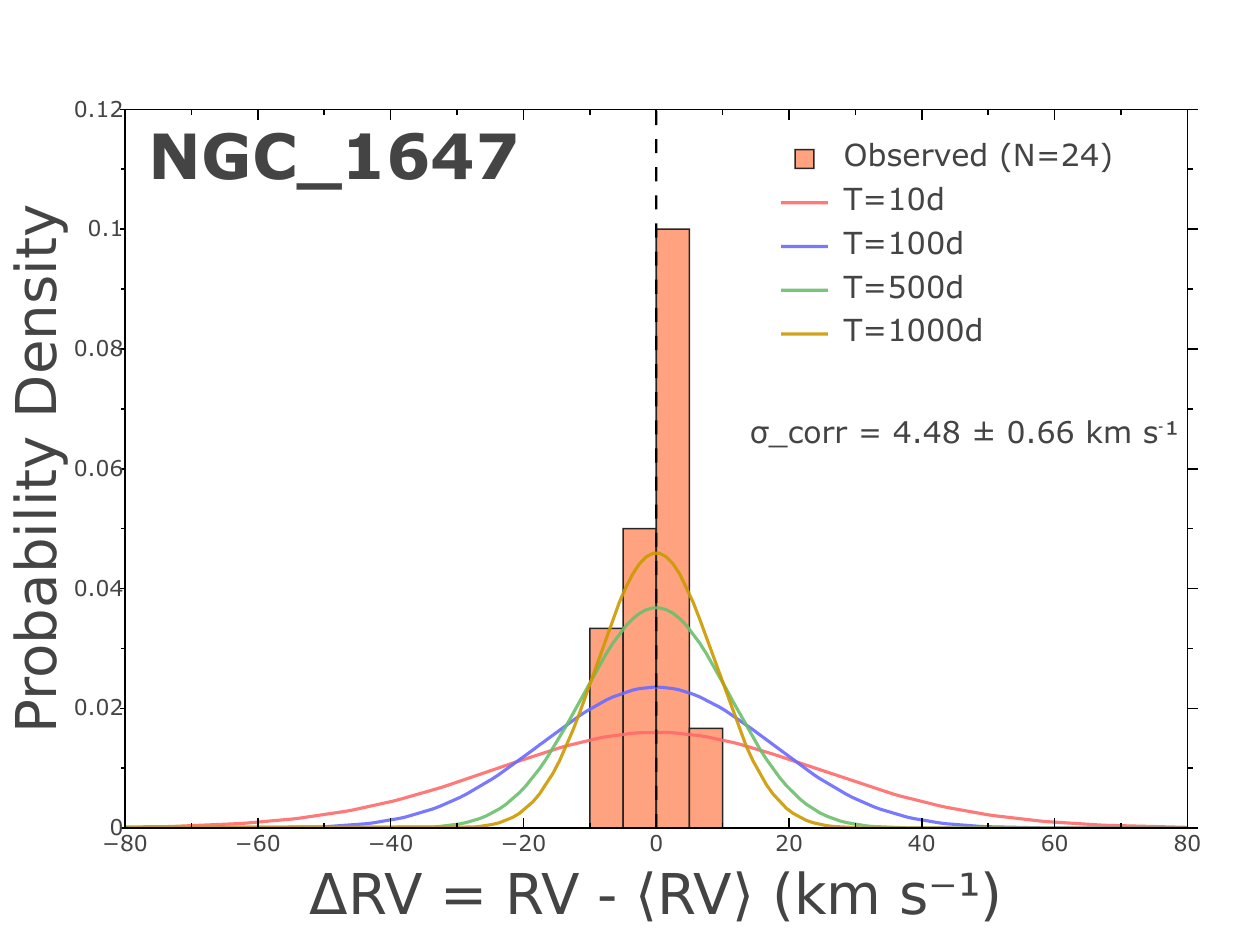}
        \includegraphics[width=0.325\hsize]{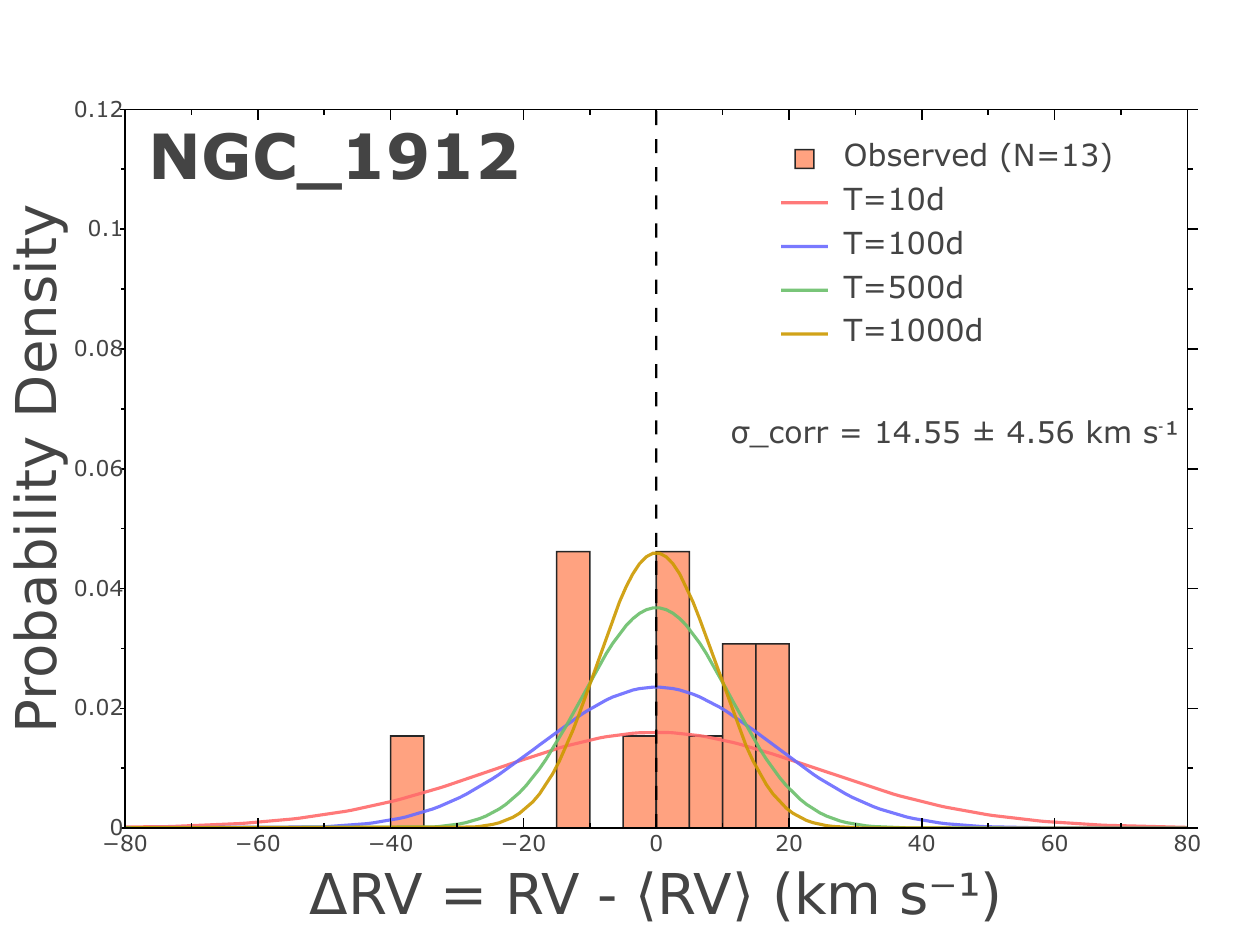}
        \includegraphics[width=0.325\hsize]{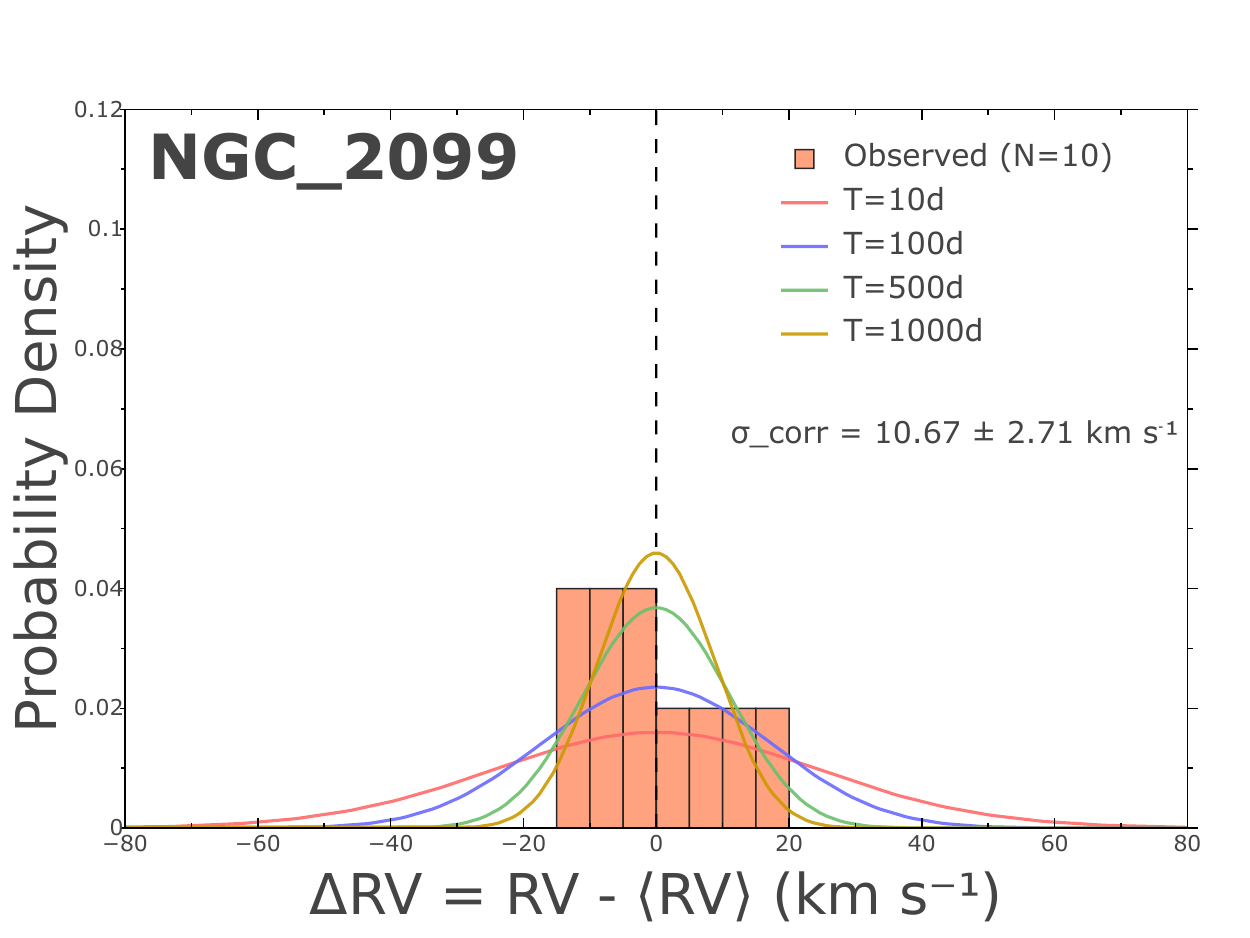}       
        \includegraphics[width=0.325\hsize]{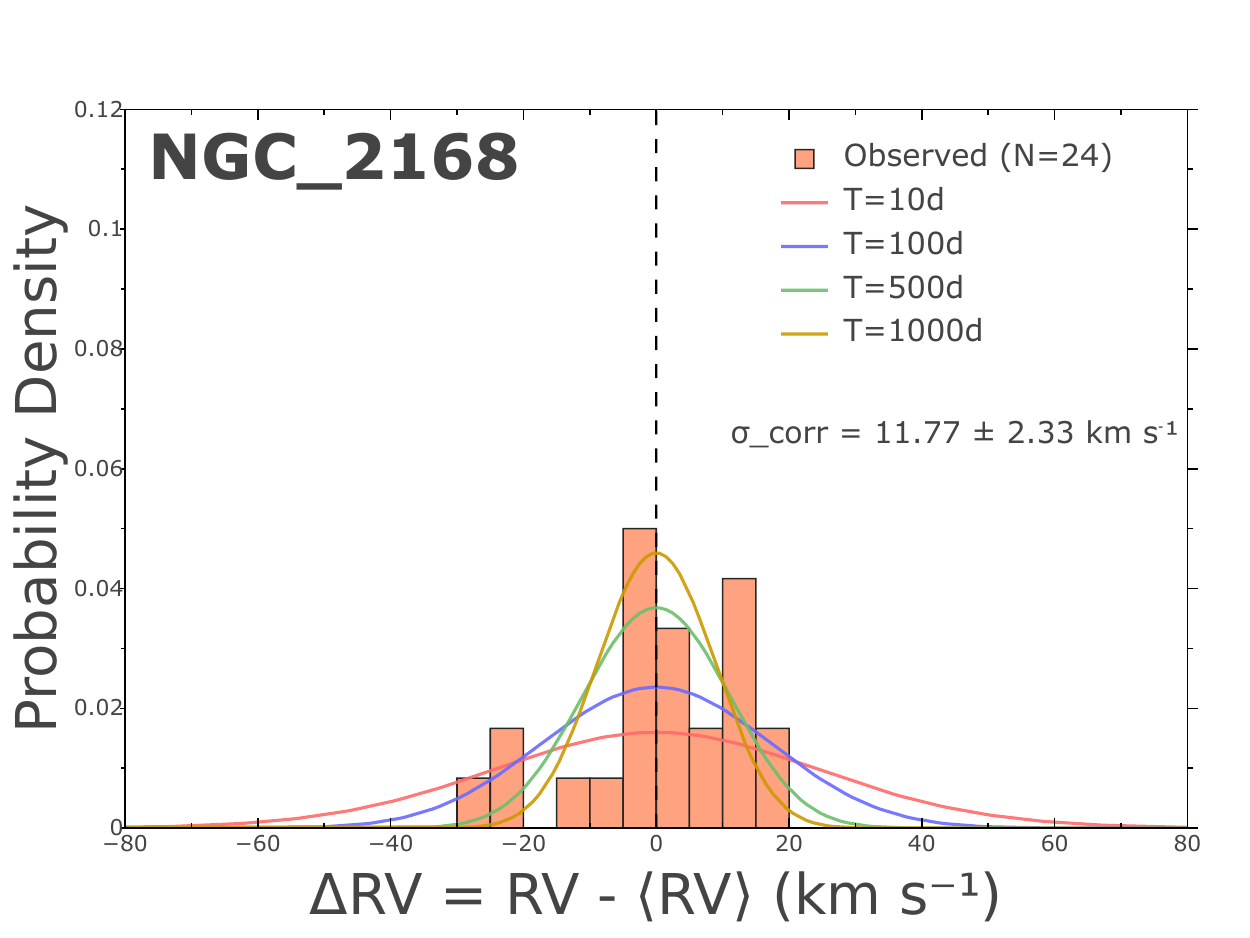}
        \includegraphics[width=0.325\hsize]{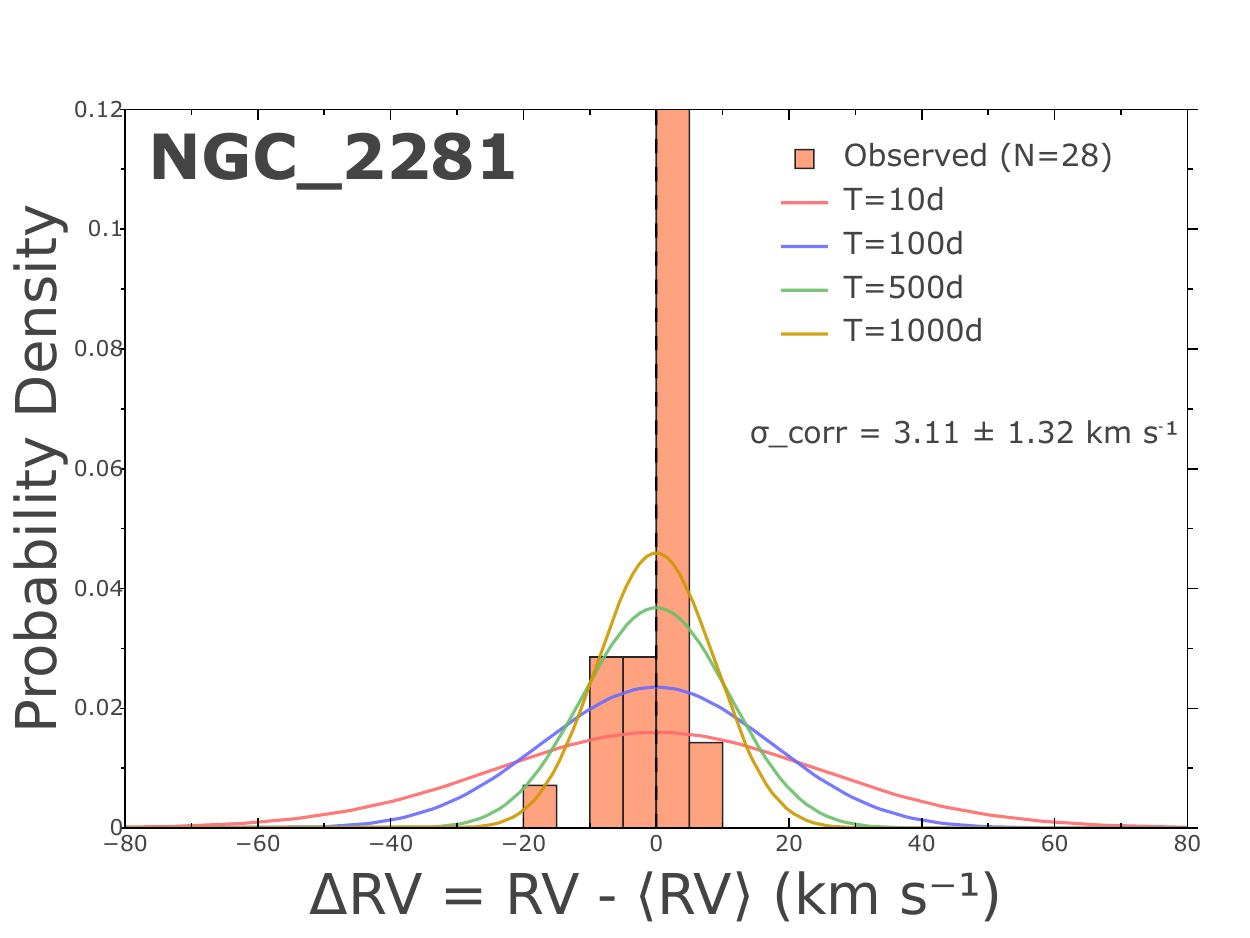}
        \includegraphics[width=0.325\hsize]{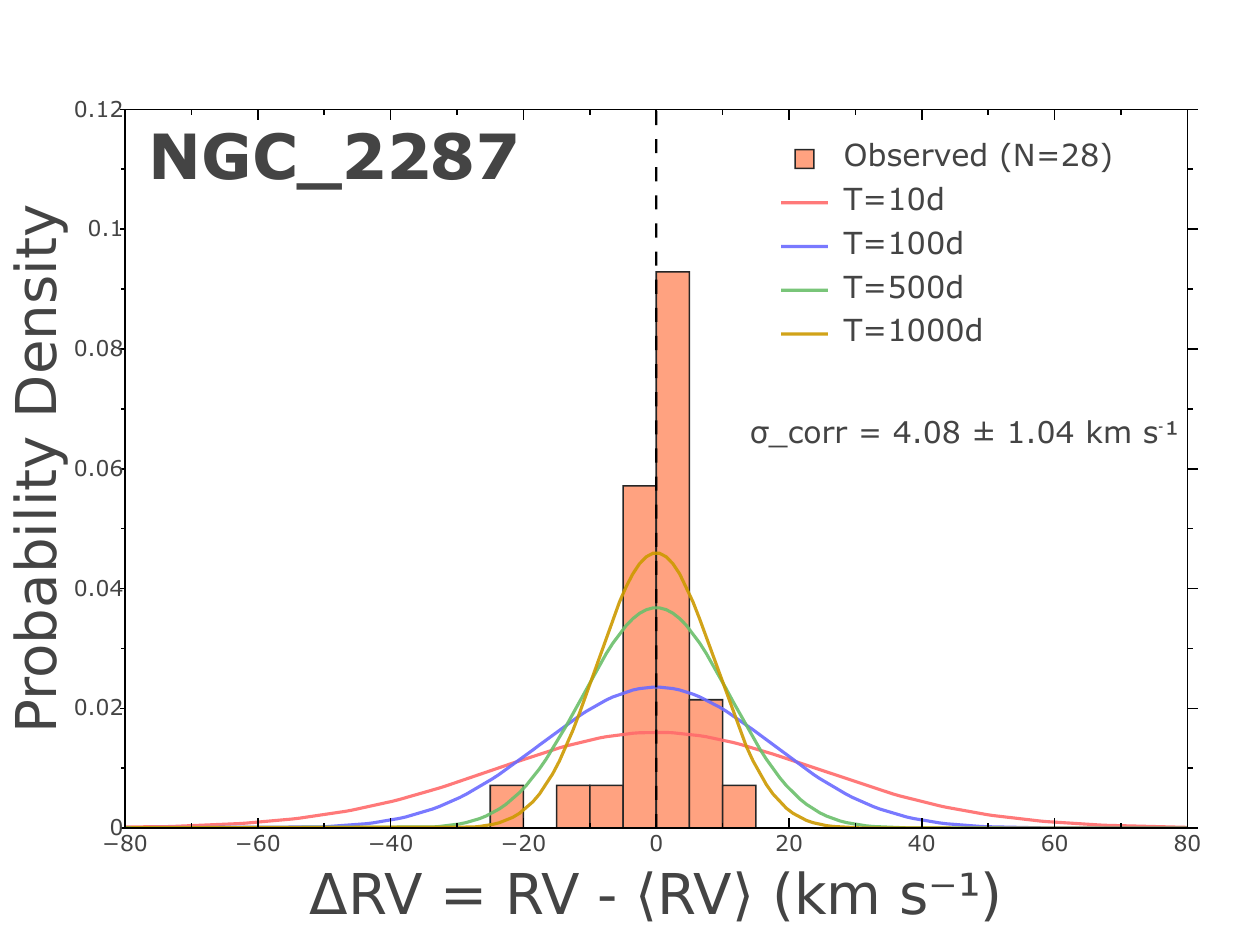}

        \caption{
        RV distribution diagrams of all 40 selected open clusters compared with tidal-locking model predictions. 
        The orange bins are the RV dispersion measured by RV from Gaia DR3, and its width is $5~\mathrm{km~s^{-1}}$. 
        The curves are as same as the curves in Figure\ref{RVdispersion_obs}.
        }        
        \label{RVdispersion_all}
    \end{figure}
    \begin{figure}
        \ContinuedFloat*
        \includegraphics[width=0.325\hsize]{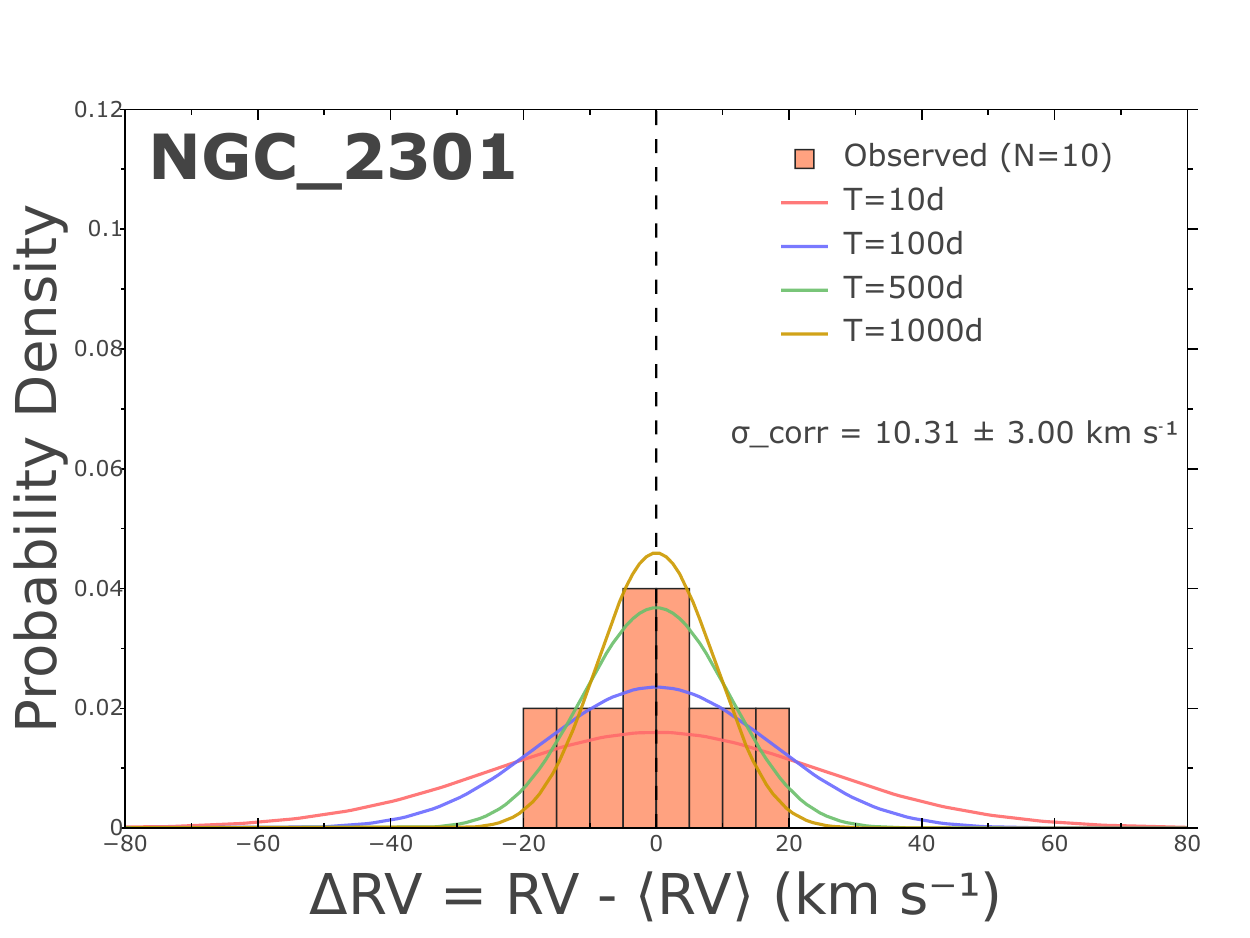}
        \includegraphics[width=0.325\hsize]{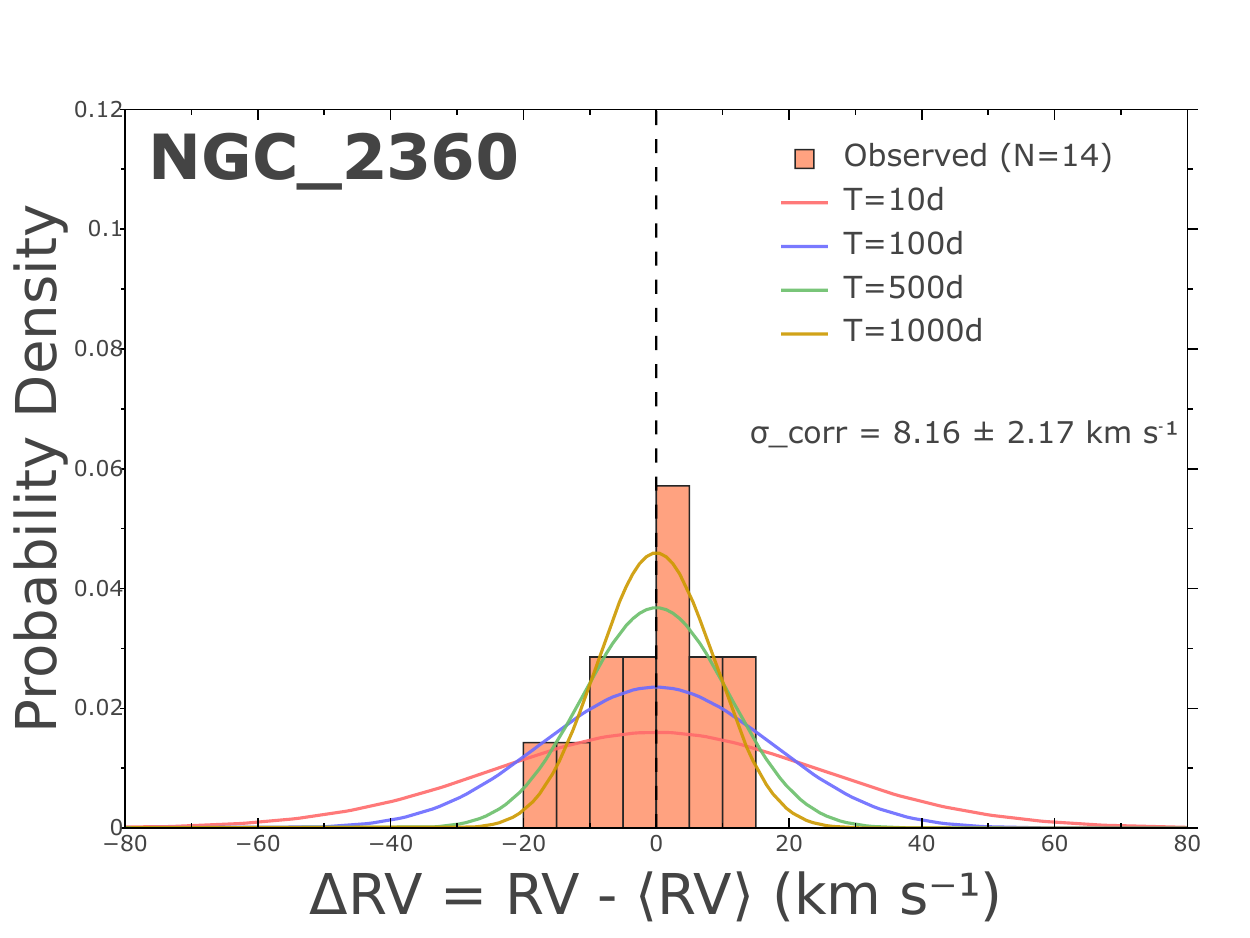}
        \includegraphics[width=0.325\hsize]{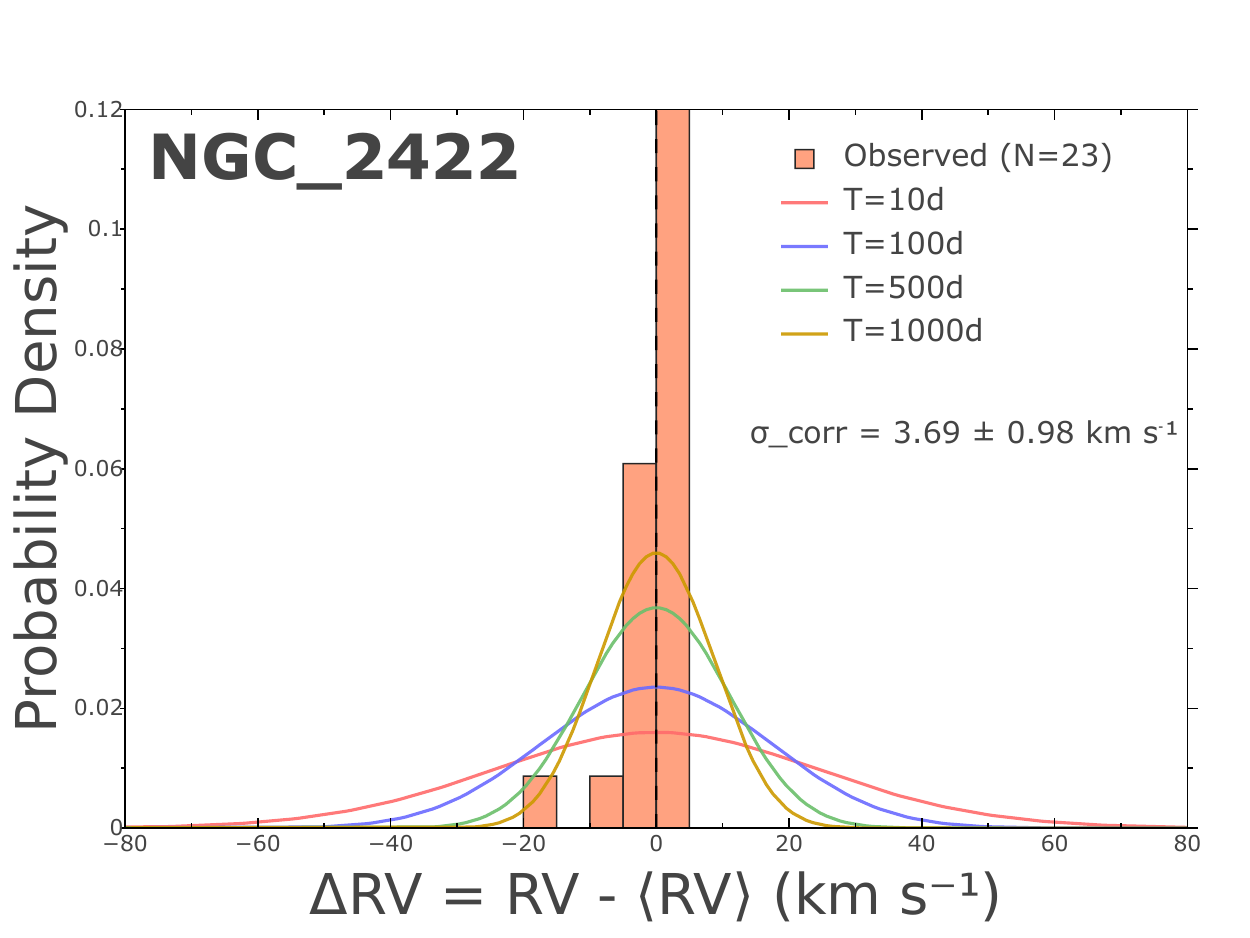}
        \includegraphics[width=0.325\hsize]{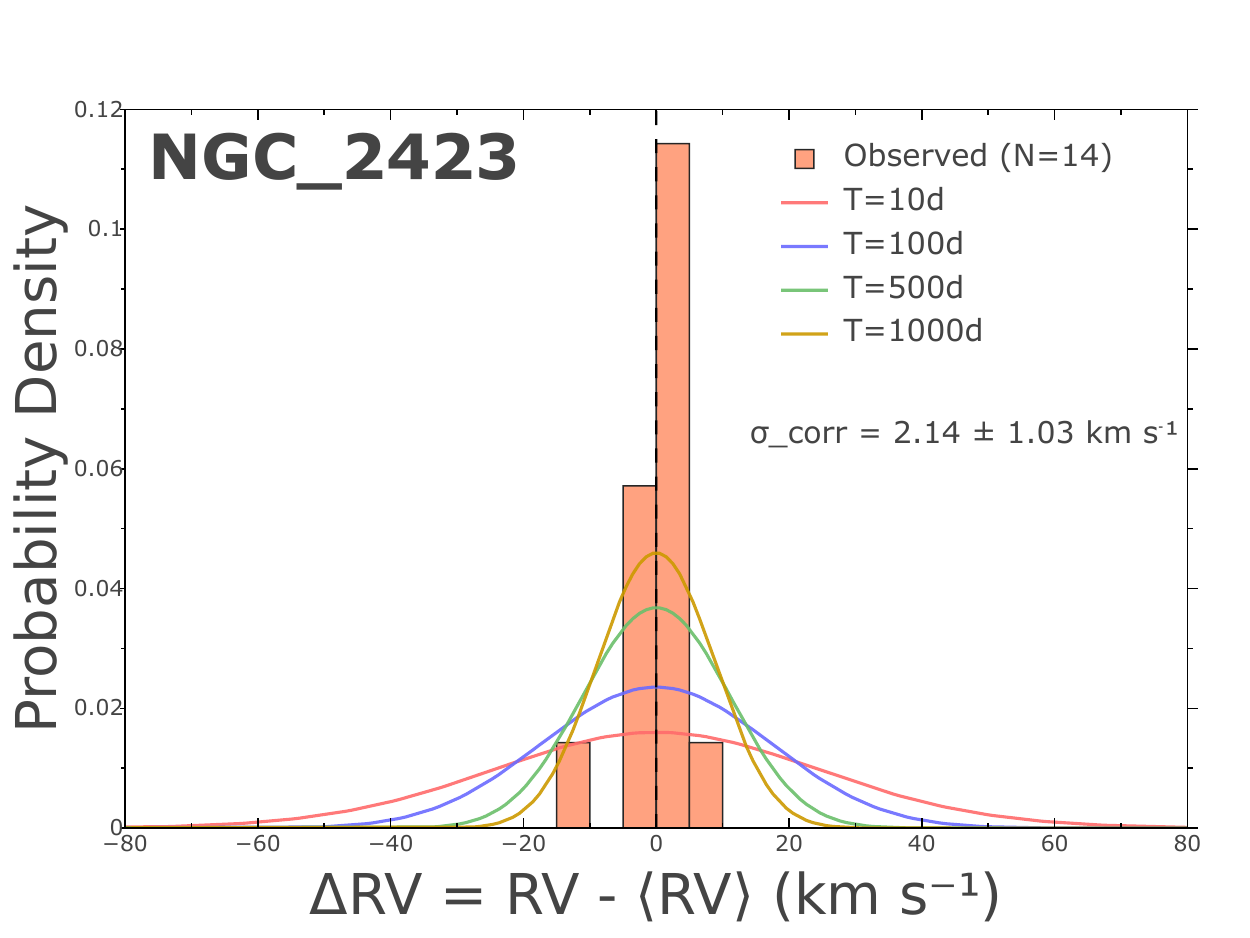}
        \includegraphics[width=0.325\hsize]{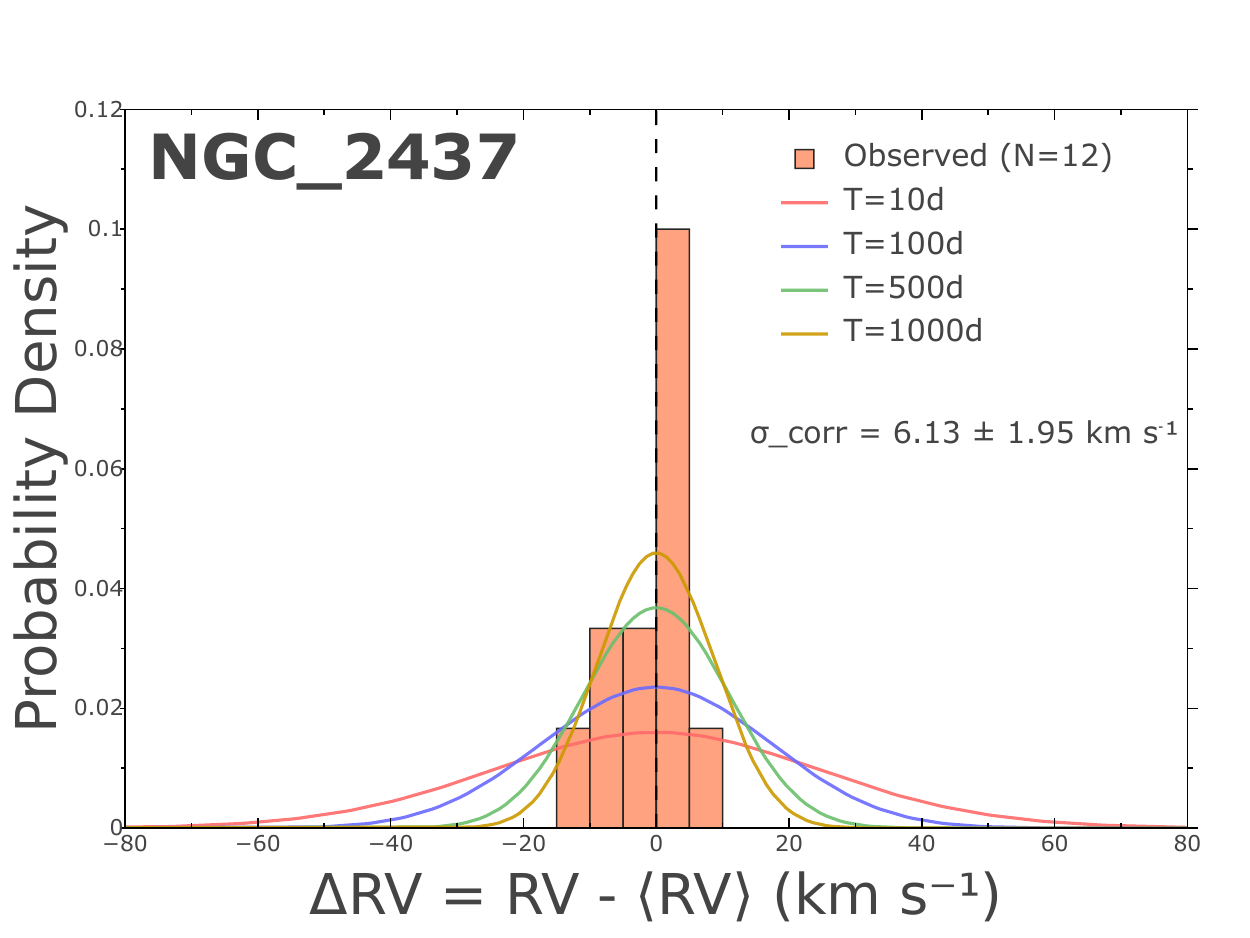}
        \includegraphics[width=0.325\hsize]{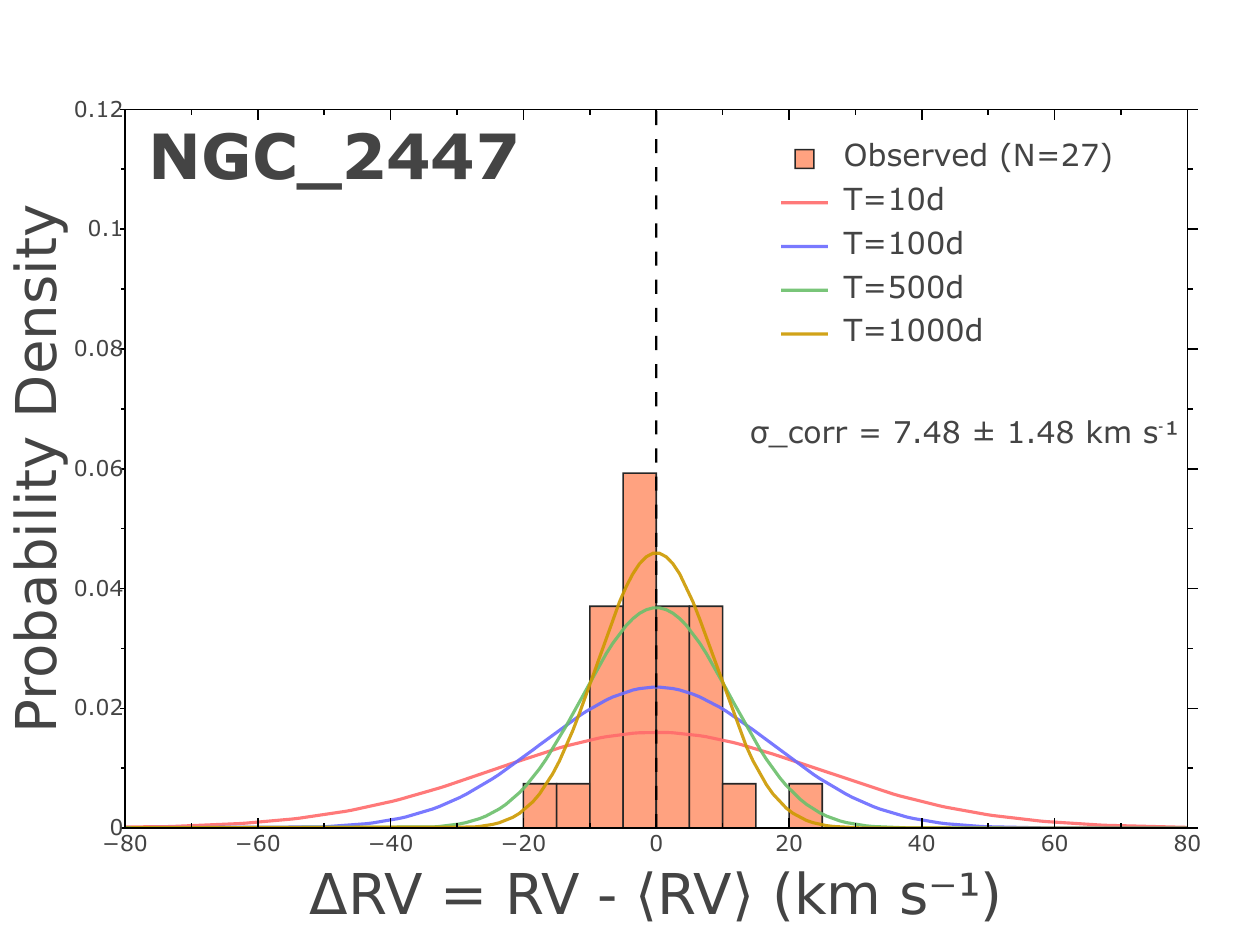}
        \includegraphics[width=0.325\hsize]{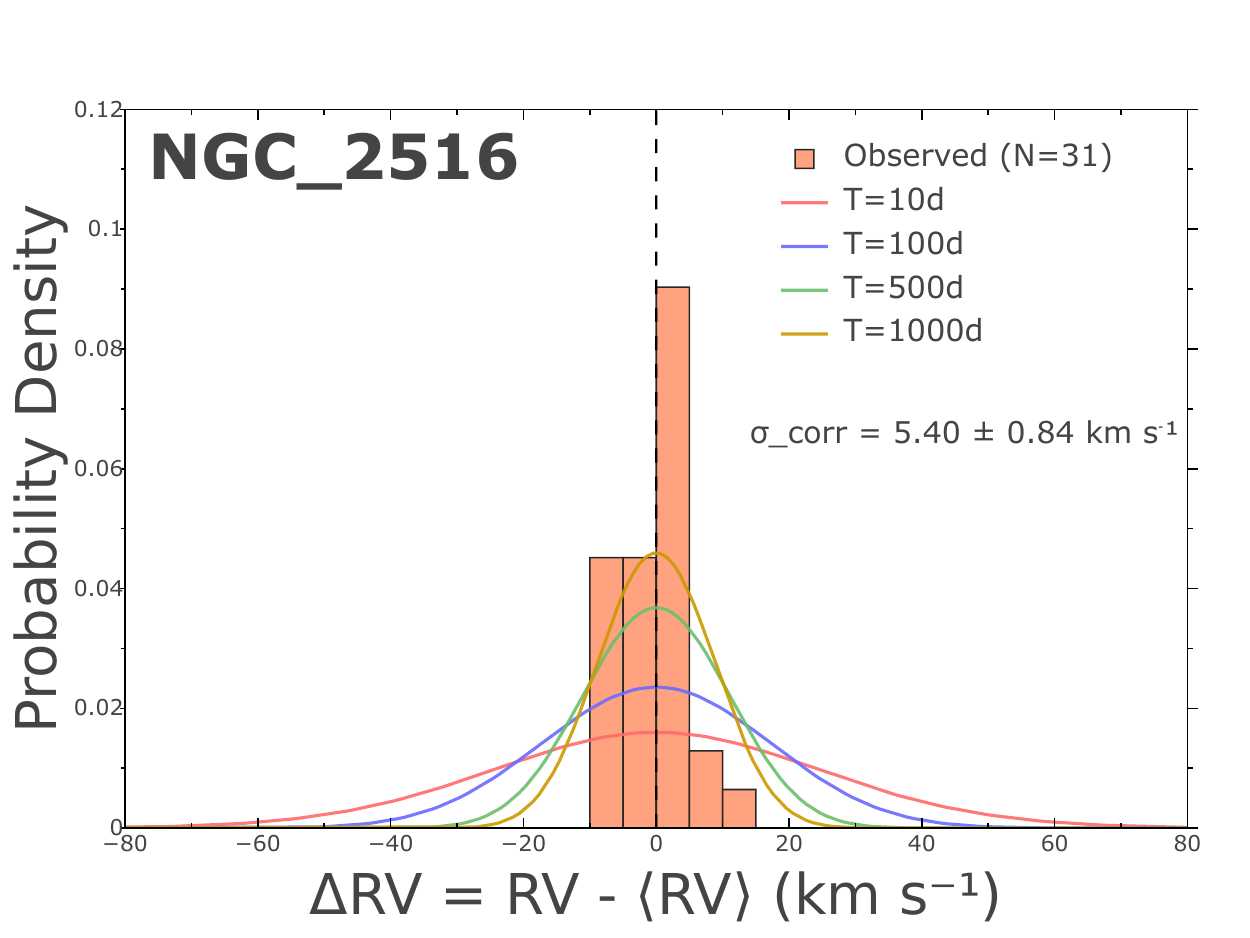}
        \includegraphics[width=0.325\hsize]{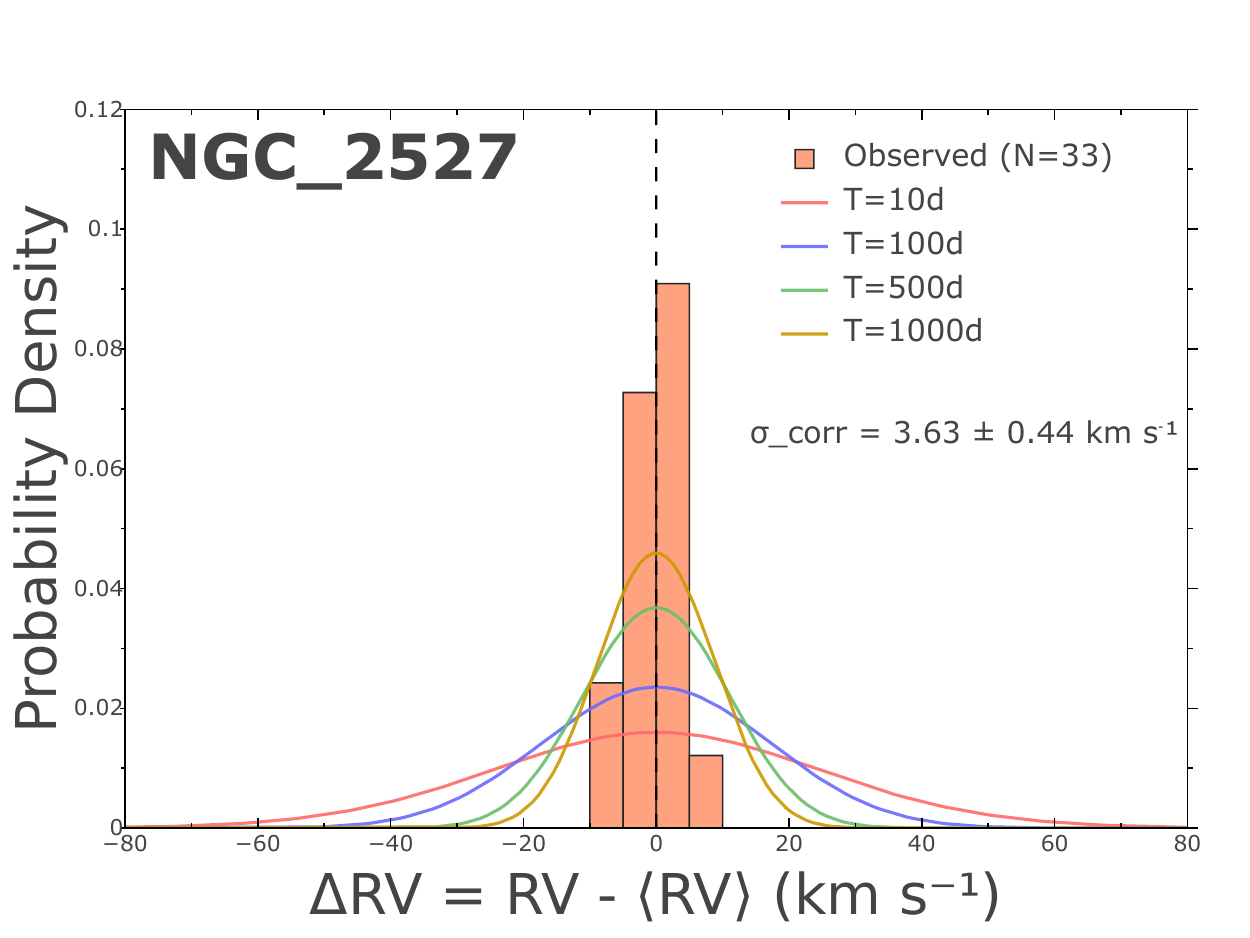}
        \includegraphics[width=0.325\hsize]{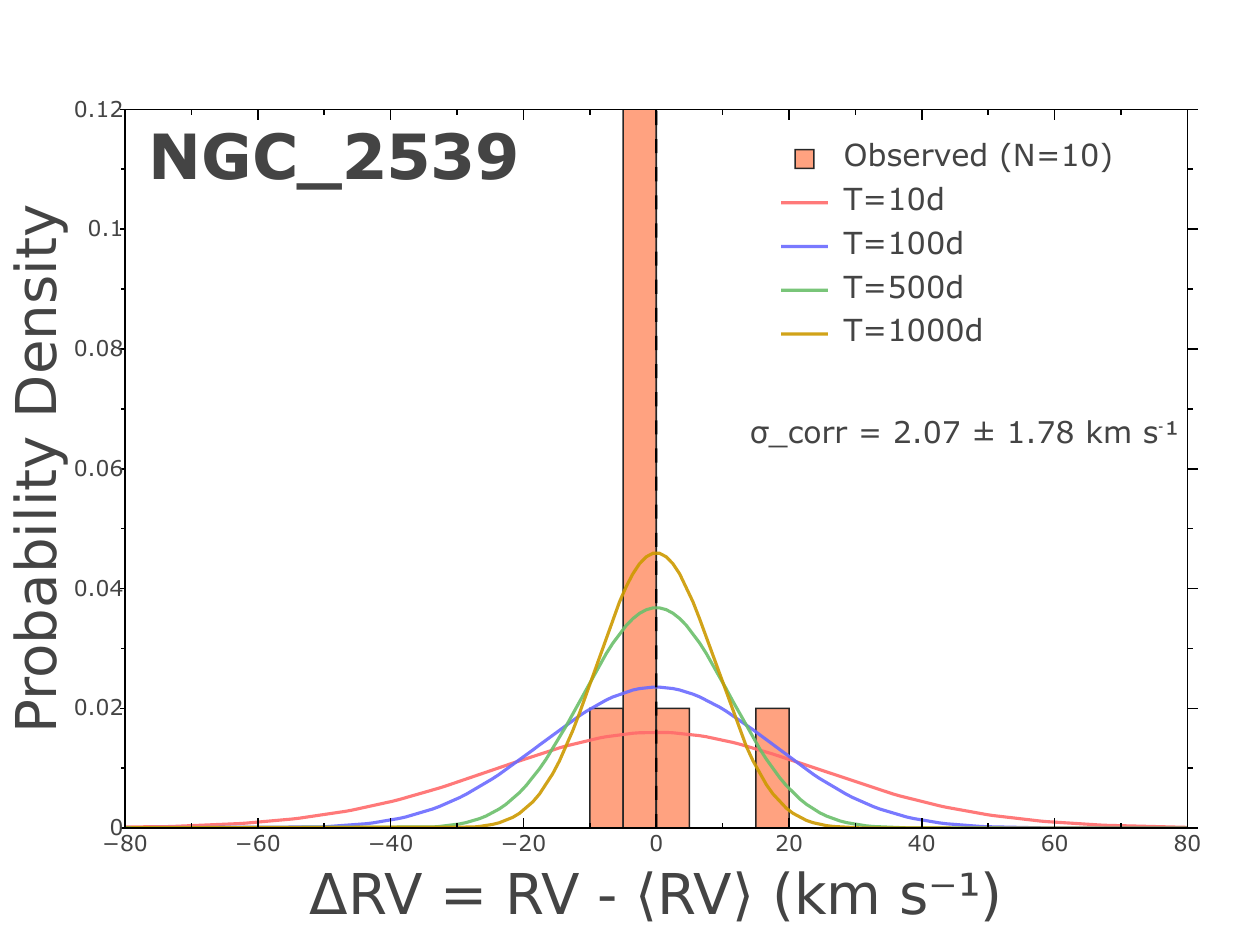}
        \includegraphics[width=0.325\hsize]{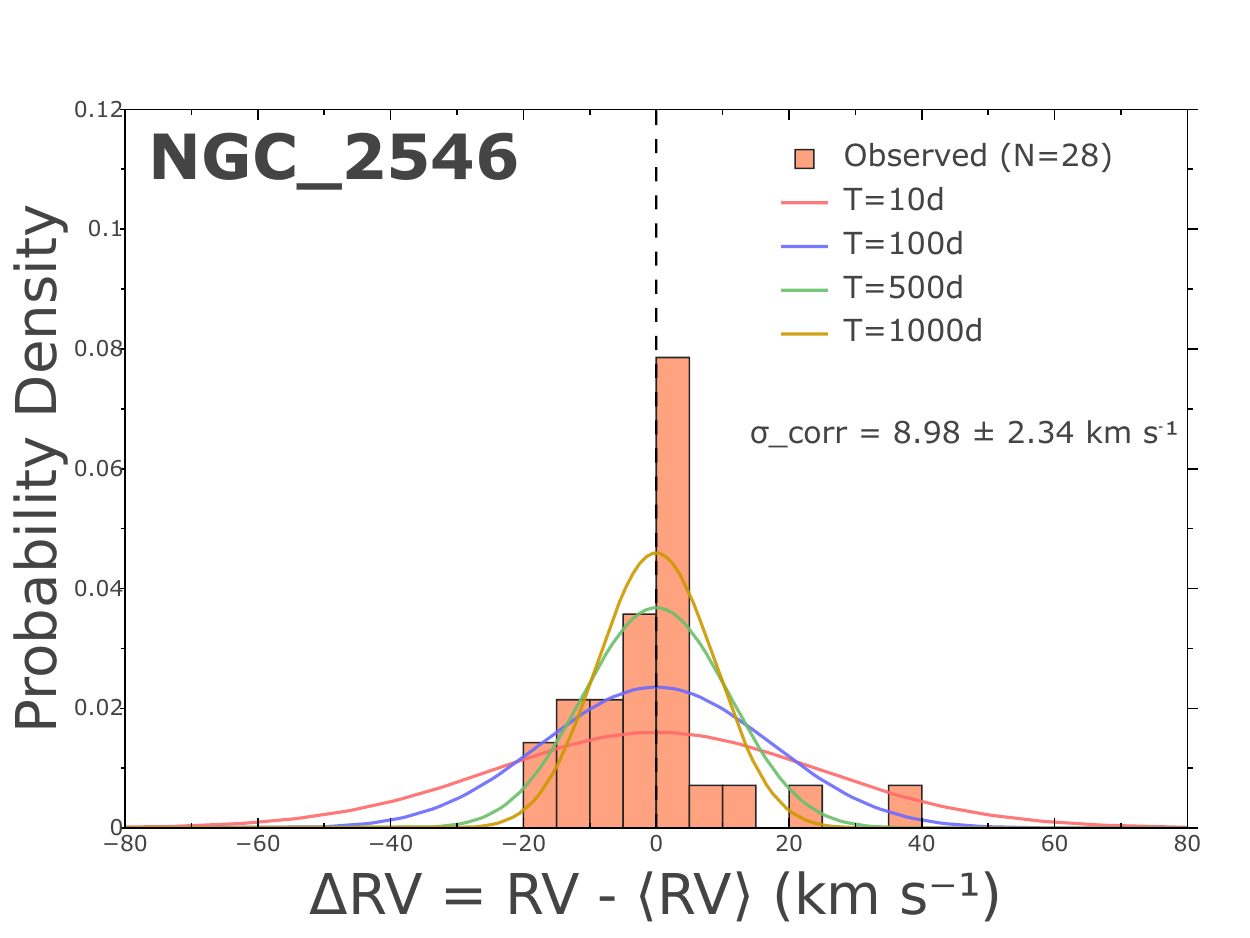}
        \includegraphics[width=0.325\hsize]{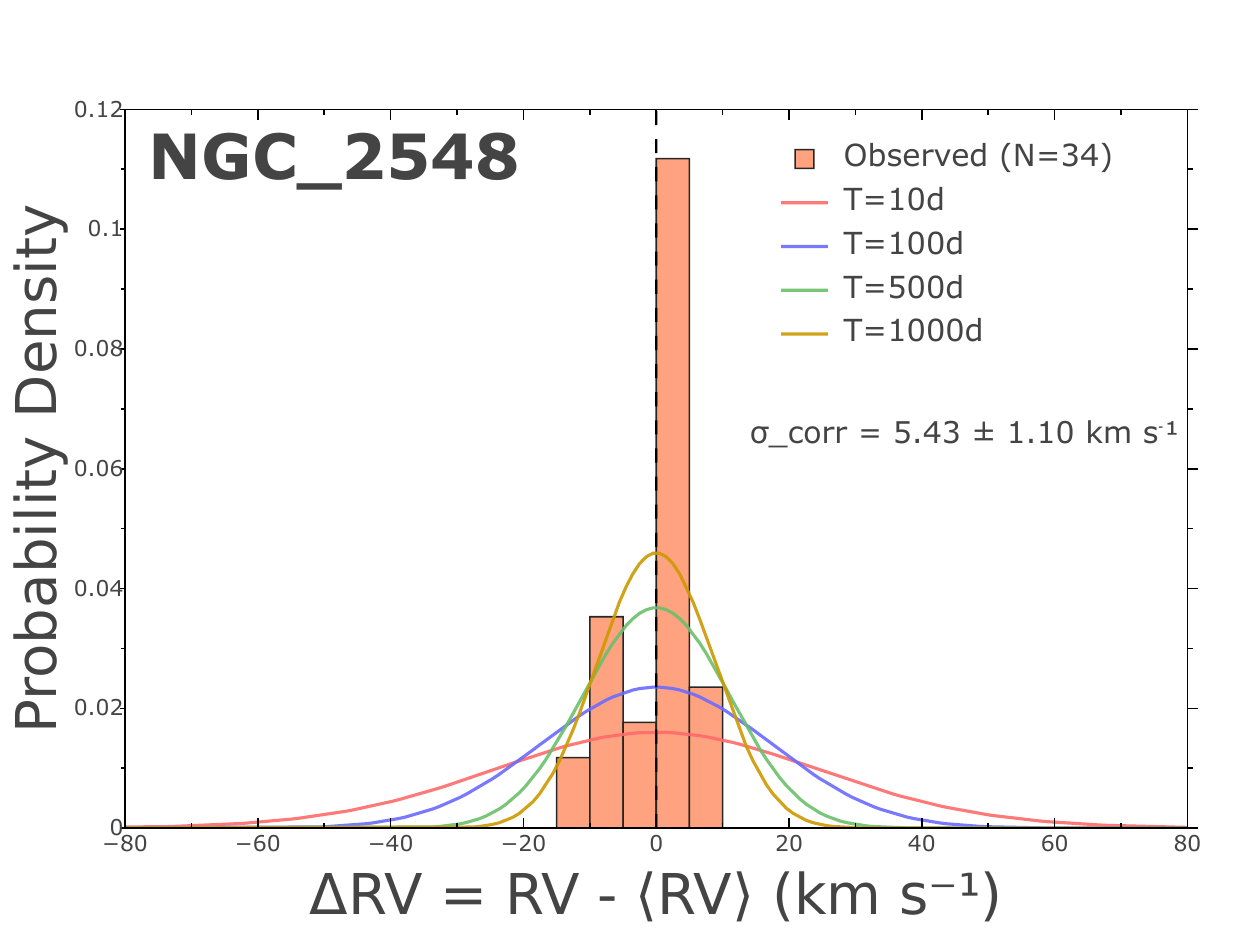}
        \includegraphics[width=0.325\hsize]{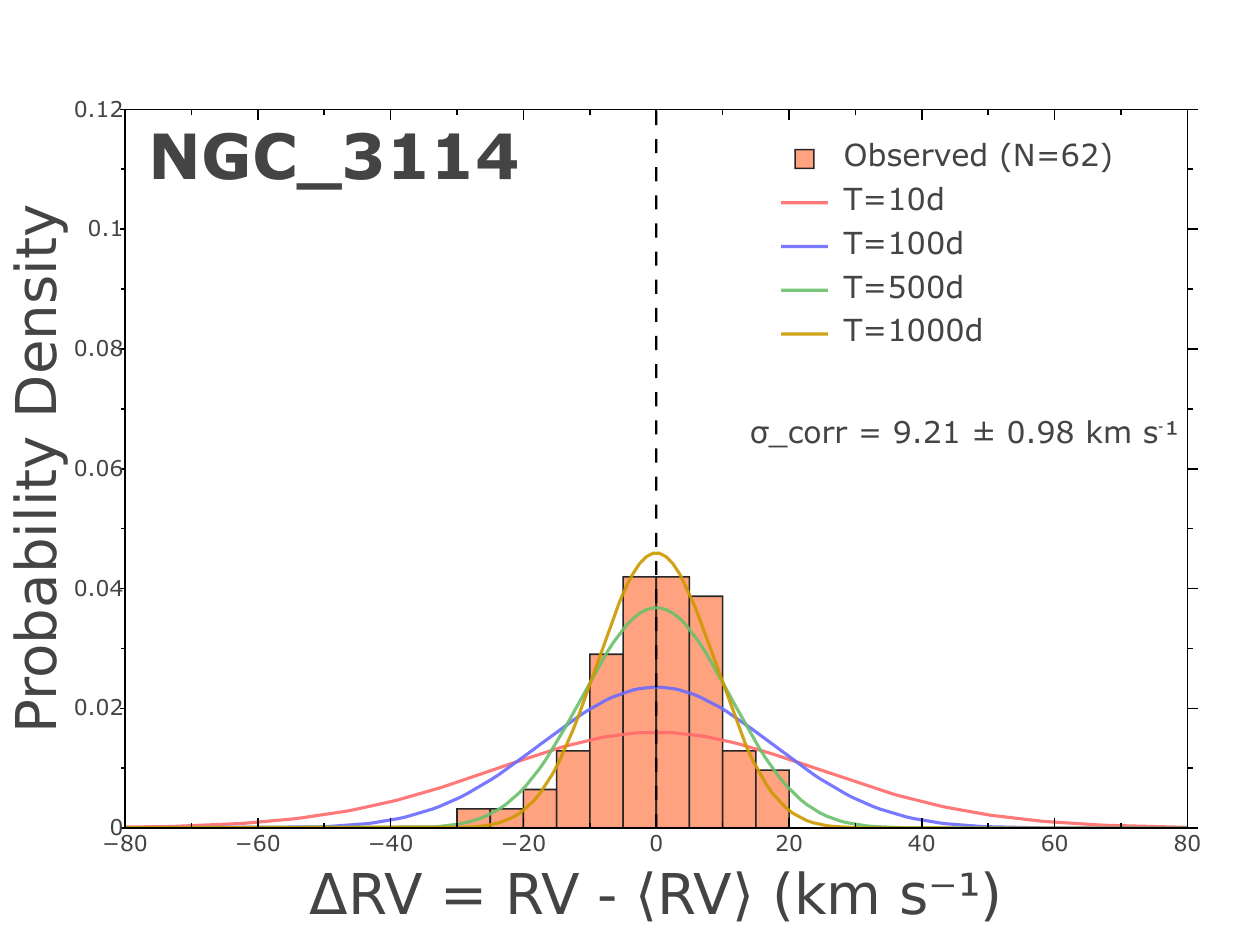}
        \includegraphics[width=0.325\hsize]{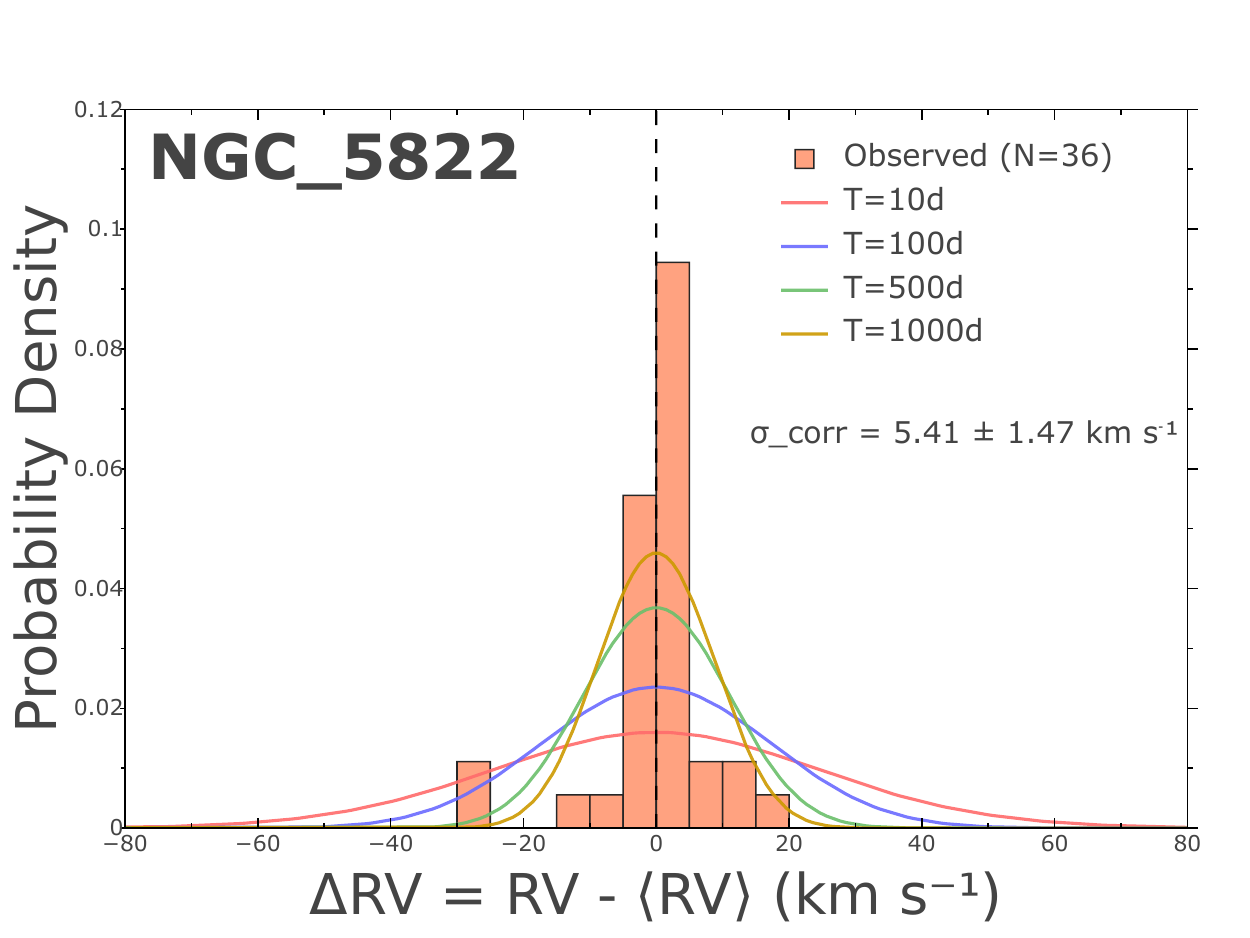}
        \includegraphics[width=0.325\hsize]{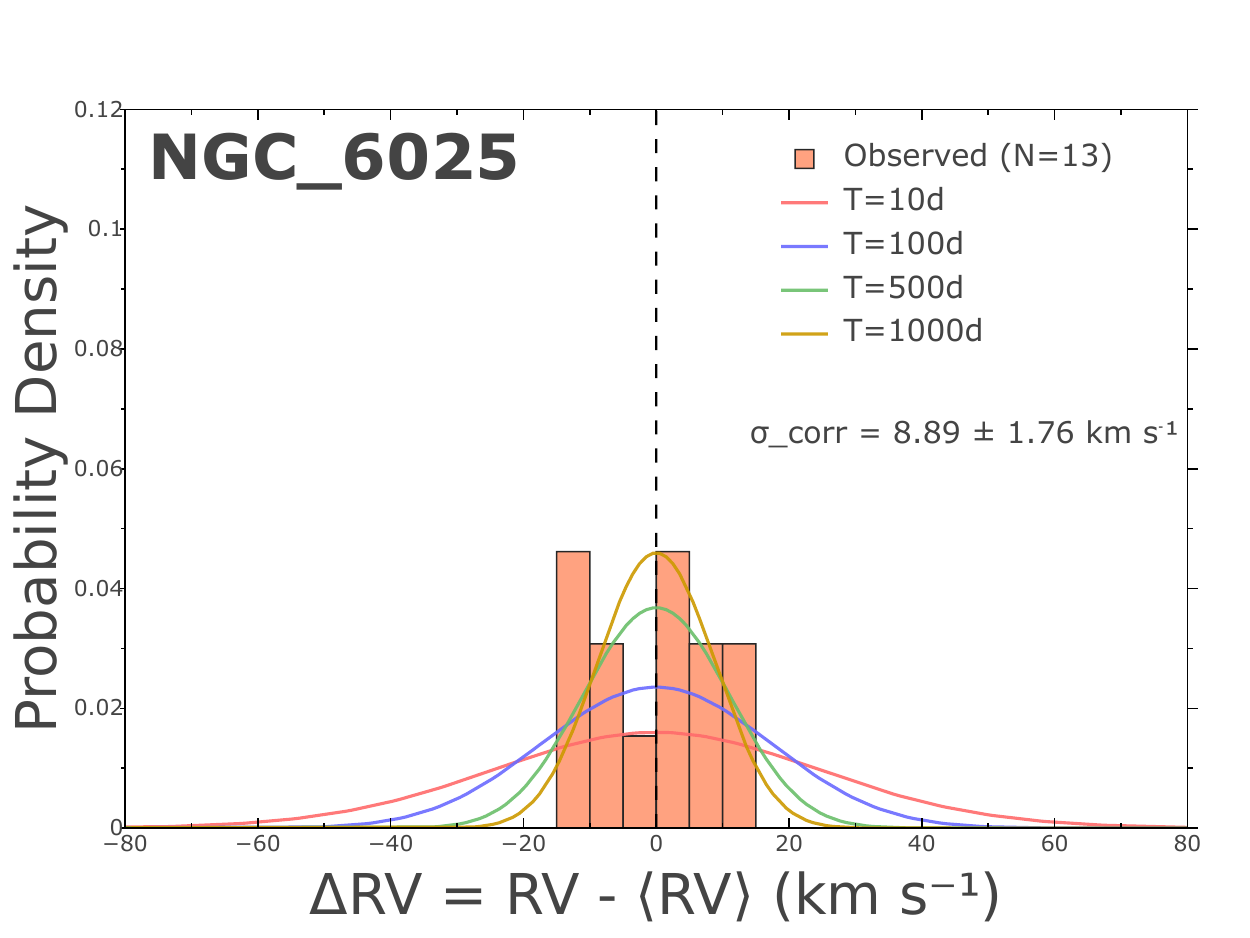}     
        \includegraphics[width=0.325\hsize]{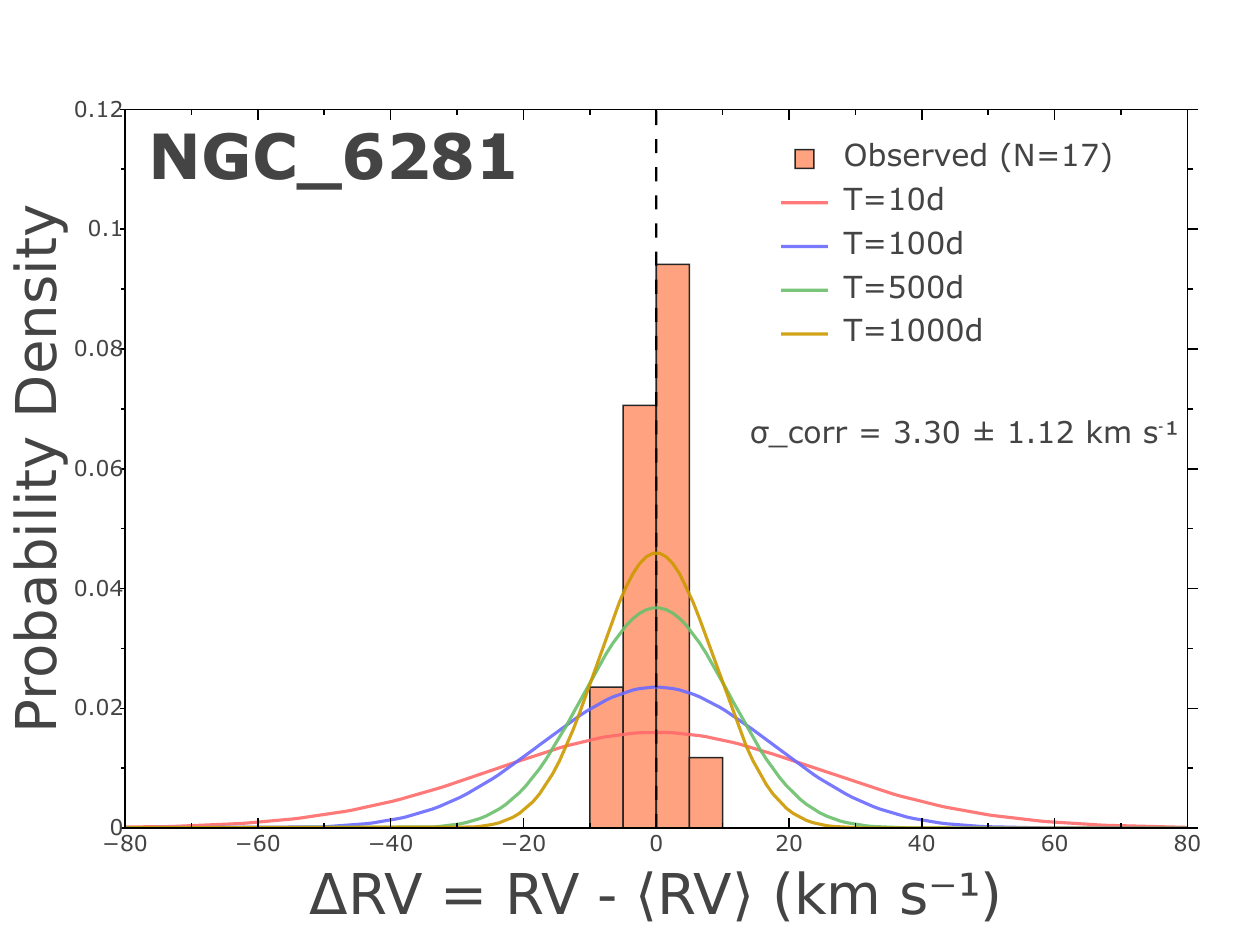}

        \caption{Figure~\ref{RVdispersion_all} continued}
    \end{figure}
    \begin{figure}
        \ContinuedFloat*  
        \centering

        \includegraphics[width=0.325\hsize]{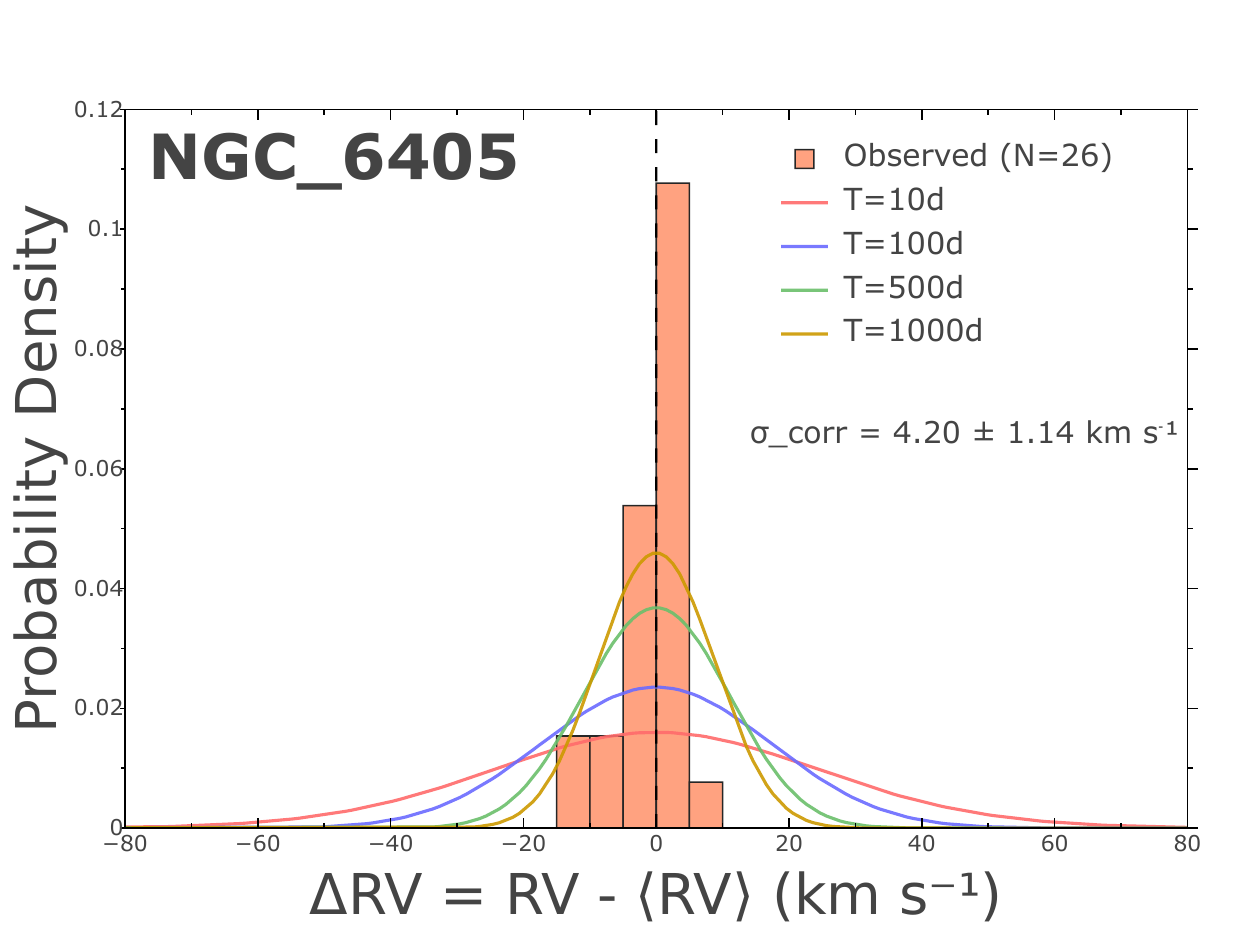}
        \includegraphics[width=0.325\hsize]{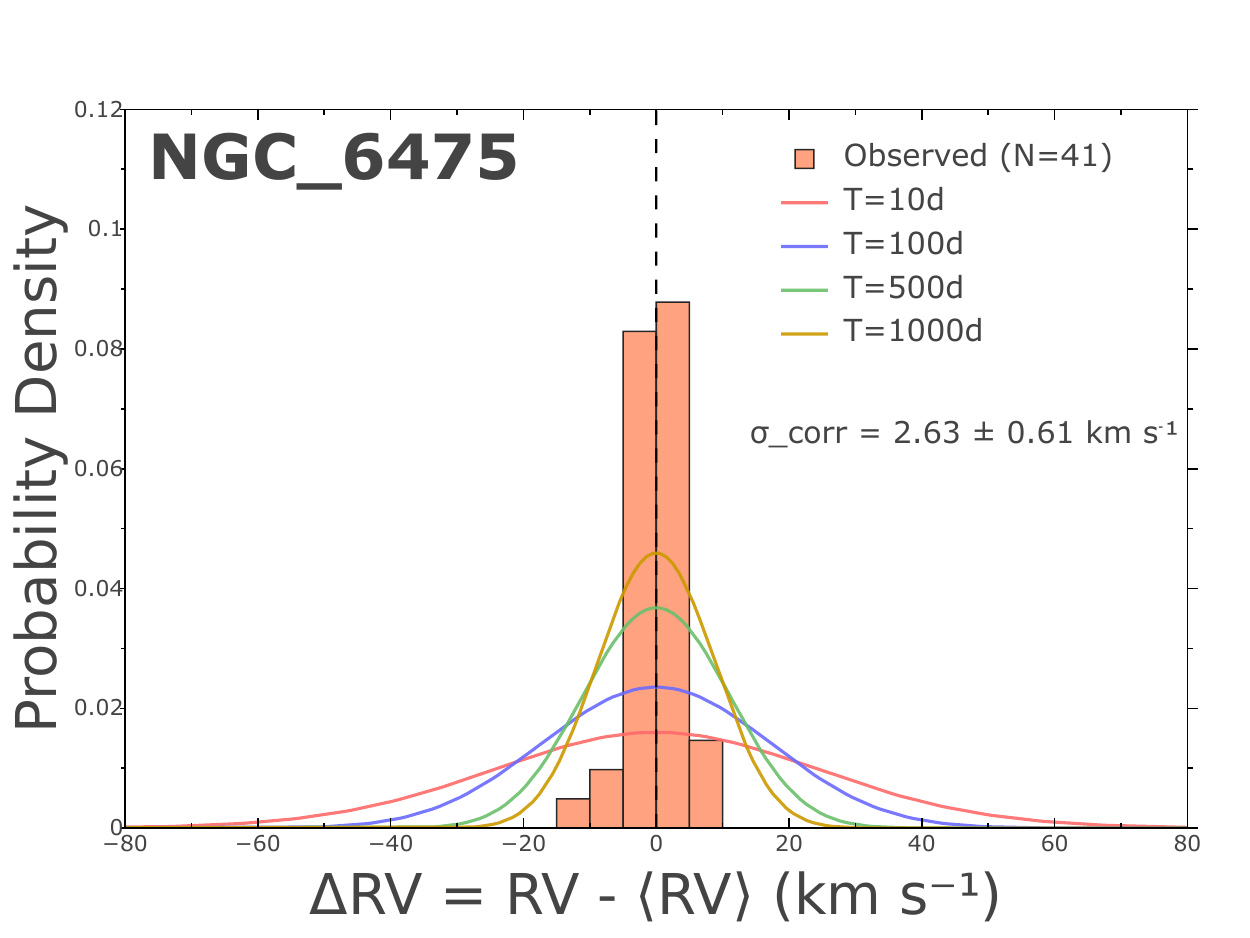}
        \includegraphics[width=0.325\hsize]{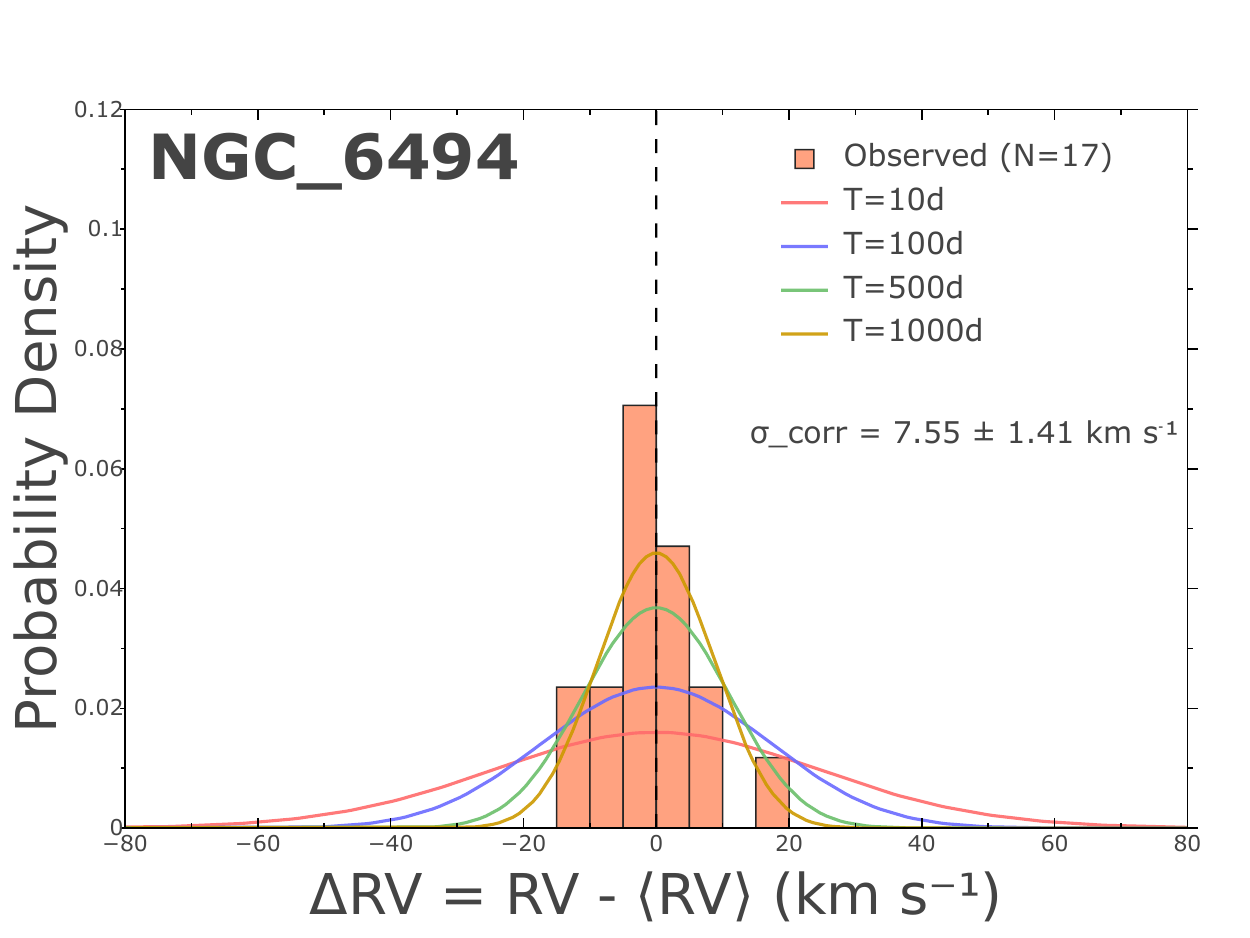}
        \includegraphics[width=0.325\hsize]{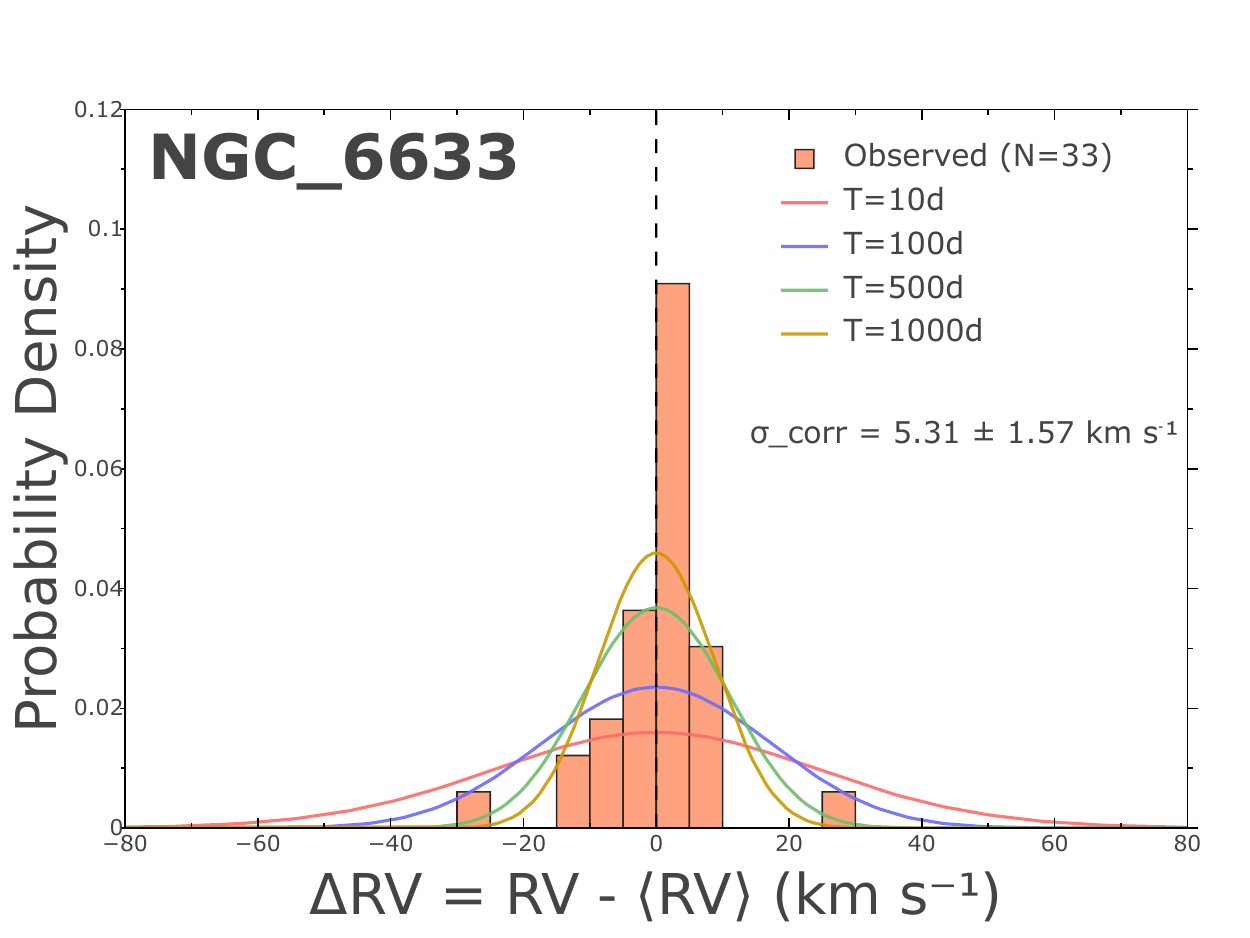}
        \includegraphics[width=0.325\hsize]{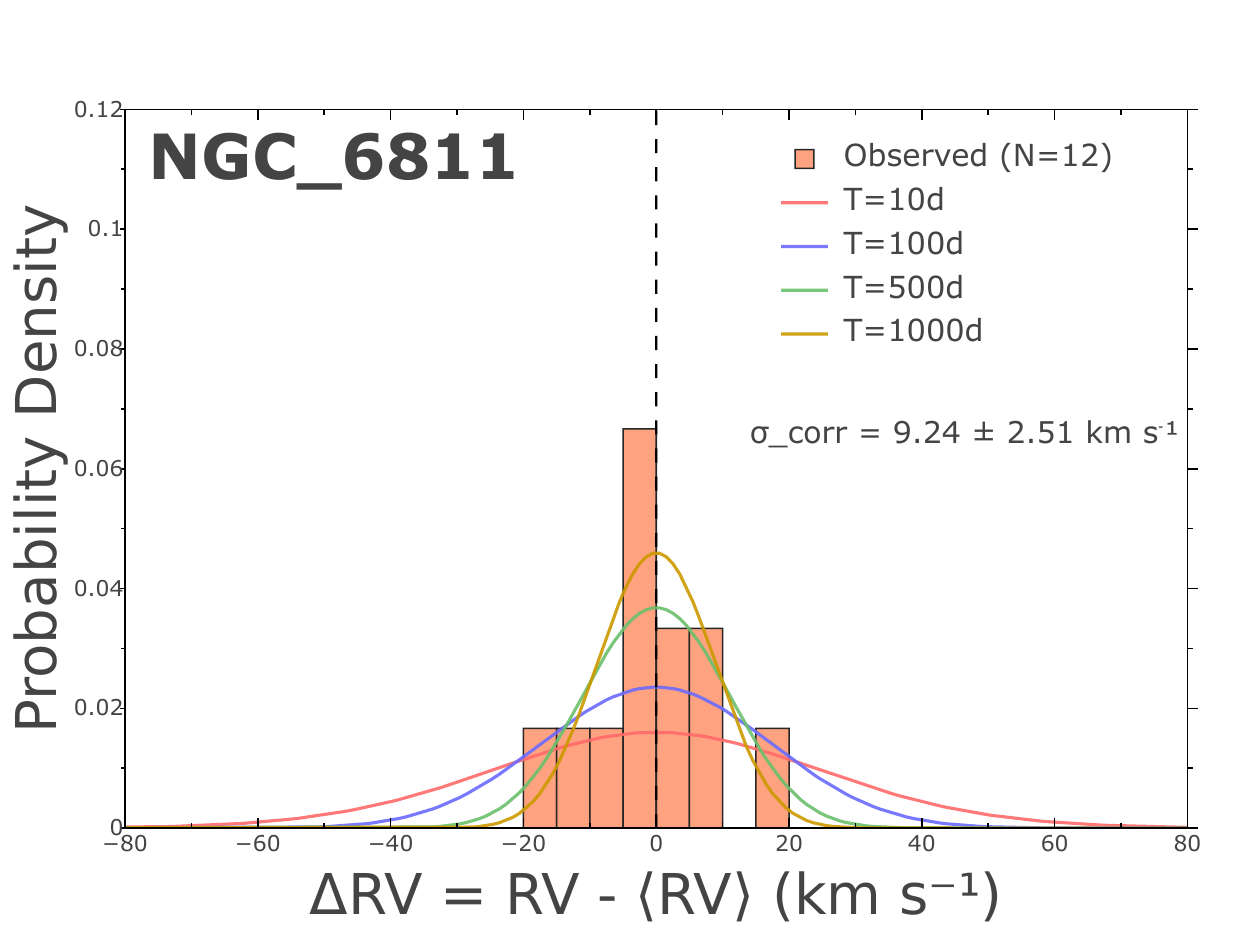}
        \includegraphics[width=0.325\hsize]{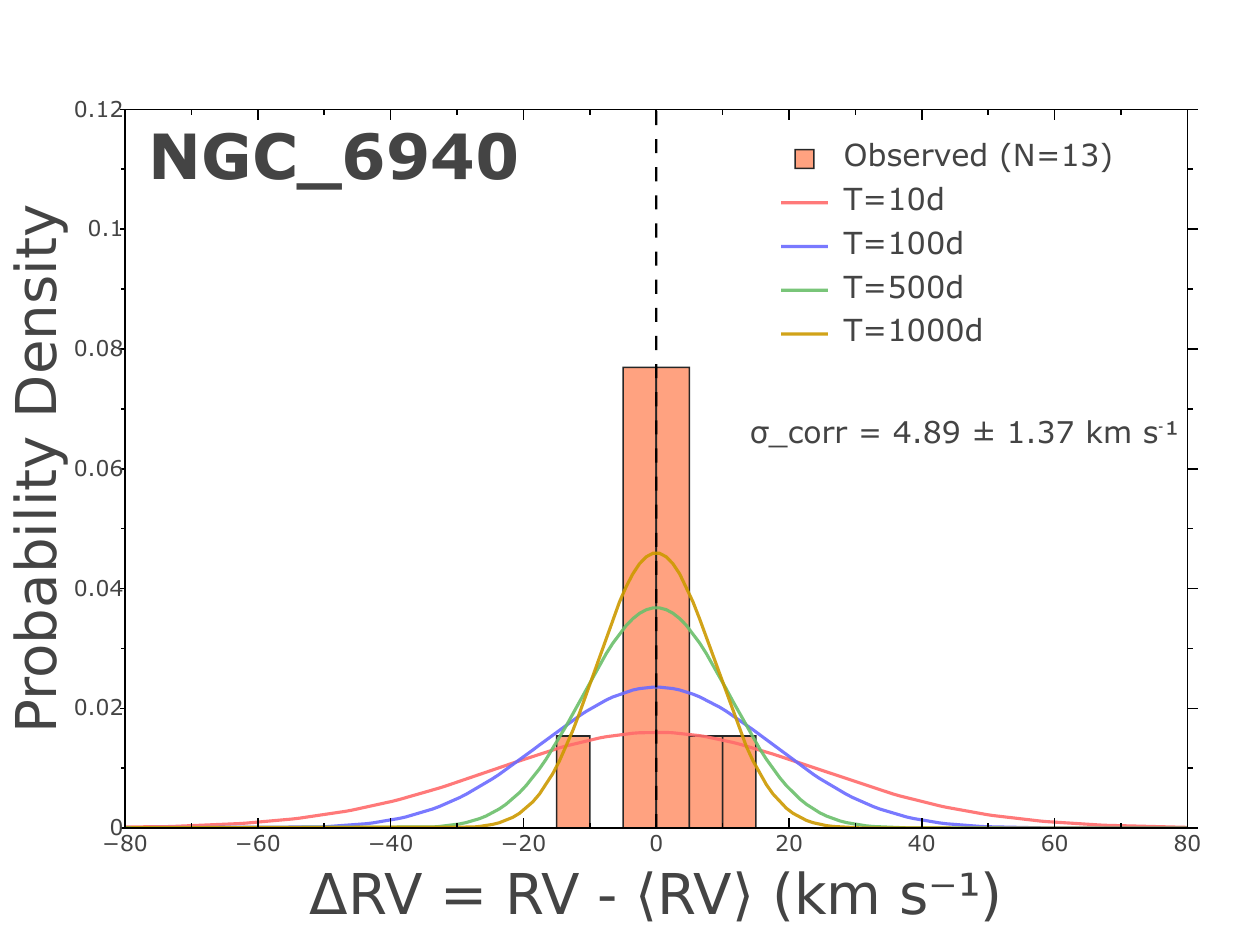}
        \includegraphics[width=0.325\hsize]{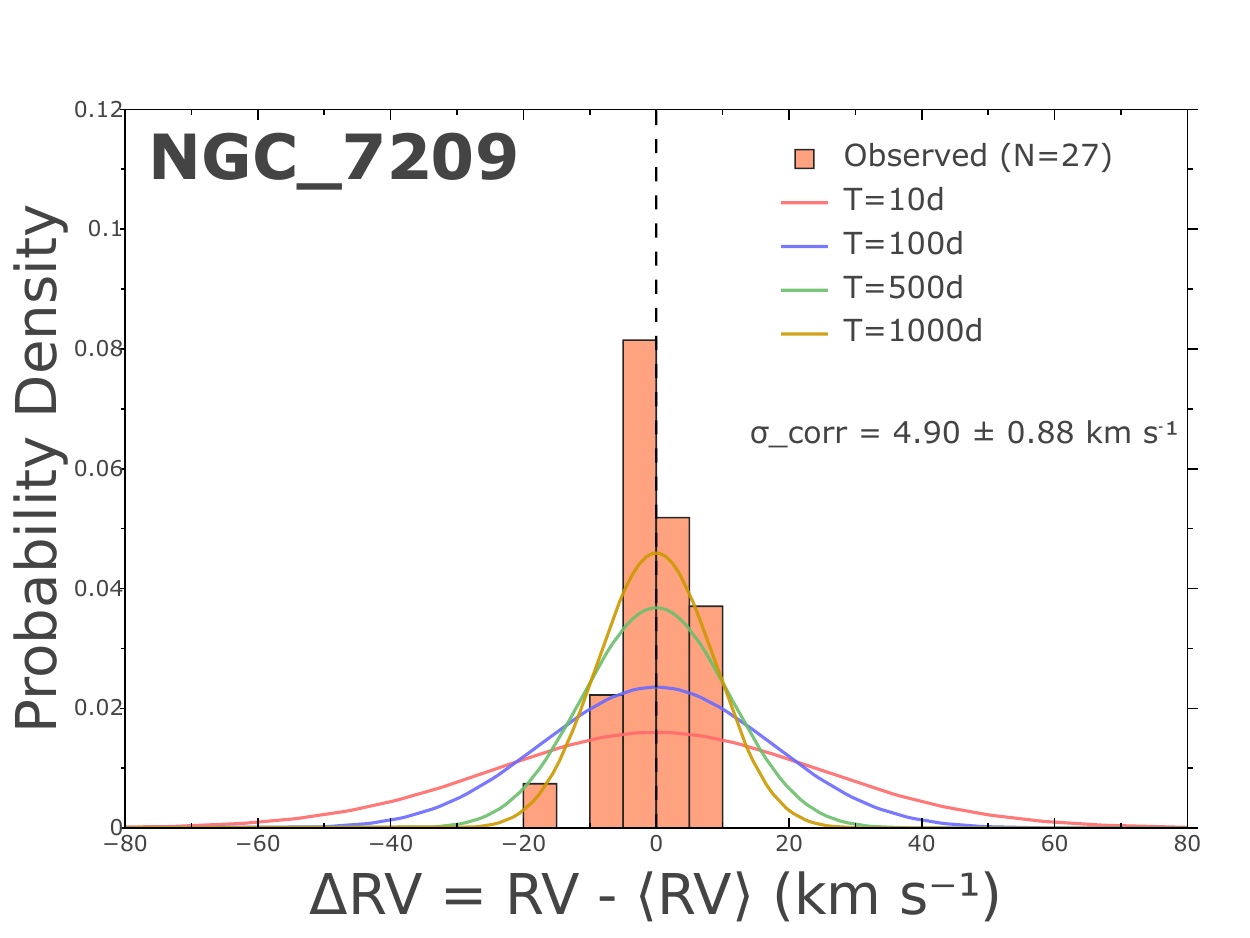}  
        \includegraphics[width=0.325\hsize]{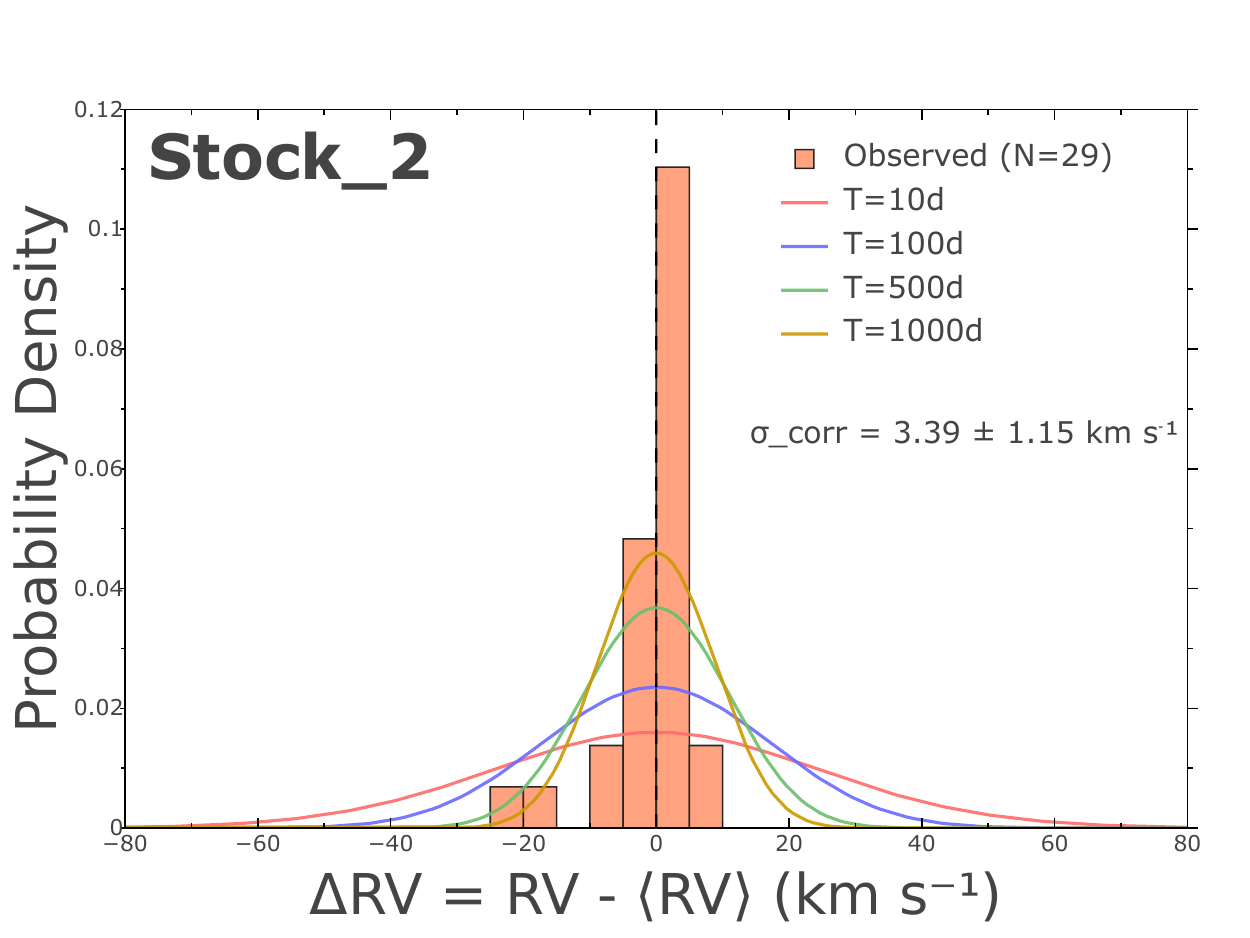}
        \includegraphics[width=0.325\hsize]{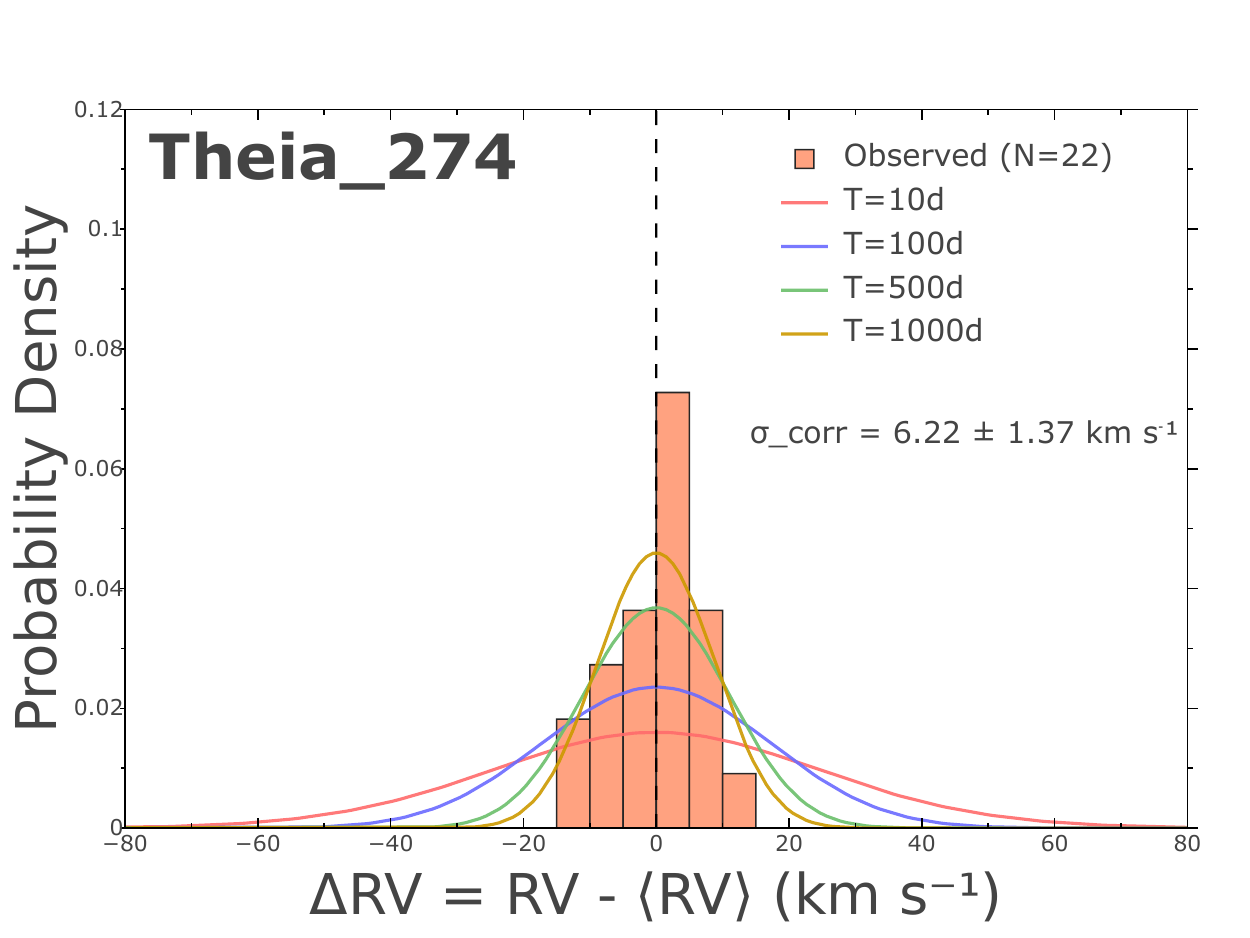}      
        \includegraphics[width=0.325\hsize]{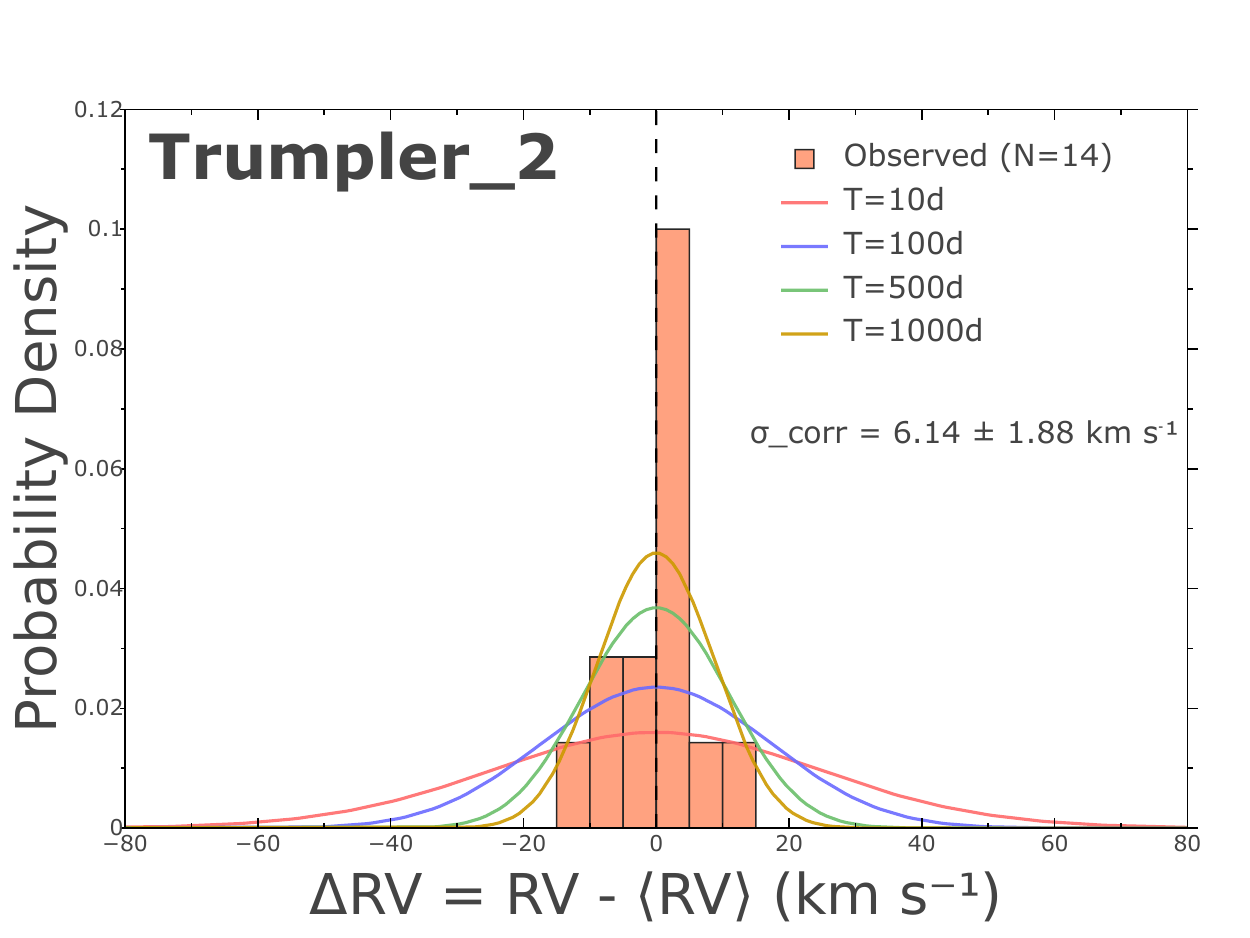}
        \caption*{Figure~\ref{RVdispersion_all} continued}

    \end{figure}
\end{appendix}

\end{document}